\documentclass[12pt]{article}
\usepackage{amsmath}
\usepackage{graphicx,psfrag,epsf}
\usepackage{enumerate}
\usepackage[colorlinks=true,citecolor=blue]{hyperref}
\usepackage{natbib}
\usepackage{amsfonts}
\usepackage{algpseudocode}
\usepackage{algorithm}
\usepackage{xcolor}
\usepackage{tabularx}
\usepackage{graphicx}
\usepackage{pdflscape}
\usepackage{subcaption}
\usepackage[affil-it]{authblk}
\usepackage{url}

\def\spacingset#1{\renewcommand{\baselinestretch}%
{#1}\small\normalsize} \spacingset{1}

\newtheorem{assumption}{Assumption}

\DeclareMathOperator*{\argmax}{argmax}
\DeclareMathOperator*{\argmin}{argmin}
\DeclareMathOperator*{\Conv}{conv}
\DeclareMathOperator*{\Dirac}{Dirac}
\DeclareMathOperator*{\expit}{expit}
\DeclareMathOperator*{\logit}{logit}

\DeclareMathOperator*{\Var}{Var}

\newcommand{\bign}{\boldsymbol{n}}
\newcommand{\bone}{\textbf{1}}
\newcommand{\pit}{\widetilde{\pi}}
\newcommand{\uS}{\overline{S}}
\newcommand{\lNu}{\underline{\mathcal{V}}}

\begin{document}

 \title{\bf Policy learning under constraint: Maximizing a primary outcome while controlling an
adverse event}

\author[1]{Laura Fuentes-Vicente\thanks{Corresponding author: \texttt{laura.fuentes-vicente@inria.fr}}}
\author[1]{Mathieu Even}
\author[2]{Gaëlle Dormion}
\author[1]{Julie Josse}
\author[3]{Antoine Chambaz}

\affil[1]{Inria PreMeDICaL, Inserm, University of Montpellier, France}
\affil[2]{Elixir Health, Paris, France}
\affil[3]{Université Paris Cité, CNRS, MAP5, F-75006 Paris, France}
\maketitle

\bigskip
\begin{abstract}
A medical policy aims to support decision-making by mapping patient
characteristics to individualized treatment recommendations.  Standard
approaches typically optimize a single outcome criterion.  For example,
recommending treatment according to the sign of the Conditional Average
Treatment Effect (CATE) maximizes the policy ``value" by exploiting
treatment effect heterogeneity.  This point of view shifts policy learning towards the
challenge of learning a reliable CATE estimator.  However, in
multi-outcome settings, such strategies ignore the risk of adverse
events, despite their relevance.
PLUC (Policy Learning Under Constraint) 
addresses this challenges by learning an estimator of the CATE that 
yields smoothed policies controlling the probability of an adverse 
event in observational settings. 
Inspired by insights from EP-learning \citep{van2024combining}, PLUC 
involves the optimization of strongly convex Lagrangian criteria over 
a convex hull of functions.  Its alternating procedure iteratively 
applies the Frank-Wolfe algorithm to minimize the current criterion, 
then performs a targeting step that updates the criterion so 
that its evaluations at previously visited landmarks become targeted 
estimators of the corresponding theoretical quantities.
An R package (\href{https://github.com/laufuentes/PLUCR.git}{PLUC-R}) 
provides a practical implementation. We illustrate PLUC's performance 
through a series of numerical experiments.

\end{abstract}

\noindent%
\vfill

\spacingset{1.3} 

\newpage

\section{\centering \textsc{Introduction}}
\label{sec:intro}
Precision medicine leverages individual characteristics to recommend tailored treatments, moving beyond the conventional uniform-treatment paradigm. This shift has fueled a growing literature on learning personalized treatment policies through the assessment of treatment heterogeneity. While some approaches frame the problem as a classification task \citep{zhang2012estimating,zhao2012estimating,athey2021policy}, indirect approaches rely on estimating the Conditional Average Treatment Effect (CATE), the expected difference in outcomes between treatment and control for individuals with a given set of covariates. Such approaches typically employ meta-learners such as the X-, DR-, and R-learners \citep{kunzel2019metalearners,nie2021quasi,kennedy2023towards}. The CATE quantifies which treatment is expected to yield the better outcome for each individual profile. As a result, it naturally leads to treatment policies that assign treatment whenever the estimated effect is positive. Among CATE-based approaches, EP-learning \citep{van2024combining} has emerged as a promising method, addressing key limitations of the DR-learner while retaining the stability of plug-in methods.

However, in personalized treatment, methods that target a single outcome remain fundamentally limited since in many real-world domains, especially healthcare, treatment decisions must be made in the presence of multiple, often competing outcomes. For example, in oncology a clinician aims to maximize tumor cell destruction while simultaneously minimizing harm to healthy tissue. 

Although several studies have addressed multiple outcome frameworks, via composite outcomes \citep{butler2018incorporating}, win-ratio based \citep{pocock2012win,even2025rethinkingwinratiocausal}, through fairness-aware criteria \citep{MR4646602, zhou2023optimal, MR4713926}, budget or resource constraints \citep{zbMATH07938665, sun2021treatment}, or balancing long and short term effects \citep{wang2024pareto}, surprisingly few works explicitly focus on adverse events. Existing contributions on adverse events are confined to randomized trials, which are typically under-powered for high-dimensional, heterogeneous settings where precision medicine thrives, and rely on assumptions that generally do not extend to observational data, thereby limiting their applicability and external validity.

Notable exceptions include \cite{MR3490046} who formulate a constrained optimization problem within SMART (Sequential Multiple Assignment Randomized Trials) designs to maximize treatment efficacy while controlling for adverse effects. However, their reliance on parametric models of conditional distributions limits practical use. Similarly, \cite{wang2018learning} proposes BR-O, a framework designed for RCTs that maximizes clinical benefit while constraining average risk. While this approach demonstrates conceptual strength, its reliance on convex programming algorithms and prior knowledge of treatment assignment mechanisms introduces computational and practical challenges, particularly for observational settings. Extensions to multiple treatments \citep{MR4186844}, multiple treatment stages \citep{liu2024learning}, and individual-level risk constraints \citep{MR4728668} continue this line of work but remain primarily tied to randomized settings. Finally, we note that similar questions have also been addressed from a reinforcement learning perspective \citep{MR3251666, kumar2021benchmark, afshari2019constrained}, often relying on Pareto optimality~\citep{MR3595145, kone2025constrained}.

In this work, we introduce a unified theoretical (Sections~\ref{sec:setup} and \ref{subsec:oracular}) and statistical (Section~\ref{subsec:statistical}) framework for learning treatment policies that maximize a primary outcome while explicitly controlling the probability of an adverse event. Adopting an indirect approach, we estimate CATEs through convex risk minimization over a convex hull of functions, and introduce a novel non-parametric and versatile class of smooth probabilistic policies. Our non-parametric policies naturally quantify treatment assignment confidence and facilitate both theoretical analysis and practical optimization.

Adverse-event control is incorporated through a Lagrangian criterion. Our proposed algorithm, PLUC (Policy Learning Under Constraint, Section~\ref{subsubsec:PLUC}) is non-parametric, highly flexible, and explicitly designed for observational settings. PLUC leverages the adaptability of super learners and an alternating optimization scheme. This procedure combines the Frank–Wolfe algorithm~\citep{frank1956algorithm}, providing efficient, projection-free exploration of the convex hull, with a targeting step~\citep{van2006targeted,TMLE2011,TMLE2018} that progressively aligns the empirical objective with its theoretical counterpart at previously visited iterates. We additionally establish convergence guarantees for the Frank–Wolfe algorithm tailored to our setting.

To facilitate practical adoption, we provide an accessible implementation of PLUC through the PLUC-R package, completed with a vignette. We evaluate the method through extensive numerical experiments on synthetic data (Section~\ref{sec:num:exp}).

\section{\centering \textsc{Problem setup}}
\label{sec:setup}

\subsection{A quick overview of policy learning}
\label{subsec:policy:learning}

\paragraph*{Statistical modeling.}

We have access to a sample of $n$ independent and identically distributed (i.i.d.) observations, $O_1,...,O_n$, drawn from a law $P_0$ on $\mathcal{O} = \mathcal{X} \times \{0,1\} \times [0,1] \times \{0,1\}$, where $\mathcal{X}$ is a subset of $[0,1]^{d}$. The law $P_{0}$ is known to  belong to a statistical model $\mathcal{M}$, a collection of laws on $\mathcal{O}$.  A generic data structure decomposes as $O = (X,A,Y,\xi)$, with $X \in \mathcal{X}$ a vector of covariates, $A \in \{0,1\}$ a binary treatment assignment indicator, $Y\in [0,1]$ a primary outcome of interest (higher values indicating better outcomes), and $\xi \in \{0,1\}$ an indicator of occurrence of an adverse event. 

By construction, for any $P\in\mathcal{M}$  and $a \in \{0,1\}$, the following
conditional expectations are  well defined on the support of  the marginal law
of $X$  under $P$ (a law  denoted as $P_{X}$): $\mu_{P}(a,X)  = E_P[Y|A=a,X]$,
$\nu_P(a,X)=E_P[\xi=1|A=a,X]$,       and        $e_P(a,X)=P(A=a|X)$.       The
infinite-dimensional features  $\mu_{P}, \nu_{P},  e_{P}$ of  $P$ will  play a
central    role    in    our    study ($e_{P}$ is 
called the propensity score under $P$).   To    avoid    cluttered    notation,
$\mu_{P_{0}},      \nu_{P_{0}},      e_{P_{0}}$     are      also      denoted
$\mu_{0}, \nu_{0}, e_{0}$.

\paragraph*{Policies and their values.}
A policy $\pit$ is an element of a user-supplied set $\Pi \subset [0,1]^{\mathcal{X}}$ (throughout the manuscript, we denote by $V^U$ the class of functions mapping $U$ to $V$). It maps any vector $x\in \mathcal{X}$ of covariates to a treatment assignment probability. The value of a policy $\pit \in \Pi$ under any $P\in\mathcal{M}$ is defined as
\begin{equation}
    \label{eq:value}
    \mathcal{V}_{P}(\pit) = E_{P} [\pit(X) \cdot \mu_{P}(1,X) + (1-\pit(X)) \cdot \mu_{P}(0,X)].
\end{equation}
It can be interpreted as the (causal) average outcome in a world where (possibly contrary to facts) $A$ would be drawn conditionally on $X$ from the Bernoulli law with parameter $\pit(X)$. This statement can be clarified in terms of potential outcomes~\citep{rubin2005causal}. Let $\mathbb{P}_{\pit}$ be a law on $\mathcal{X} \times [0,1]^{2} \times \{0,1\}^{2} \times \{0,1\}$ from which a generic sample decomposes as $(X, Y(0), Y(1), \xi(0), \xi(1), A)$. Here, $Y(a)$ and $\xi(a)$ are interpreted as the outcome and adverse event indicator when treatment $a$ is imposed, for each $a \in \{0,1\}$. It is further assumed that, under $\mathbb{P}_{\pit}$, \textit{(a)} the marginal law of $X$ equals $P_{X}$; \textit{(b)} the conditional law of $A$ given $X$ is the Bernoulli law with parameter $\pit(X)$; \textit{(c)} the conditional law of $A\cdot Y(1) + (1-A) \cdot Y(0)$ given $(A,X)$ coincides with that of $Y$ given $(A,X)$ under $P$. Then, the value of $\pit$ under $P$ defined in~\eqref{eq:value} satisfies  
\begin{equation*}
     \mathcal{V}_{P}(\pit) = E_{\;\mathbb{P}_{\pit}}[Y(A)].
\end{equation*} 
 
The causal interpretation of the value of $\pit$ under $P$ justifies the interest in policies that maximize $\pit \mapsto \mathcal{V}_{P}(\pit)$ over $\Pi$. Such policies are said value-optimal. Note that, if
\begin{equation}
    \label{eq:CATE:Y}
    \Delta\mu_{P}(\cdot) = \mu_{P}(1,\cdot) - \mu_{P}(0,\cdot)   
\end{equation}
(for any $P \in \mathcal{M}$, and also denoted by $\Delta\mu_{0}$ if $P=P_{0}$), then maximizing  $\pit \mapsto \mathcal{V}_{P}(\pit)$ is equivalent to maximizing  $\pit \mapsto E_{P} [\pit(X) \cdot \Delta\mu_{P}(X)]$. Note that any $\Delta\mu_{P}$ takes its values in $[-1,1]$.
 
\paragraph*{Direct and indirect policy learning.}

Learning a value-optimal policy can be performed directly or indirectly. Both kinds of approaches hinge on the fact that
\begin{equation}
    \label{eq:pi:star}
    \pi_{0}: x \mapsto \bone\{\Delta\mu_{0}(x) > 0\}
\end{equation}
maximizes $\pit \mapsto \mathcal{V}_{P_{0}}(\pit)$ over $[0,1]^{\mathcal{X}}$.

The direct approach consists of performing a classification task, aiming to predict  $\bone\{\Delta\mu_{0}(X)>0\}$ given $X$. This can be done, for instance, via optimal classification \citep{zhang2012estimating}, outcome weighted learning \citep{zhao2012estimating,zhou2017residual,montoya2023optimal}, weighted classification \citep{athey2021policy}.

Alternatively, the indirect approach focuses on the estimation of the 
heterogeneous treatment effect $\Delta\mu_{0}$ \citep{wager2018estimation, 
kunzel2019metalearners, nie2021quasi,kennedy2023towards} and uses a 
plug-in estimator of \eqref{eq:pi:star}. We recall that, given a 
parameter $\Phi$, viewed as a functional from the statistical
model $\mathcal{M}$ to a parameter space $\Theta$, a substitution (or
plug-in) estimator of $\Phi(P)$ for $P \in \mathcal{M}$ is simply
obtained by replacing the unknown law $P$ with some $Q \in \mathcal{M}$ 
whose $\Phi$-relevant nuisance components have been estimated from data
sampled under $P$. Importantly, $Q$ need not be fully specified; only
those parts of it that matter for evaluating $\Phi$ must be estimated.

In general, learning $\Delta\mu_{0}$  poses challenges such as the need to carry out non-convex optimization or the existence of estimates of $\Delta\mu_{0}(x)$ beyond the interval $[-1,1]$. \cite{van2024combining} have introduced EP-learning, a novel approach based on an efficient plug-in (EP) risk estimator for directly estimating infinite-dimensional features of $P_{0}$ identified as risk minimizers. For instance, EP-learning can be developed to estimate $\Delta\mu_{0}$ from i.i.d.\ data drawn from $P_{0}$~\cite[][Example~1]{van2024combining}. The simpler version of EP-learning can be summarized as follows in the context of our study. One chooses a convex working model $\Psi  \subset [-1,1]^{\mathcal{X}}$ and decides to target $\Delta\mu_{0}$ by estimating its ``projection'' on $\Psi$. The projection is characterized as the minimizer over $\Psi$ of the 1-strongly convex risk functional $R_{0} = R_{P_{0}}$, where, for any $P \in \mathcal{M}$, 
\begin{equation}
    \label{eq:OTR_risk}
    \psi \mapsto R_{P}(\psi) = E_{P}\left[\psi(X)^2 - 2\psi(X)\cdot\Delta\mu_{P}(X)\right].
\end{equation} 
The 1-strong convexity of $R_{0}$ is shown in Section~\ref{subsec:more:arguments}. The risk functional must be estimated. An initial estimator of $\Delta\mu_{0}$ (not necessarily an element of $\Psi$) is updated in such a way that the resulting plug-in estimator $R_{n}$ of $R_{0}$ is uniformly targeted towards $R_{0}$, in the sense that, under mild assumptions, $R_{n}(\psi)$ is an efficient estimator of $R_{0}(\psi)$ uniformly in $\psi\in\Psi$. The updating procedure involves a cross-fitting scheme. 

The main objective of our study is to learn an optimal policy. We adopt an indirect policy learning approach. Therefore, the estimation of $\Delta\mu_{0}$ is an important secondary objective. Although we do not straightforwardly apply EP-learning for this purpose, our methodology is inspired from its core ideas.

\subsection{Policy learning under constraint}
\label{subsec:pluc}

The optimal policy that we aim to learn is defined as a maximizer of the value function subject to a constraint. Fix $\alpha \in [0,\tfrac{1}{2}]$. For any $P\in\mathcal{M}$ and policy $\pit\in [0,1]^{\mathcal{X}}$, let $\Delta\nu_{P}(\cdot) = \nu_{P}(1,\cdot) - \nu_{P}(0,\cdot)$, and 
\begin{equation}
    \label{eq:S}
    S_{P}(\pit) = E_{P} \left[\pit(X) \cdot \Delta\nu_{P}(X)\right] - \alpha.
\end{equation}
The ideal optimal policy of interest is defined as any element of
\begin{equation}
    \label{eq:ideal:policy}
    \argmax\left\{\mathcal{V}_{P_{0}}(\pit) : \pit \in [0,1]^{\mathcal{X}} \text{ s.t. } S_{P_{0}}(\pit) \leq 0\right\}.
\end{equation}
Let $S_{0}=S_{P_{0}}$. We note that, if $\pit$ maps $\mathcal{X}$ to $\{0\}$, then $S_{0}(\pit) = -\alpha \leq 0$. Therefore, the $\argmax$ is not the empty set. 

The meaning of the constraint in \eqref{eq:ideal:policy} is easily explained in terms of potential outcomes. For any $\pit \in [0,1]^{\mathcal{X}}$,
\begin{equation*}
    S_{P}(\pit) = \mathbb{P}_{\pit} [\xi(A)=1] - \mathbb{P}_{\pit} [\xi(0)=1] - \alpha. 
\end{equation*}
Therefore, requesting that $S_{P}(\pit) \leq 0$ in \eqref{eq:ideal:policy} can be interpreted as follows: we focus on policies such that the probability that the adverse event occur in a world where $A$  would be drawn  conditionally on $X$ from the Bernoulli law with parameter $\pit(X)$ does not exceed $\alpha$ plus the baseline probability that the adverse event occurs in a world where treatment $a=0$ is imposed. 

From now on, we make the following assumption.
\begin{assumption}
    \label{assum:delet_effect}
    For any $P \in \mathcal{M}$, $\Delta\nu_{P}\geq 0$.
\end{assumption}
In terms of potential outcomes, a sufficient condition for Assumption~\ref{assum:delet_effect} to hold is that $\mathbb{P}_{\pit} [\xi(1)=1| X] \geq \mathbb{P}_{\pit} [\xi(0)=1|X]$ $P$-almost surely: given $X$, the conditional probability of occurrence of the adverse event is larger when $a=1$ is imposed compared to when $a=0$ is imposed. Therefore, another sufficient condition  is that $\xi(1) \geq \xi(0)$ $\mathbb{P}_{\pit}$-almost surely: if the adverse event occurs in a world where one imposes $a=0$, then it also does in a world where $a=1$ is imposed. 

\subsection{The addressed statistical challenge}
\label{subsec:challenge}

We address a statistical challenge inspired by \eqref{eq:ideal:policy}, with a reformulation guided by insights from EP-learning. The claims we make in this section are proved in Section~\ref{subsec:more:arguments}.

We choose the working model defined as the following closure of a convex hull:
\begin{equation}
    \label{eq:convex:hull}
    \Psi = \overline{\Conv}\left(\{x\mapsto 2\expit(\theta^\top x)-1: \theta \in \mathbb{R}^{d}\}  \cup \{-1\}\right) \subset [-1,1]^{\mathcal{X}}
\end{equation} 
with -1 the function that maps $\mathcal{X}$ to \{-1\}. The closure is taken with respect to (w.r.t.) the $\|\cdot\|_{2,P_{0}}$-norm. For any $\beta > 0$, let $\sigma_{\beta}:[-1,1]\to [0,1]$ be the strongly convex scaling transformation characterized by 
\begin{equation*}
    \sigma_{\beta}(u) = c(\beta)^{-1}\log\left(\frac{1+e^{\beta u}}{1+ e^{-\beta}}\right),
\end{equation*} 
where $c(\beta)^{-1}= \log(1+ e^\beta)-\log(1+e^{-\beta})$. In addition, let $\sigma_{0}:[-1,1]\to [0,1]$ be the linear scaling transformation characterized by $\sigma_{0}(u) = \tfrac{1}{2}(1+u)$. Note that $\sigma_{0}$ is the pointwise limit of $\sigma_{\beta}$ when $\beta$ goes to 0. The transformations $\sigma_{\beta}$ are used to map elements of $\Psi$ to policies, as illustrated next.

Fix an arbitrary $\beta \geq 0$. We now define the actual (as opposed to ideal) $\beta$-specific optimal policy of interest. It is the composition of $\sigma_{\beta}$ with \textit{the unique solution} to the following optimization problem: 
\begin{equation}
    \label{eq:actual:policy}
    \argmin \{R_{0}(\psi) : \psi \in \Psi \text{ s.t. } S_{0}(\sigma_{\beta}\circ \psi ) \leq 0\}.
\end{equation}
This formulation should be read in relation to \eqref{eq:ideal:policy}. Minimizing $\psi \mapsto R_{0}(\psi)$ targets a feature $\psi_{0}$ of $P_{0}$ defined as the $L^{2}(P_{0})$-projection of $\Delta\mu_{0}$ onto $\Psi_{0,\beta} = \{\psi \in \Psi : S_{0}(\sigma_{\beta} \circ \psi) \leq 0\}$, a convex  subclass of $\Psi$. Here, $\sigma_{\beta} \circ \psi$ plays the role of a smooth and convex approximation of the hard decision rule $x \mapsto \bone\{\psi(x) > 0\}$. Since $\pi_{0} : x\mapsto \bone\{\Delta\mu_{0}(x) > 0\}$ maximizes  $\pit \mapsto \mathcal{V}_{P_{0}} (\pit)$ over $[0,1]^{\mathcal{X}}$,  \eqref{eq:actual:policy} can  be interpreted as an indirect way of solving \eqref{eq:ideal:policy}. 

In this light, we focus on policies that are elements of 
\begin{equation}
    \label{eq:Pi:beta}
    \Pi_{\beta}= \{ \pit_{\beta, \psi} = \sigma_{\beta} \circ \psi : \psi \in \Psi \} \subset [0,1]^{\mathcal{X}}.
\end{equation}
For   a   fixed    $\psi   \in   \Psi$   and   any    $x   \in   \mathcal{X}$,
$\beta   \mapsto   \sigma_{\beta}   \circ  \psi(x)$   is   decreasing.   Under
Assumption~\ref{assum:delet_effect},
$\beta \mapsto S_{0}(\sigma_{\beta} \circ  \psi)$ is therefore non-increasing.
Consequently,     for      $0     \leq     \beta_{1}      \leq     \beta_{2}$,
$\Psi_{0,\beta_{2}} \subset \Psi_{0,\beta_{1}}$. The $L^{2}(P_{0})$-projection
of   $\Delta\mu_{0}$  onto   the   larger   class  $\Psi_{0,\beta_{1}}$,   say
$\psi_{0,\beta_{1}}$, is  closer to  $\Delta\mu_{0}$ than the  projection onto
$\Psi_{0,\beta_{2}}$,   say  $\psi_{0,\beta_{2}}$.   However,  this   gain  in
proximity        is        offset        by        the        fact        that
$\sigma_{\beta_{1}}   \circ   \psi_{0,\beta_{1}}$   is   further   away   from
$x      \mapsto       \bone\{\psi_{0,\beta_{1}}(x)      >       0\}$      than
$\sigma_{\beta_{2}}       \circ        \psi_{0,\beta_{2}}$       is       from
$x \mapsto \bone\{\psi_{0,\beta_{2}}(x) > 0\}$.


\section{\centering \textsc{PLUC: learning a constrained policy}}
\label{sec:PLUC}

We now present PLUC (Policy Learning Under Constraint), a methodology 
designed to empirically approximate a solution to \eqref{eq:actual:policy}. 
Section~\ref{subsec:oracular} takes an oracular perspective, focusing on 
solving \eqref{eq:actual:policy} itself. By \textit{oracular} we mean that 
we proceed as if certain parameters $\Phi(P)$ of the true law $P$ were 
known. This idealized viewpoint serves to clarify the logic of the 
method before addressing the practical constraint that $P$ is unknown. 
In the implementable procedure, we replace these unattainable quantities 
$\Phi(P)$ with estimators constructed from data sampled under $P$. 
Section~\ref{subsec:statistical} extends this approach into a practical 
statistical procedure.

\subsection{Oracular viewpoint}
\label{subsec:oracular}

\paragraph{Pseudo-Lagrangian formulation. }
\label{subsec:Lagrangian}

Solving \eqref{eq:actual:policy} can be addressed via the method of Lagrange multipliers. For any $\beta\geq 0$, let  $\mathcal{L}_{0}(\cdot;\beta)$ be characterized on  $\Psi \times \mathbb{R}_+$ by
\begin{equation}
    \label{eq:Lagrange}
    \mathcal{L}_{0}(\psi,\lambda;\beta) = R_{0}(\psi) + \lambda S_{0}(\pit_{\beta,\psi}).
\end{equation}
For fixed $\lambda \geq 0$, the mapping $\psi \mapsto \mathcal{L}_{0}(\psi,\lambda;\beta)$ is  strongly convex over the convex and closed set  $\Psi$. A straightforward adaptation of the proof of the existence of a unique solution to~\eqref{eq:actual:policy} (see Section~\ref{subsec:more:arguments}) ensures that $\psi \mapsto \mathcal{L}_{0}(\psi,\lambda;\beta)$ admits a unique minimizer over $\Psi$. When $\lambda = 0$, the unique minimizer is the projection of $\Delta\mu_0$ onto $\Psi$ w.r.t.\ the $\|\cdot\|_{2,P_{0}}$-norm. As $\lambda > 0$ increases, the constraint $S_{0}(\pit_{\beta,\psi})$ has greater weight in the Lagrangian criterion, and the unique minimizer gets closer to the constant function -1. 

For fixed $\lambda \geq 0$, minimizing $\psi \mapsto \mathcal{L}_{0}(\psi,\lambda;\beta)$ over $\Psi$ is non-trivial, since $\Psi$ is defined as the closure of a  convex hull of functions. To avoid computing projections onto $\Psi$, 
 we employ the Frank-Wolfe algorithm~\citep{frank1956algorithm, jaggi2013revisiting}. The appeal of this method lies in its  computational tractability, as it reduces the search for the unique minimizer in the large set $\Psi$ to an iterative approximation using convex combinations of $\Psi$'s extreme points. 
 Further details, including the adaptation of the convergence proof \citep{jaggi2013revisiting}, are given in Section~\ref{subsec:algo:fw}.

Let $\Lambda \subset \mathbb{R}_{+}$ and $B \subset \mathbb{R}_{+}$ be two user-supplied finite sets of candidates values for $\lambda$ and $\beta$.  Each pair $(\lambda, \beta) \in \Lambda \times B$ yields a minimizer in $\psi \in \Psi$ of \eqref{eq:Lagrange}. We now need to carefully identify and select  a best solution in the resulting family of candidate solutions. 

\paragraph{Identification of the best policy.}
\label{sec:hyperparam}

For every $(\lambda,\beta) \in \Lambda\times B$, let $\psi_{\lambda,\beta} \in \Psi$ be the approximate minimizer of $\psi \mapsto \mathcal{L}_{0}(\psi, \lambda; \beta)$ output by the Frank-Wolfe algorithm. Each $\psi_{\lambda,\beta}$ yields a policy $\pit_{\lambda,\beta} = \sigma_{\beta} \circ \psi_{\lambda,\beta} \in [0,1]^{\mathcal{X}}$. If $\min \{S_{0}(\pit_{\lambda,\beta}) : (\lambda,\beta) \in \Lambda \times B\} > 0$, then no candidate policy is eligible. We cannot do better than recommending to never treat. Otherwise, the best policy is indexed by any element of 
\begin{equation*}
    \argmax \left\{\mathcal{V}_{P_{0}}(\pit_{\lambda,\beta}) : (\lambda,\beta) \in \Lambda \times B \text{ s.t. } S_{0}(\pit_{\lambda,\beta}) \leq 0\right\}.    
\end{equation*}


\subsection{Statistical viewpoint}
\label{subsec:statistical}

In this section, we derive a practical statistical procedure from the oracular
one   described  in   Section~\ref{subsec:oracular}.    We   proceed  in   two
steps. Section~\ref{subsubsec:PLUC:0}  describes a procedure which  involves a
naive      plug-in     estimator      of     the      Lagrangian     criterion
$\mathcal{L}_{0}(\cdot, \lambda;\beta)$. Section~\ref{subsubsec:PLUC} enhances
the  naive procedure,  by considering  targeted estimators  of the  Lagrangian
criterion. \textit{Targeting} an initial  substitution estimator $\Phi(Q)$ of a 
particular parameter of interest $\Phi(P)$ refers to updating
this possibly inconsistent or inefficient estimator into $\Phi(Q')$ so
that it aligns with and focuses on $\Phi(P)$. In the sense used in the
targeted minimum loss estimation (TMLE) paradigm, targeting modifies
only the components of $Q$ that influence the estimation of the
parameter, leaving all other aspects unchanged. The goal is to produce
an estimator whose behavior is optimally tuned to the parameter
$\Phi(P)$ under consideration.

Throughout, cross-fitting is used to avoid overfitting issues. Let $\{1, \ldots
, n\} = \bign_{1} \cup \bign_{2} \cup \bign_{3}$ with mutually disjoint sets 
$\bign_{1}, \bign_{2}, \bign_{3}$ of the same cardinality up to one. Other 
partitions could be preferred on a case-by-case basis. 

\subsubsection{Naive PLUC}
\label{subsubsec:PLUC:0}

\paragraph{Naive empirical counterpart of the Lagrangian formulation.} 
\label{subsec:lagrangian:naive}

Fix $(\lambda, \beta) \in \Lambda \times B$. We first deal with the estimation
of the Lagrangian  criterion $\mathcal{L}_{0}(\cdot,\lambda;\beta)$ using only
data  from  $\{O_{i} :  i  \in  \bign_{1}\cup \bign_{2}\}$.  Let  $\mu_{\bign_{1}}^{0}$  and
$\nu_{\bign_{1}}^{0}$ be  initial estimators  of $\mu_{0}$ and  $\nu_{0}$. The
0~superscript is a visual reminder  of the fact that $\mu_{\bign_{1}}^{0}$ and
$\nu_{\bign_{1}}^{0}$  are \textit{initial}  estimators (which  will be  later
updated). They are  typically obtained by aggregating  several algorithms, for
instance by super learning~\citep{van2007super}. Let
$\Delta\mu_{\bign_{1}}^{0}(\cdot)     =     \mu_{\bign_{1}}^{0}(1,\cdot)     -
\mu_{\bign_{1}}^{0}(0,\cdot)$                                              and
$\Delta\nu_{\bign_{1}}^{0}(\cdot)=
\nu_{\bign_{1}}^{0}(1,\cdot)-\nu_{\bign_{1}}^{0}(0,\cdot)$  be  the  resulting
estimators of $\Delta\mu_{0}$ and $\Delta\nu_{0}$.  We assume that, by design,
$\mu_{\bign_{1}}^{0}$  and $\nu_{\bign_{1}}^{0}$  both  take  their values  in
$[0,1]$. Then, for any $\psi \in \Psi$,
\begin{align}
    \label{eq:initial:Rn}
    R_{\bign_{1}\cup  \bign_{2}}^0(\psi)  &= \frac{3}{n}\sum_{i\in  \bign_{2}}
                                          \left[\psi(X_{i})^{2}
                                            -2\psi(X_{i})\cdot
                                            \Delta\mu_{\bign_{1}}^{0}(X_{i})\right],\\ 
    \label{eq:initial:Sn}
    S_{\bign_{1}\cup\bign_{2}}^0(\pit_{\beta,\psi})  &=  \frac{3}{n}\sum_{i\in
                                                      \bign_{2}}
                                                       [\pit_{\beta,\psi}(X_{i})\cdot
                                                       \Delta\nu_{\bign_{1}}^0(X_{i})]
                                                       - \alpha 
\end{align}
are      natural     plug-in      estimators     of      $R_{0}(\psi)$     and
$S_{0}(\pit_{\beta,\psi})$. We  combine them  to obtain the  following natural
plug-in estimator of $\mathcal{L}_{0}(\psi,\lambda;\beta)$: 
\begin{equation}
  \label{eq:initial:lagrangian}
    \mathcal{L}_{\bign_{1}\cup               \bign_{2}}^0(\psi,\lambda;\beta)=
    R_{\bign_{1}\cup    \bign_{2}}^0(\psi)    +    \lambda    S_{\bign_{1}\cup
      \bign_{2}}^0(\pit_{\beta, \psi}).  
\end{equation} 
Specifically, assuming that the statistical model $\mathcal{M}$ contains a law
$P_{\bign_{1}     \cup     \bign_{2}}^{0}    \in\mathcal{M}$     such     that
$\Delta\mu_{P_{\bign_{1}  \cup  \bign_{2}}^{0}} =  \Delta\mu_{\bign_{1}}^{0}$,
$\Delta\nu_{P_{\bign_{1} \cup  \bign_{2}}^{0}} =  \Delta\nu_{\bign_{1} }^{0}$,
and
$P_{\bign_{1}   \cup  \bign_{2},X}^{0}   =   (3/n)   \sum_{i  \in   \bign_{2}}
\Dirac(X_{i})$    (that    is,    the    marginal    law    of    $X$    under
$P_{\bign_{1} \cup  \bign_{2}}^{0}$ is the $\bign_{2}$-specific  empirical law
of           $X$),           then            it           holds           that
$R_{\bign_{1} \cup \bign_{2}}^{0}(\psi) = R_{P_{\bign_{1} \cup \bign_{2}}^{0}}
(\psi)$,
$S_{\bign_{1} \cup  \bign_{2}}^{0}(\pit_{\beta\,\psi}) =  S_{P_{\bign_{1} \cup
    \bign_{2}}^{0}}                  (\pit_{\beta,\psi})$,                 and
$\mathcal{L}_{\bign_{1}     \cup      \bign_{2}}^{0}(\psi,\lambda;\beta)     =
\mathcal{L}_{P_{\bign_{1} \cup \bign_{2}}^{0}}(\psi,\lambda;\beta)$.

The empirical Lagrangian criterion $\mathcal{L}_{\bign_{1} \cup \bign_{2}}^0(\cdot,\lambda;\beta)$ is a convex function. Again, we can rely on the Frank-Wolfe algorithm to approximate a minimizer by iteratively forming convex combinations of extreme points of $\Psi$ (see Sections~\ref{subsec:algo:fw} and \ref{subsec:more:arguments} for details). 
\paragraph{Identification of the best policy.} 

Let $\psi_{\bign_{1} \cup \bign_{2},\lambda,\beta}\in \Psi$ be the approximate
minimizer                                                                   of
$\psi \mapsto  \mathcal{L}_{\bign_{1} \cup \bign_{2}}^{0}(\psi,\lambda;\beta)$
as     output      by     the     Frank-Wolfe     algorithm      for     every
$(\lambda,\beta)        \in        \Lambda        \times        B$        (see
Section~\ref{subsec:algo:fw}).                                            Each
$\psi_{\bign_{1}    \cup    \bign_{2},\lambda,\beta}$    yields    a    policy
$\pit_{\bign_{1}   \cup   \bign_{2},\lambda,\beta}  =   \sigma_{\beta}   \circ
\psi_{\bign_{1}  \cup \bign_{2},\lambda,\beta}  \in [0,1]^{\mathcal{X}}$.  The
identification        of         the        best         solution        among
$\{\pit_{\bign_{1}  \cup  \bign_{2},\lambda,\beta}   :  (\lambda,  \beta)  \in
\Lambda  \times   B\}$  hinges  on   estimators  of  the   corresponding  sets
$\{S_{0}(\pit_{\bign_{1} \cup \bign_{2},\lambda,\beta}) : (\lambda, \beta) \in
\Lambda                   \times                    B\}$                   and
$\{\mathcal{V}_{P_{0}}(\pit_{\bign_{1}   \cup    \bign_{2},\lambda,\beta})   :
(\lambda, \beta)  \in \Lambda  \times B\}$.  These estimators  are constructed
based on $\{O_{i} : i \in \bign_{3}\}$.

Let  $\mu_{\bign_{3}}^{0}$ and  $\nu_{\bign_{3}}^{0}$  be additional,  initial
estimators    of   $\mu_{0}$    and   $\nu_{0}$    using   only    data   from
$\{O_{i}         :         i          \in         \bign_{3}\}$.         Define
$\Delta\mu_{\bign_{3}}^{0}(\cdot)=\mu_{\bign_{3}}^{0}(1,\cdot)-\mu_{\bign_{3}}^{0}(0,\cdot)$
and
$\Delta\nu_{\bign_{3}}^{0}(\cdot)=\nu_{\bign_{3}}^{0}(1,\cdot)-\nu_{\bign_{3}}^{0}(0,\cdot)$. Then,
for every $(\lambda,\beta) \in \Lambda \times B$,
\begin{align*}
    S_{\bign_{3}}^{0}(\pit_{\bign_{1}\cup\bign_{2},\lambda,\beta}) & = \frac{3}{n}\sum_{i\in \bign_{3}}[\pit_{\bign_{1}\cup\bign_{2},\lambda,\beta}(X_{i})\cdot \Delta\nu_{\bign_{3}}^{0}(X_{i})] - \alpha \quad \text{and}\\
    \mathcal{V}_{\bign_{3}}^{0}(\pit_{\bign_{1}\cup \bign_{2},\lambda,\beta}) &= \frac{3}{n}\sum_{i \in \bign_{3}}\pit_{\bign_{1}\cup\bign_{2},\lambda,\beta}(X_{i})\cdot \mu_{\bign_{3}}^{0}(1,X_{i}) + (1-\pit_{\bign_{1}\cup \bign_{2},\lambda,\beta}(X_{i}))\cdot \mu_{\bign_{3}}^{0}(0,X_{i})
\end{align*} 
are            natural             plug-in            estimators            of
$S_{0}(\pit_{\bign_{1}\cup\bign_{2},\lambda,\beta})$                       and
$\mathcal{V}_{P_{0}}(\pit_{\bign_{1}\cup\bign_{2},\lambda,\beta})$,
respectively.

Fix $(\lambda,\beta) \in \Lambda \times B$. Following the paradigm of targeted
learning~\citep{van2006targeted,TMLE2011,TMLE2018},  we   update  the  initial
estimator  $\nu_{\bign_{3}}^{0}$ into  $\nu_{\bign_{3},\lambda,\beta}^{*}$ and
define
$\Delta\nu_{\bign_{3}\lambda,\beta}^{*}(\cdot)=\nu_{\bign_{3},\lambda,\beta}^{*}(1,\cdot)-\nu_{\bign_{3},\lambda,\beta}^{*}(0,\cdot)$
so that
\begin{equation}
\label{eq:updated:S}
    S_{\bign_{3}}^{*}(\pit_{\bign_{1}\cup\bign_{2},\lambda,\beta}) = \frac{3}{n}\sum_{i\in \bign_{3}}[\pit_{\bign_{1}\cup\bign_{2},\lambda,\beta}(X_{i})\cdot \Delta\nu_{\bign_{3}\lambda,\beta}^{*}(X_{i})] - \alpha
\end{equation} 
is              a              targeted         estimator              of
$S_{0}(\pit_{\bign_{1}\cup\bign_{2},\lambda,\beta})$. Likewise,  we update the
initial            estimator             $\mu_{\bign_{3}}^{0}$            into
$\mu_{\bign_{3},\lambda,\beta}^{*}$ so that
\begin{align}
\label{eq:updated:V}
        &\mathcal{V}_{\bign_{3}}^{*}(\pit_{\bign_{1}\cup \bign_{2},\lambda,\beta}) \nonumber \\
        &= \frac{3}{n}\sum_{i \in \bign_{3}}\pit_{\bign_{1}\cup\bign_{2},\lambda,\beta}(X_{i})\cdot \mu_{\bign_{3}, \lambda, \beta}^{*}(1,X_{i}) + (1-\pit_{\bign_{1}\cup \bign_{2},\lambda,\beta}(X_{i}))\cdot \mu_{\bign_{3}, \lambda, \beta}^{*}(0,X_{i})
\end{align}
is              a              targeted              estimator              of
$\mathcal{V}_{P_{0}}(\pit_{\bign_{1}\cup\bign_{2},\lambda,\beta})$.        Let
$\uS_{\bign_{3}}^{*}(\pit_{\bign_{1}\cup\bign_{2},\lambda,\beta})$         and
$\lNu_{\bign_{3}}^{*}(\pit_{\bign_{1}\cup\bign_{2},\lambda,\beta})$   be   the
95\%-confidence          upper-          and         lower-bounds          for
$S_{0}(\pit_{\bign_{1}\cup\bign_{2},\lambda,\beta})$                       and
$\mathcal{V}_{P_{0}}(\pit_{\bign_{1}\cup\bign_{2},\lambda,\beta})$     derived
from     the     above     targeted     estimators     (see     details     in
Section~\ref{subsec:TMLE:uS:lV}).

If
\begin{equation*}
    \min\left\{\uS_{\bign_{3}}^{*}(\pit_{\bign_{1}\cup \bign_{2},\lambda,\beta}) : (\lambda, \beta)\in \Lambda \times B\right\} > 0, 
\end{equation*}
then no candidate solution is deemed admissible, and the optimal recommendation is to never treat. Otherwise, the best policy is indexed by any element of 
\begin{equation*}
    \argmax \left\{ \lNu_{\bign_{3}}^{*}(\pit_{\bign_{1}\cup \bign_{2},\lambda,\beta}) : (\lambda, \beta)\in \Lambda \times B \text{ s.t. } \uS_{\bign_{3}}^{*}(\pit_{\bign_{1}\cup \bign_{2},\lambda,\beta})\leq 0 \right\}.
\end{equation*}
In     words,    we     restrict    attention     to    candidate     policies
$\pit_{\bign_{1}\cup  \bign_{2},\lambda,\beta}$ that  are unlikely  to violate
the                              constraint                              (since
$\uS_{\bign_{3}}^{*}(\pit_{\bign_{1}\cup \bign_{2},\lambda,\beta})\leq 0$) and
select  among them  the  one whose  value  is likely  the  largest (since  its
lower-bound             is             larger             than             any
$\lNu_{\bign_{3}}^{*}(\pit_{\bign_{1}\cup  \bign_{2},\lambda,\beta})$). It  is
worth   emphasizing   that   we   rely  on   upper-   and   lower-bounds   for
$S_{0}(\pit_{\bign_{1}\cup\bign_{2},\lambda,\beta})$                       and
$\mathcal{V}_{P_{0}}(\pit_{\bign_{1}\cup\bign_{2},\lambda,\beta})$ rather than
pointwise estimators,  and that we do  not adjust their confidence  levels for
the multiplicity.

The Naive PLUC procedure is summarized in Algorithm~\ref{algo:pluc}.

\subsubsection{PLUC}
\label{subsubsec:PLUC}

\paragraph{Why Naive PLUC is potentially flawed.}

For  fixed $(\lambda,\beta)\in  \Lambda \times  B$  and $\psi  \in \Psi$,  the
empirical                                                            criterion
$\mathcal{L}_{\bign_{1}\cup  \bign_{2}}^{0}(\psi,\lambda;\beta)$ viewed  as an
estimator  of  $\mathcal{L}_{0}(\psi,  \lambda;\beta)$   is  afflicted  by  an
intrinsic  bias.  This  bias is  fundamentally characterized  in terms  of the
efficient influence  curve of 
\begin{equation}
\label{eq:lagrangian}
  P  \mapsto \mathcal{L}_{P}(\psi,\lambda;\beta) = R_{P}(\psi) + \lambda S_{P}(\pit_{\beta, \psi})
\end{equation}
at   $P_{0}$,  which   we  denote   by  $D_{\psi,\lambda,\beta}(P_{0})$   (see \eqref{eq:OTR_risk}, \eqref{eq:S}, and 
Section~\ref{app:proof:EIC}). The  objective of  this section  is to
correct  the bias  while preserving  convexity of  the corrected  estimator of
$\psi\mapsto \mathcal{L}_{0}(\psi, \lambda;\beta)$.

A  first  natural  path  is  to   implement  a  one-step  correction  of  each
$\mathcal{L}_{\bign_{1}\cup
  \bign_{2}}^{0}(\psi,\lambda;\beta)$~\citep{Pfanzagl82,tlride}.   Recall  the
definition  of $P_{\bign_{1}\cup  \bign_{2}}^{0}  \in \mathcal{M}$  introduced
right after  \eqref{eq:initial:lagrangian}, and let $e_{\bign_{1}}^{0}$  be an
estimator     of    $e_{0}$     constructed    using     only    data     from
$\{O_{i}  : i  \in  \bign_{1}\}$ (typically  obtained  by aggregating  several
estimators, like  $\mu_{\bign_{1}}^{0}$ and  $\nu_{\bign_{1}}^{0}$).  Assuming
in  addition that  $P_{\bign_{1}\cup \bign_{2}}^{0}  \in \mathcal{M}$  is such
that   the   $P_{\bign_{1}\cup   \bign_{2}}^{0}$-specific   propensity   score
$e_{P_{\bign_{1}\cup \bign_{2}}^{0}}$ equals $e_{\bign_{1}}^{0}$, the one-step
correction            can            consist           of            replacing
$\mathcal{L}_{\bign_{1}\cup \bign_{2}}^{0}(\psi,\lambda;\beta)$ by
\begin{equation*}
  \mathcal{L}_{\bign_{1}\cup \bign_{2}}^{0}(\psi,\lambda;\beta)  + \frac{3}{n}
  \sum_{i      \in     \bign_{2}}      D_{\psi,\lambda,\beta}(P_{\bign_{1}\cup
    \bign_{2}}^{0})(O_{i}).
\end{equation*}
Unfortunately,  the  corrected empirical  criterion  viewed  as a  real-valued
mapping defined on $\Psi$ is not convex anymore, because the second summand is
not a convex function of $\psi$. 

Adopting a  targeted learning perspective,  another natural path is  to update
the initial estimators  $\Delta\mu_{\bign_{1}}$ and $\Delta\nu_{\bign_{1}}$ of
$\Delta\mu_{0}$     and     $\Delta\nu_{0}$     in     order     to     target
$\psi \mapsto \mathcal{L}_{0}(\psi,\lambda;\beta)$ (\textit{simultaneously for
  all} $\psi  \in \Psi$), thus  automatically preserving the convexity  of the
corrected  empirical criterion.  This approach  proved very  challenging, both
theoretically and computationally.

Instead, we follow a third path inspired by the targeted learning paradigm. It
is presented next.

\paragraph{Targeted empirical counterpart of the Lagrangian formulation.}

Fix arbitrarily  $(\lambda, \beta)  \in \Lambda \times  B$.  The  targeted (as
opposed to naive)  empirical counterpart of the  Lagrangian formulation yields
an alternating iterative procedure where  each step embeds two sub-steps.  The
initial                                                              estimator
$\mathcal{L}_{\bign_{1}\cup\bign_{2}}^{0}(\cdot,\lambda;\beta)$     of     the
Lagrangian  criterion $\mathcal{L}_{0}(\cdot,\lambda;\beta)$  is alternatively
updated and minimized.

More  specifically,  at  the  $k$-th  iteration,  given  the  $k$  approximate
minimizers     $\psi_{\bign_{1}\cup\bign_{2},\lambda,\beta}^{0}$,    $\ldots$,
$\psi_{\bign_{1}\cup\bign_{2},\lambda,\beta}^{k-1}$ from the previous steps,
\begin{enumerate}
\item  we  update  $P_{\bign_{1}\cup   \bign_{2}}^{0}  \in  \mathcal{M}$  into
  $P_{\bign_{1}\cup  \bign_{2}}^{k}  \in  \mathcal{M}$  in  such  a  way  that
  $\mathcal{L}_{\bign_{1}    \cup     \bign_{2}}^{k}(\cdot,\lambda;\beta)    =
  \mathcal{L}_{P_{\bign_{1}     \cup     \bign_{2}}^{k}}(\cdot,\lambda;\beta)$
  \textit{simultaneously targets the $k$ landmark values}
  \begin{equation*}
    \mathcal{L}_{0}(\psi_{\bign_{1}\cup\bign_{2},\lambda,\beta}^{0},\lambda;\beta),
    \ldots,
    \mathcal{L}_{0}(\psi_{\bign_{1}\cup\bign_{2},\lambda,\beta}^{k-1},\lambda;\beta); 
  \end{equation*}
\item     we     compute      the     $(k+1)$-th     approximate     minimizer
  $\psi_{\bign_{1}\cup\bign_{2},\lambda,\beta}^{k}$  by minimizing  the convex
  mapping
  $\psi             \mapsto            \mathcal{L}_{\bign_{1}             \cup
    \bign_{2}}^{k}(\psi,\lambda;\beta)$ using the Frank-Wolfe algorithm.
\end{enumerate}
The        hope        is         that,        as        $k$        increases,
$\psi \mapsto  \mathcal{L}_{\bign_{1} \cup \bign_{2}}^{k}(\psi,\lambda;\beta)$
becomes         a          more         reliable          estimator         of
$\psi \mapsto \mathcal{L}_{0}(\psi,\lambda;\beta)$ over the whole class $\Psi$
by targeting  an increasing number  of relevant landmark values  thereof.  The
alternating iterative procedure terminates when either a stopping criterion is
satisfied or  the maximum number of  iterations is reached.  The  algorithm is
detailed in Section~\ref{app:algo:ap}.

\paragraph{Identification of the best policy.}

For     every    $(\lambda,\beta)     \in    \Lambda     \times    B$,     let
$\pit_{\bign_{1}  \cup  \bign_{2},\lambda,\beta}^{*}  =  \sigma_{\beta}  \circ
\psi_{\bign_{1}  \cup  \bign_{2},\lambda,\beta}^{*}\in  \Psi$  be  the  policy
output by the alternating iterative  procedure (note the $*$-superscript).  As
in                     Section~\ref{subsubsec:PLUC:0},                     let
$\uS_{\bign_{3}}^{*}(\pit_{\bign_{1}\cup\bign_{2},\lambda,\beta}^{*})$     and
$\lNu_{\bign_{3}}^{*}(\pit_{\bign_{1}\cup\bign_{2},\lambda,\beta}^{*})$     be
95\%-confidence          upper-          and         lower-bounds          for
$S_{0}(\pit_{\bign_{1}\cup\bign_{2},\lambda,\beta})$                       and
$\mathcal{V}_{P_{0}}(\pit_{\bign_{1}\cup\bign_{2},\lambda,\beta})$    derived,
for every $(\lambda,\beta) \in \Lambda \times B$,
using  only  data  from  $\{O_{i}  :   i  \in  \bign_{3}\}$  (see  details  in
Section~\ref{subsec:TMLE:uS:lV}). If
\begin{equation*}
    \min\left\{\uS_{\bign_{3}}^{*}(\pit_{\bign_{1}\cup
        \bign_{2},\lambda,\beta}^{*})  :  (\lambda, \beta)\in  \Lambda  \times
      B\right\} > 0, 
\end{equation*}
then   no  candidate   solution  is   deemed  admissible,   and  the   optimal
recommendation is to never treat. Otherwise, the best policy is indexed by any
element of
\begin{equation*}
  \argmax           \left\{           \lNu_{\bign_{3}}^{*}(\pit_{\bign_{1}\cup
      \bign_{2},\lambda,\beta}^{*})  : (\lambda,  \beta)\in  \Lambda \times  B
    \text{        s.t.        }        \uS_{\bign_{3}}^{*}(\pit_{\bign_{1}\cup
      \bign_{2},\lambda,\beta})\leq 0 \right\}. 
\end{equation*}
Similarly  to the  naive PLUC  procedure, we  restrict attention  to candidate
policies $\pit_{\bign_{1}\cup \bign_{2},\lambda,\beta}^{*}$  that are unlikely
to             violate             the            constraint             (since
$\uS_{\bign_{3}}^{*}(\pit_{\bign_{1}\cup \bign_{2},\lambda,\beta}^{*})\leq 0$)
and select  among them the  one whose value is  likely the largest  (since its
lower-bound             is             larger             than             any
$\lNu_{\bign_{3}}^{*}(\pit_{\bign_{1}\cup \bign_{2},\lambda,\beta}^{*})$).

The PLUC procedure is summarized in Algorithm~\ref{algo:pluc}.


\subsection{Making recommendations based on policies}
\label{subsec:binarized}

In Section~\ref{subsec:policy:learning}, a policy is  defined as an element of
a     user-supplied     $\Pi      \subset     [0,1]^{\mathcal{X}}$.      Given
$x,  x' \in  \mathcal{X}$, any  policy  $\pit \in  \Pi$  can be  used to  make
recommendations  by  independently  sampling   two  random  variable  $A_{x}$,
$A_{x'}$ from the Bernoulli laws with parameters $\pit(x)$ and $\pit(x')$.

Of course, this procedure does not  guarantee that if $\pit(x) \geq \pit(x')$,
then  $A_{x} \geq  A_{x'}$.   One way  to enforce  this  property consists  of
selecting a  threshold $t \in  [0,1]$ and making recommendations  by comparing
$\pit(x)$ and $\pit(x')$ to $t$.

For    any    $\pit    \in    \Pi$    and    $t    \in    [0,1]$,    introduce
$\pi^{t} : x  \mapsto \bone\{\pit(x) \geq t\}$. From  an oracular perspective,
any element of
\begin{equation*}
  \argmax\{   \mathcal{V}_{P_{0}}(\pi^{t}):  t   \in  [0,1]   \text{  s.t.   }
  S_{0}(\pi^{t})\leq 0 \}
\end{equation*}
can be considered as an optimal (oracular) threshold.

From a  statistical viewpoint,  given 95\%-confidence upper-  and lower-bounds
$\uS_{\bign_{3}}^{*}(\pi^{t})$    and   $\lNu_{\bign_{3}}^{*}(\pi^{t})$    for
$S_{0}(\pi^{t})$  and $\mathcal{V}_{P_{0}}(\pi^{t})$,  as  those developed  in
Section~\ref{subsec:statistical}, any element of
\begin{equation*}
  \argmax \left\{  \lNu_{\bign_{3}}^{*}(\pi^{t})  : t\in
    [0,1]    \text{     s.t.     }     \uS_{\bign_{3}}^{*}(\pi^{t})\leq 0   \right\}.  
\end{equation*}
can be considered as an optimal threshold.

\section{\centering \textsc{Numerical experiments}}
\label{sec:num:exp}

We assessed the statistical algorithms introduced in Section~\ref{subsec:statistical} 
on synthetic data. We additionally implemented Oracular PLUC from Section~\ref{subsec:oracular},
leveraging available counterfactuals. We designed a set of controlled simulation scenarios
featuring diverse treatment effect patterns for both the primary outcome ($Y\in [0,1]$) 
and adverse events ($\xi\in \{0,1\}$), as well as  varying sample sizes,  
enabling a thorough assessment of methodological flexibility. We also developed a more
realistic setting based on real primary outcomes ($Y\in \mathbb{R}$) and covariates 
outside the normalized $[0,1]$ range, thereby extending the evaluation beyond 
idealized conditions. For clarity and brevity, we report results for one 
representative synthetic scenario in the main text. Results of two additional representative 
synthetic scenarios, as well as for the realistic setting, are provided in Section~\ref{app:about:num:exp}.

\begin{figure}[H]
\centering
        \includegraphics[width=0.7\textwidth]{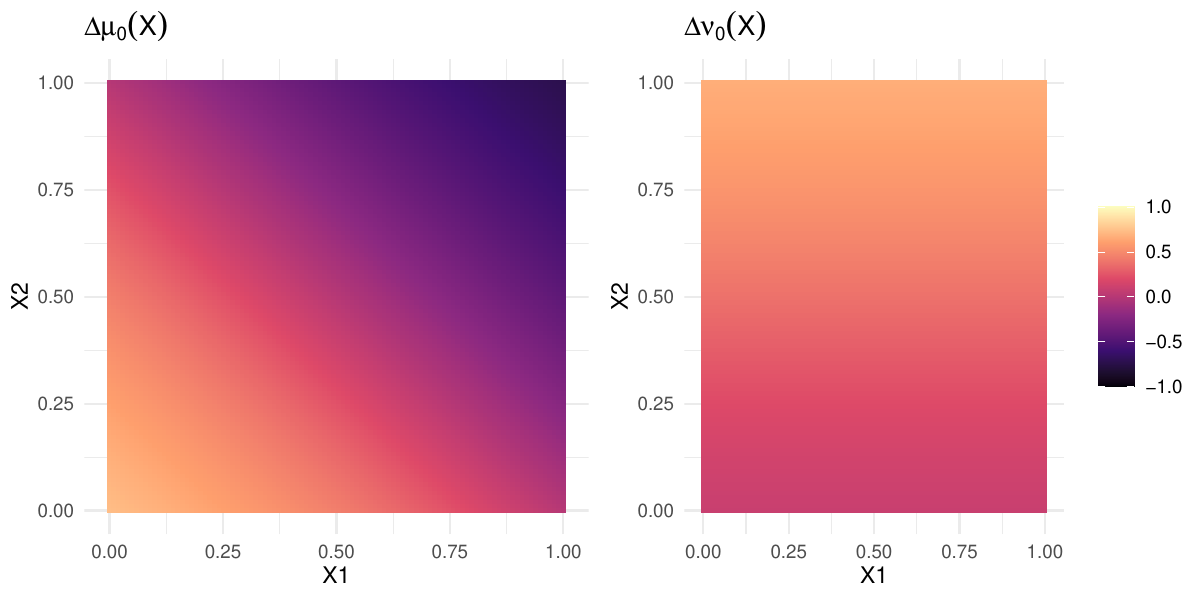}
    \caption{Treatment effect patterns, with the primary outcome displayed on the left and the adverse event displayed on the right.}
    \label{fig:scenarios}
\end{figure}

\paragraph{Simulation scenario. }
The experiment  was conducted using  a covariate vector  uniformly distributed
over  $[0,1]^{10}$.   The primary  outcome $Y$  featured a
linear  treatment effect  driven by  two covariates,  while the  adverse event
$\xi$ was modeled with a linear effect based on a single covariate. A detailed
description  of  the   synthetic  data  generation  process   is  provided  in
Section~\ref{app:controlled:settings}. Figure~\ref{fig:scenarios} illustrates the
treatment effect patterns in a two-dimensional representation based on the two
relevant covariates.   The color gradient,  ranging from dark to  light tones,
reflects the transition from negative to positive treatment effects. Note that
the constraint $\Delta\nu_0(X)  \ge 0$ (see Assumption~\ref{assum:delet_effect})
restricts the treatment effect for the  adverse event to the interval $[0,1]$,
where  lower values  are desirable,  while  higher treatment  effects for  the
primary outcome correspond to better expected results. 

We considered three sample sizes $n \in \{3{,}000, 6{,}000, 18{,}000\}$,
generating 50 replicates per size.  For each replicate, we partitioned the index set $\{1, \ldots, n\}$ into three disjoint subsets $\bign_{1}, \bign_{2}, \bign_{3}$ of equal size. Each subset $\bign_{k}$ identifies the observations assigned to the corresponding data split.  The constraint 
parameter was set to $\alpha=0.1$, and the candidate sets were defined as 
$\Lambda = \{1,\ldots,10\}$ and $B=\{0,0.05,0.1,0.25,0.5\}$. The three 
algorithms were run with parameters detailed in 
Section~\ref{app:exp:hyperaparam}. 

For each replicate, estimation was performed on $\{O_{i}:i\in   
\bign_{1}\}$ and $\{O_{i}:i\in \bign_{3}\}$ via the super learner   
ensemble method \citep{van2007super,SuperLearner}, which leverages 
cross-validation to aggregate candidate algorithms from a library. The \texttt{SuperLearner}'s library included mean imputation, generalized linear models (GLMs)
\citep{nelder1972generalized},         random         forests         (ranger)
\citep{wright2017ranger},      generalized      random      forests      (GRF)
\citep{wager2018estimation},    and   gradient    boosted   trees    (XGBoost)
\citep{chen2016xgboost}.    The   super   learner  was   configured   with   a
\texttt{gaussian()}   family  for   continuous   outcomes   in  $[0,1]$ (and thresholding to enforce that the super learner takes its values in $[0,1]$),   and
\texttt{binomial()} family for binary outcomes. Cross-validation was performed
using the setting \texttt{SuperLearner.CV.control(V = 2L)}. 
 
The outputs of the three algorithms were evaluated via Monte Carlo approximation 
of the oracular policy value $\pit \mapsto \mathcal{V}_{P_{0}}(\pit)$, and the 
constraint, $\pit \mapsto S_{0}(\pit)$, using an additional dataset of 
$1{,}000{,}000$ complete observations including potential outcomes. In practical settings, without the potential outcomes, the 
evaluation would instead rely on $\{O_{i}: i \in \bign_{3}\}$. Finally, 
recommendations from the three output policies, as described in Section~\ref{subsec:binarized}, 
were evaluated using the same procedure. 

\paragraph{Competitor. }

We compared the performance of the recommendations induced by Oracle PLUC, Naive 
PLUC and PLUC, with that of BR-O~\citep{wang2018learning}, 
which can be viewed as  a special case of the framework of~\citep{liu2024learning} under a static treatment regime. 

From an oracular viewpoint, BR-O algorithm seeks to maximize 
$E_\mathbb{P}[Y(\pi(X))]$ with respect to $\pi \in \{0,1\}^{
\mathcal{X}}$, while constraining $E_\mathbb{P}[\xi(\pi(X))]$, to 
remain below a pre-specified threshold $\tau \geq 0$.  
Although $\tau$ is generally arbitrary, setting
$\tau = \alpha + E_{\mathbb{P}}[\xi(0)]$ 
renders BR-O's constraint oracularly equivalent to ours,
\begin{equation*}
    E_\mathbb{P}[\pi(X)\cdot(\xi(1) - \xi(0))]  \leq  \alpha.
\end{equation*}
The summand $E_\mathbb{P}[\xi(0)]$ in the definition of $\tau$ is approximated by Monte Carlo using the same $1{,}000{,}000$ complete observations as before. It follows that BR-O is an oracular algorithm.  

The ``BR–O" was originally developed for randomized controlled trials (RCTs), 
where propensity scores are known and sample sizes are moderate. In our 
setting, we provided BR-O with the true propensity scores, further reinforcing its oracular status. To handle the non-convexity of binary treatment policies,
the approach explicitly relies on the Difference-of-Convex (DC) algorithm, combined 
with \texttt{solve.QP} in R \citep{goldfarb1983numerically}, to solve the optimization 
problem. While computationally efficient in RCT settings, this algorithm becomes 
impractical and prone to non-convergence for large observational datasets. In our 
experiments, this limitation meant that the competitor could only be computed for 
the smallest sample size, $n=3{,}000$.  

Finally, following the authors’ implementation guidelines, we set $\delta = 0.02$, 
performed the grid search $\mathtt{C.grid}$ to the range $\{2^{i} : -10 \le i \le 10\}$ 
and chose a linear kernel.

\paragraph{Empirical findings from simulations.}

\begin{figure}[H]
    \centering
        \begin{minipage}[t]{0.48\textwidth}
            \centering
            \includegraphics[width=\textwidth]{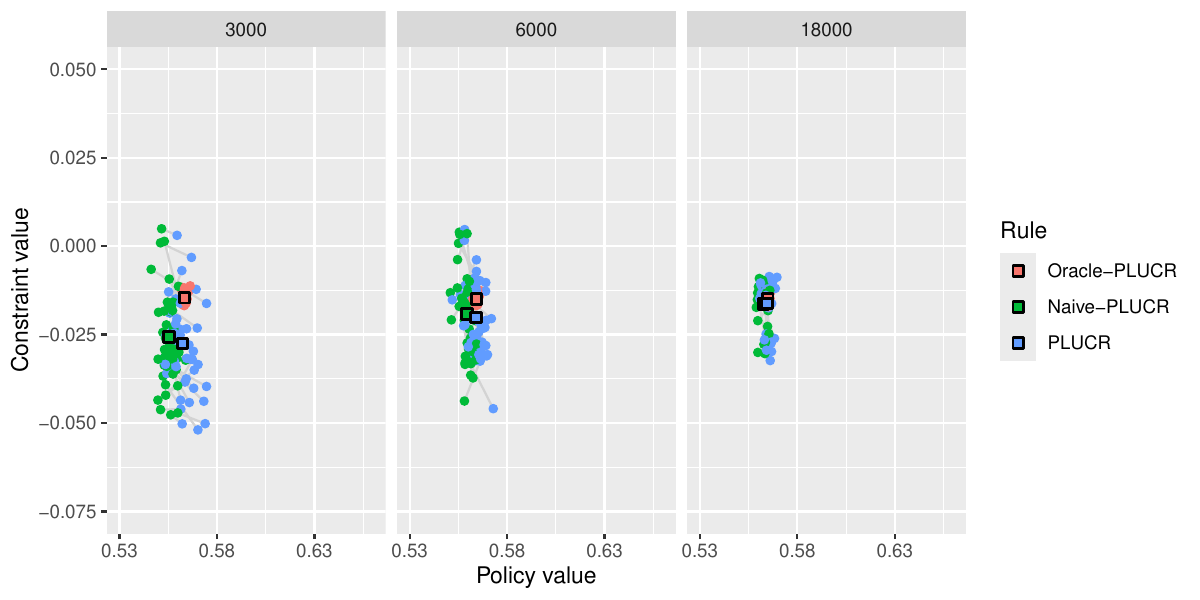}
            \subcaption{Policies}
            \label{fig:qq:linear:proba}
        \end{minipage}
        \hfill
        \begin{minipage}[t]{0.48\textwidth}
            \centering
            \includegraphics[width=\textwidth]{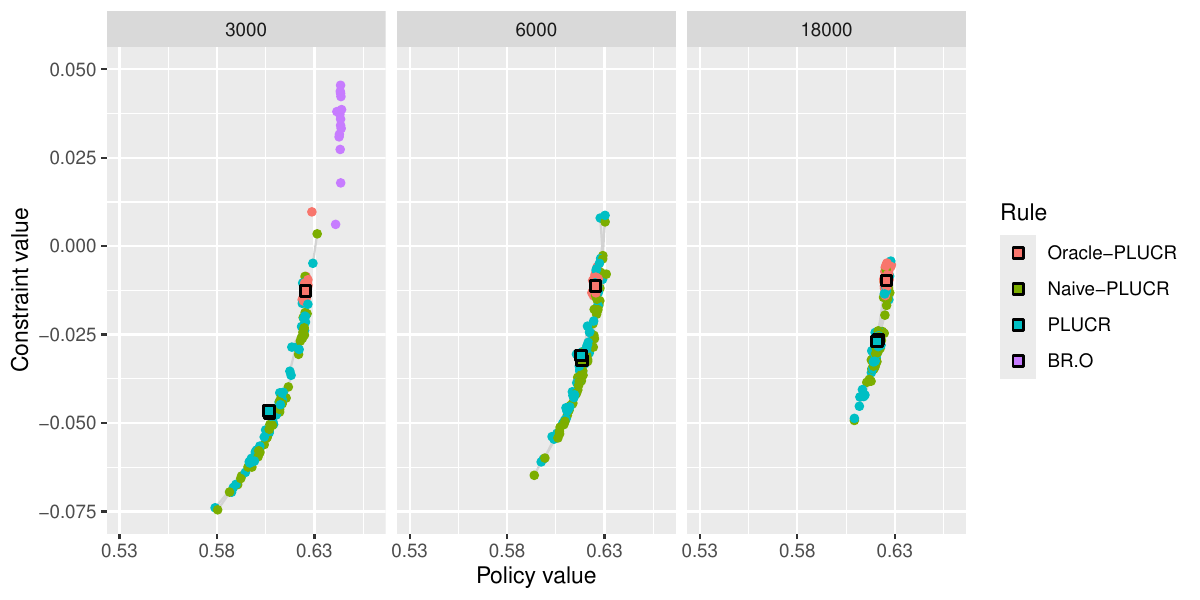}
            \subcaption{Decisions}
            \label{fig:qq:linear:decision}
         \end{minipage}
        \caption{Policy value-constraint plot for policies (a) and recommendations (b).}
    \label{fig:linear:QQ}
\end{figure}

We first present the policy-constraint plot 
(Figure~\ref{fig:linear:QQ})  for the linear case, which illustrates 
key properties of the proposed algorithms. We then comment 
on  the main findings of the complete  simulation study,  reported in 
Section~\ref{app:full:simulation:results}.

Specifically, in Figure \ref{fig:linear:QQ}, each point corresponds to a policy, with color indicating the generation technique. The $x$-axis reports the policy value, while the $y$-axis displays the corresponding constraint 
value. For each technique, the barycenter of the associated policies is  marked by a square of the same color, summarizing the mean policy and constraint values.
Policies with better performance appear further to the right, 
while feasibility corresponds to non-positive constraint values. Gray line segments connect 
 policies obtained via Naive PLUC (green) to their counterparts produced by PLUC (blue). 
A rightward shift from green to blue indicates an improvement in policy value, provided 
feasibility is preserved. 

As a first initial observation, Oracle PLUC (red) exhibits superior performance compared to Naive 
PLUC (green) and PLUC (blue). In Figure~\ref{fig:linear:QQ}, the red policies  consistently lie further to the right, achieving  higher policy values while  
satisfying the constraint with low variability. As the sample size 
increases, the policies produced by PLUC and Naive PLUC progressively concentrate around the oracle, indicating convergence toward the ideal benchmark. Although PLUC 
does not uniformly improve the Naive PLUC policies at the individual level (Figures~\ref{fig:qq:linear:proba}, 
\ref{fig:qq:threshold:proba}, \ref{fig:qq:satisfied:proba}, \ref{fig:qq:realistic:proba}), 
its average policy value is consistently equal to or  higher than that of Naive PLUC.  
This improvement at the aggregate level is reflected in barycenters that are either closely aligned or shifted to the right, while reliably preserving constraint satisfaction. 

Despite differences in the underlying policies, the recommendations derived 
from both algorithms exhibit statistically similar average performance
(Figures~\ref{fig:qq:linear:decision}, \ref{fig:qq:threshold:decision}, 
\ref{fig:qq:satisfied:decision}, \ref{fig:qq:realistic:decision}). 
This similarity arises because both policies capture the same underlying decision pattern, 
differing primarily by a scaling transformation. These differences 
are subsequently absorbed through the choice of decision 
thresholds when deriving recommendations. As a result, the induced recommendations often achieve higher policy 
values than the corresponding probabilistic policies themselves, while consistently satisfying the constraint, at the cost of increased variance (Figures \ref{fig:boxplot:linear:pv}, \ref{fig:boxplot:linear:constraint}, 
\ref{fig:boxplot:threshold:pv}, \ref{fig:boxplot:threshold:constraint}, 
\ref{fig:boxplot:realistic:pv}, \ref{fig:boxplot:realistic:constraint}).
Using probabilities rather than binary recommendations provides richer information,
yielding both a quantitative measure of confidence and a flexible representation capable of  capturing complex 
patterns (Figure~\ref{fig:linear:treatment}). This richer information allows 
 policies to be used either directly for  assessment or subsequently translated into recommendations.

BR-O, who directly outputs treatment recommendations, exhibited marked instability 
across the full simulation study, frequently failing to converge (33 out of 
50 runs in the linear setting). This behavior  likely reflects  a mismatch between BR-O's framework and the policy-learning setting considered here.
In particular, while the DC algorithm underlying BR-O is well suited to RCTs, its application in larger-sample of more complex settings appears  prone to  convergence 
issues. 

BR-O occasionally attains higher policy values (Figures \ref{fig:qq:linear:decision}, 
\ref{fig:qq:realistic:decision}), but at the expense of  violating the constraint. 
We observe that BR-O frequently closely aligns with the unconstrained policy value 
maximizer $x\mapsto \bone\{\Delta\mu_{0}(x)>0\}$ (Figure \ref{fig:linear:boxplot}), 
which is consistent with the observed constraint violations. This behavior likely reflects inaccuracies 
in the estimation of  $E_{\mathbb{P}}[\xi(0)]$, 
which directly affects the learning of the constraint parameter $\tau$ required to enforce feasibility.

Despite these methodological limitations, BR-O produced policies comparable to those obtained with our algorithm in several scenarios
(Figure~\ref{fig:qq:threshold:decision}), and even outperformed it in settings where 
the constraint is systematically satisfied (Figure~\ref{fig:qq:satisfied:decision}).



\section{\centering \textsc{Conclusion}}
\label{sec:conclusion}

\paragraph*{Summary.}
We have investigated PLUC, a policy-learning approach designed to enforce explicit control of adverse-event risks. Building on insights from EP-learning~\citep{van2024combining}, the method relies on strongly convex Lagrangian criteria defined over a convex hull of functions (Section~\ref{sec:PLUC}), which characterize a novel non-parametric class of policies which is both theory and optimization-friendly.
We further established the existence and uniqueness of associated optimizers (Sections~\ref{subsec:algo:fw} and~\ref{app:proofs:fw}).
The algorithm carefully combines Frank–Wolfe optimization~\citep{frank1956algorithm} over the non-parametric policy class with a targeting step (Section~\ref{app:algo:ap}) that progressively aligns the criterion with its oracular counterpart at previously visited landmarks. Our convergence arguments, for the Frank-Wolfe algorithm, build on an adaptation of~\citet{jaggi2013revisiting} (Section~\ref{app:proof:fw:jaggi}). A practical implementation is available in the PLUC-R package, and our numerical experiments underscore the method's empirical effectiveness.  

\paragraph*{Perspectives.}
Future directions might include pairing policies with confidence scores, providing a principled measure of uncertainty in treatment assignment. An avenue of future research is to go beyond binary treatments. Finally, incorporating a broader class of constraints into the learning problem offers another natural and potentially impactful extension. 

\section*{\centering \textsc{Acknowledgments}}
\label{acknowledgements}

This work was supported by the project ``IA pour la Santé" funded by Aniti and the Occitanie region. 
We thank Mochuan Liu for sharing and commenting on the code for BR-O as well as for helpful discussions.

\bibliographystyle{apalike}
\setcitestyle{authoryear}
\bibliography{bibliography}
\bigskip
\newpage
\begin{center}
\appendix
{\large\bf SUPPLEMENTARY MATERIALS}
\end{center}
\section{\centering Algorithms}
\label{sec:app:algo}
\subsection{The Frank-Wolfe algorithm}
\label{subsec:algo:fw}

Claims made in this section are proved in Section~\ref{app:proofs:fw}. 

\subsubsection{Oracular criteria} 

\paragraph{The Frank-Wolfe algorithm.} 

The   Frank–Wolfe    algorithm   \citep{frank1956algorithm}    introduced   in
Section~\ref{subsec:oracular}  is among  the  simplest  iterative methods  for
constrained convex optimization  over compact convex sets. It is  used in that
section to minimize oracular Lagrangian criteria of the form
\begin{equation}
    \label{eq:oracular:Lagrangian}
    \psi \mapsto \mathcal{L}_{0}(\psi,\lambda;\beta) = R_{0}(\psi) + \lambda S_{0}(\pit_{\beta,\psi}) 
\end{equation}
over  $\Psi$ \eqref{eq:convex:hull},  with $\lambda,  \beta \geq  0$. The  set
$\Psi$ is convex and weakly compact.
The Lagrangian criteria are convex functions, since each of them is the sum of
two convex functions. These statements are proved in Section~\ref{subsec:more:arguments}.

Algorithm~\ref{algo:FW} iteratively constructs  convex combinations of extreme
points of $\Psi$, yielding feasible solutions and thereby obviating projection
steps. In view  of the convergence analysis presented below,  the precision is
proportional to the inverse of the number of iterations.

Introduce the set
\begin{equation*}
    \mathcal{E}  =  \{x \mapsto  s_{\theta}(x)  =  2\expit(x^\top \theta)-1  :
    \theta \in \mathbb{R}^{d}\} \cup \{-1\}. 
\end{equation*}
The  extreme points  of  $\Psi$ lie  in  the $\|\cdot\|_{2,P_{0}}$-closure  of
$\mathcal{E}$. Starting from the initial $\psi^{0} = -1$, the $j$-th iteration
consists  of  defining  $\psi^{j+1}  \in  \Psi$ as  a  convex  combination  of
$\psi^{j}$  and an  element $s_{j}$  of $\mathcal{E}$.   The search  direction
$s_{j}$ is selected by (approximately) solving the linearized subproblem
\begin{equation}
    \label{eq:linearized}
    \argmin_{\psi   \in   \Psi}    \left\{E_{P_{0}}   \left[   \psi(X)   \cdot
        \nabla\mathcal{L}_{0}(\psi^{j}, \lambda; \beta)(X) \right] \right\} 
\end{equation}
where, for any $\psi \in \Psi$, 
\begin{equation}
    \label{eq:nabla:Lagrangian}
    \nabla \mathcal{L}_{0}(\psi,  \lambda;\beta) = 2 (\psi  - \Delta\mu_{0}) +
    \lambda \sigma_{\beta}' \circ \psi \cdot \Delta\nu_{0}. 
\end{equation}
By definition of $\Psi$ and  because the criterion in \eqref{eq:linearized} is
linear      and      $\|\cdot\|_{2,P_{0}}$-continuous,      minimizers      in
\eqref{eq:linearized}    lie   in    the   $\|\cdot\|_{2,P_{0}}$-closure    of
$\mathcal{E}$. Therefore,  to compute  $s_{j}$ we perform  stochastic gradient
descent (SGD) to numerically solve 
\begin{equation}
    \label{eq:SGD}
    \argmin_{\theta \in \mathbb{R}^{d}}  \left\{E_{P_{0}} \left[ s_{\theta}(X)
        \cdot   \nabla\mathcal{L}_{0}(\psi^{j},  \lambda;   \beta)(X)  \right]
    \right\}. 
\end{equation}
Given   the  approximate   solution  $s_{j}$,   $\psi^{j+1}$  is   defined  as
\begin{equation*}
  \psi^{j+1} = (1-\gamma_{j}) \psi^{j} + \gamma_{j} s_{j},
\end{equation*}
with  $\gamma_{j}  =  2/(2+j)$.   After $(J+1)$  iterations,  the  approximate
minimizer is $\psi^J = 2\sum_{j=1}^{J} j s_{j-1}/[J(J+1)]$ (see Section~\ref{app:closed:form}).

\begin{algorithm}
\caption{Frank-Wolfe algorithm under $P_{0}$}
\label{algo:FW}
\begin{algorithmic}
\State \textbf{require:} number of iterations $J$
  
\State \textbf{initialize:} $\psi^0 = -1$

\For{$j = 0,1,\dots, J$}
    \State set  $\gamma_{j} = 2/(2+j)$
    \State  derive $s_{j}$ by SGD, an approximate solution of \eqref{eq:linearized}
    \State update $\psi^{j+1} = (1-\gamma_{j}) \psi^{j} + \gamma_{j} s_{j}$
\EndFor
 \State \Return $\psi^{J}$
\end{algorithmic}
\end{algorithm}

\paragraph{Convergence analysis.}

Fix arbitrarily $\lambda,\beta \geq 0$.  Let $\psi_{\lambda,\beta}^{\flat}$ be
the unique  minimizer of  $\psi \mapsto  \mathcal{L}_{0}(\psi, \lambda;\beta)$
over $\Psi$ (unicity holds because  the criterion is strongly convex).  Define
$C =  4 [1+\frac{\lambda}{2}\sigma_{\beta}''(1)]$  and consider  the following
assumption:
\begin{assumption}
  \label{assum:FW}
  There exists $\delta \geq 0$ such that, at each iteration $0 \leq j \leq J$,
  the   output  $s_{j}$   of   the  SGD   designed   to  solve   approximately
  \eqref{eq:linearized} satisfies
  \begin{equation}
    \label{eq:approx:SGD}
    E_{P_{0}}   \left[s_{j}(X)  \cdot   \nabla\mathcal{L}_{0}(\psi^{j},  \lambda;
      \beta)(X)\right]  \leq  \min_{\psi\in  \Psi}  E_{P_{0}}\left[  \psi(X)\cdot
      \nabla\mathcal{L}_{0}(\psi^{j}, \lambda;  \beta)(X) \right]  + \frac{\delta
      C}{j+2}.
  \end{equation} 
\end{assumption}
Under Assumption~\ref{assum:FW}, for every $0 \leq j < J$,
\begin{equation}
    \label{eq:theoretical:bound}
    \mathcal{L}_{0}(\psi^{j+1},\lambda;\beta)                                  -
    \mathcal{L}_{0}(\psi_{\lambda,\beta}^{\flat},\lambda;\beta)           \leq
    \frac{4C(1+\frac{\delta}{2})}{j+3}. 
\end{equation} 
The     proof     of      \eqref{eq:theoretical:bound}     is     given     in
Section~\ref{app:proof:fw:jaggi}. It is a simple  adaptation of the proof of a
similar result by~\cite{jaggi2013revisiting}.

\subsubsection{Empirical Frank-Wolfe algorithm}

In practice, we repeatedly use the Frank-Wolfe algorithm to minimize empirical
(as opposed to oracular) Lagrangian criteria of the form
\begin{equation}
  \label{eq:empirical:Lagrangian}
  \psi \mapsto \mathcal{L}_{\bign_{1}\cup \bign_{2}}^{k}(\psi,\lambda;\beta) =
  R_{\bign_{1}\cup\bign_{2}}^{k}(\psi)                +                \lambda
  S_{\bign_{1}\cup\bign_{2}}^{k}(\pit_{\beta,\psi}) 
\end{equation}
over       the       set       $\Psi$       \eqref{eq:convex:hull},       with
$(\lambda,\beta)  \in  \Lambda\times   B$  and  $k\geq  0$   an  integer  (see
Section~\ref{subsec:statistical}).  The above  presentation of the Frank-Wolfe
algorithm    can    be    easily    adapted    to    the    minimization    of
\eqref{eq:empirical:Lagrangian} over $\Psi$.

To   do   so,    one   needs   to   introduce    the   empirical   counterpart
$\nabla  \mathcal{L}_{\bign_{1}  \cup\bign_{2}}^{k}(\cdot, \lambda;\beta)$  to
$\nabla \mathcal{L}_{0}(\cdot, \lambda; \beta)$ given by
\begin{equation}
  \label{eq:nabla:Lagrangian:emp}
  \nabla  \mathcal{L}_{\bign_{1} \cup\bign_{2}}^{k}(\psi,  \lambda;\beta) =  2
  (\psi -  \Delta\mu_{\bign_{1} \cup\bign_{2}}^{k}) +  \lambda \sigma_{\beta}'
  \circ \psi \cdot \Delta\nu_{\bign_{1} \cup\bign_{2}}^{k}
\end{equation}
for any $\psi \in \Psi$. Then, one simply replaces \eqref{eq:linearized} with
\begin{equation*}
  \argmin_{\psi   \in   \Psi}    \left\{E_{P_{\bign_{1}\cup\bign_{2}}^{k}}   \left[   \psi(X)   \cdot
      \nabla\mathcal{L}_{\bign_{1}    \cup\bign_{2}}^{k}(\psi^{j},    \lambda;
      \beta)(X) \right] \right\} 
\end{equation*}
and \eqref{eq:SGD} with 
\begin{equation*}
  \argmin_{\theta                      \in                     \mathbb{R}^{d}}
  \left\{E_{P_{\bign_{1}\cup\bign_{2}}^{k}} \left[ s_{\theta}(X) 
      \cdot  \nabla\mathcal{L}_{\bign_{1}\cup\bign_{2}}^{k}(\psi^{j}, \lambda;
      \beta)(X) \right] 
  \right\}. 
\end{equation*}
The convergence analysis  can be easily adapted likewise.   We briefly discuss
some technical details in Section~\ref{app:proof:fw:jaggi}.

\subsection{Alternating procedure algorithm}
\label{app:algo:ap}

This section details the alternating, iterative, targeted procedure introduced
in Section~\ref{subsubsec:PLUC}.   Each iteration  consists of  a minimization
step        followed       by        a        targeting       step,        see
Algorithm~\ref{algo:app:alternated:proced}. 

Fix arbitrarily $\lambda,\beta  \geq 0$. Recall the  initial plug-in estimator
of                $\mathcal{L}_{0}(\cdot,\lambda,\beta)$               defined
in~\eqref{eq:initial:lagrangian}:
\begin{equation*}
  \mathcal{L}_{\bign_{1}\cup        \bign_{2}}^{0}(\cdot,\lambda;\beta)       =
  \mathcal{L}_{P_{\bign_{1}    \cup    \bign_{2}}^{0}}(\cdot,    \lambda;\beta).
\end{equation*}
At the $k$-th  iteration, we have access to $k$  first approximate minimizers,
denoted   as  $\psi_{\bign_{1}\cup\bign_{2},   \lambda,\beta}^{0}$,  $\ldots$,
$\psi_{\bign_{1}\cup\bign_{2}, \lambda, \beta}^{k-1}$.

\paragraph{Correction step.} 

This     sub-step    updates     $P_{\bign_{1}\cup    \bign_{2}}^{0}$     into
$P_{\bign_{1}\cup \bign_{2}}^{k} \in \mathcal{M}$ defined such that
\begin{equation*}
  \mathcal{L}_{\bign_{1}\cup \bign_{2}}^{k}(\cdot,\lambda;\beta) = 
  \mathcal{L}_{P_{\bign_{1} \cup \bign_{2}}^{k}}(\cdot, \lambda;\beta) 
\end{equation*}
targets             the             $k$            landmark             values
$\mathcal{L}_{0}(\psi_{\bign_{1}\cup\bign_{2},\lambda,\beta}^{0},\lambda;\beta)$,
$\ldots$,
$\mathcal{L}_{0}(\psi_{\bign_{1}\cup\bign_{2},\lambda,\beta}^{k-1},\lambda;\beta)$
simultaneously.

To this  end, we introduce  two parametric  models $\{\mu_{\bign_{1}}^{k}(\epsilon) :  \epsilon \in \mathbb{R}^{k}\}$
and
$\{\nu_{\bign_{1}}^{k}(\epsilon) :  \epsilon \in \mathbb{R}^{k}\}$ that fluctuate  the initial
estimators  $\mu_{\bign_{1}}^{0}$ and  $\nu_{\bign_{1}}^{0}$ of  $\mu_{0}$ and
$\nu_{0}$.         Their        definitions
depend on the  initial estimator $e_{\bign_{1}}^{0}$ of  the propensity score
$e_{0}$, for any $\epsilon \in \mathbb{R}^{k}$,
\begin{align}
  \label{eq:mu:update}
  \mu_{\bign_{1}}^{k}(\epsilon)(A,X)
  &=      \expit\left(\logit\left(      \mu_{\bign_{1}}^{0}(A,X)\right)      +
    \tfrac{2A-1}{e_{\bign_{1}}^{0}(A,X)}                                  \cdot
    \sum_{\ell=0}^{k-1}\epsilon_{\ell} \cdot 
    \psi_{\bign_{1}\cup \bign_{2},\lambda, \beta}^{\ell}(X) \right),\\ 
  \nu_{\bign_{1}}^{k}(\epsilon)(A,X)
  \label{eq:nu:update}
  &=     \expit\left(    \logit\left(     \nu_{\bign_{1}}^{0}(A,X)\right)    +
    \tfrac{2A-1}{e_{\bign_{1}}^{0}(A,X)}        \cdot       \sum_{\ell=0}^{k-1}
    \epsilon_{\ell}     \cdot      \sigma_{\beta}\circ     \psi_{\bign_{1}\cup
    \bign_{2},\lambda,\beta}^{\ell}(X)\right). 
\end{align}
The     optimal     fluctuation      parameters     are     any     minimizers
$\epsilon_{\mu,\bign_{1}\cup\bign_{2}}$                                    and
$\epsilon_{\nu,\bign_{1}\cup\bign_{2}}$     of     the     convex     criteria
$\ell_{\mu,\bign_{1}  \cup   \bign_{2}}^{k}$  and   $\ell_{\nu,\bign_{1}  \cup
  \bign_{2}}^{k}$ respectively given by 
\begin{align}
  \label{eq:l:Y}
  \epsilon
  &\mapsto    
    -\frac{3}{n}\sum_{i\in \bign_{2}}\left\{Y_i\cdot 
               \log[\mu_{\bign_{1}}^{k}(\epsilon)(A_{i},X_{i})] + (1-Y_i)\cdot
    \log[1-\mu_{\bign_{1}}^{k}(\epsilon)(A_{i},X_{i})]\right\},\\
  \label{eq:l:xi} 
  \epsilon
  &\mapsto    
    -\frac{3}{n}\sum_{i\in \bign_{2}}\left\{\xi_i\cdot 
    \log[\nu_{\bign_{1}}^{k}(\epsilon)(A_{i},X_{i})]      +     (1-\xi_i)\cdot
    \log[1-\nu_{\bign_{1}}^{k}(\epsilon)(A_{i},X_{i})]\right\}. 
\end{align}
We then define
\begin{align*}
  \Delta\mu_{\bign_{1}\cup\bign_{2}}^{k} (\cdot)
  &=  \mu_{\bign_{1}}^{k} (\epsilon_{\mu,\bign_{1}\cup\bign_{2}})
    (1, \cdot) - \mu_{\bign_{1}}^{k} (\epsilon_{\mu,\bign_{1}
    \cup\bign_{2}}) (0, \cdot),\\
  \Delta\nu_{\bign_{1}\cup\bign_{2}}^{k} (\cdot)
  &=\nu_{\bign_{1}}^{k}   (\epsilon_{\nu,\bign_{1}\cup\bign_{2}})
    (1, \cdot) -  \nu_{\bign_{1}}^{k} (\epsilon_{\nu,\bign_{1} \cup\bign_{2}})
    (0, \cdot), 
\end{align*}
and  assume   that  the  statistical   model  $\mathcal{M}$  contains   a  law
$P_{\bign_{1}     \cup     \bign_{2}}^{k}    \in\mathcal{M}$     such     that
$\Delta\mu_{P_{\bign_{1}  \cup  \bign_{2}}^{k}}  =  \Delta\mu_{\bign_{1}  \cup
  \bign_{2}}^{k}$,
$\Delta\nu_{P_{\bign_{1}  \cup  \bign_{2}}^{k}}  =  \Delta\nu_{\bign_{1}  \cup
  \bign_{2}}^{k}$, $e_{P_{\bign_{1} \cup \bign_{2}}^{k}} = e_{\bign_{1}}^{0}$,
and
$P_{\bign_{1}   \cup  \bign_{2},X}^{k}   =   (3/n)   \sum_{i  \in   \bign_{2}}
\Dirac(X_{i})$.

\paragraph{Minimization step.}

This  sub-step  consists of  computing  the  $(k+1)$-th approximate  minimizer
$\psi_{\bign_{1}\cup         \bign_{2},\lambda,         \beta}^{k}$         of
$\psi \mapsto  \mathcal{L}_{\bign_{1}\cup \bign_{2}}^{k}(\psi,\lambda;\beta)$,
by using the Frank-Wolfe algorithm.

\begin{algorithm}[H]
    \caption{Alternating procedure for policy estimation}
    \begin{algorithmic}
      \State \textbf{require:} $\lambda>0$, $\beta \geq 0$,  tolerance parameter $\gamma>0$,
      maximum number of iterations $K$
      \State \textbf{initialization:} set $k=0$, $\psi_{\bign_{1}\cup \bign_{2},
        \lambda,\beta}^{-1}=-1$     and compute   $\psi_{\bign_{1}\cup     \bign_{2},
        \lambda,\beta}^0$, an approximate minimizer of 
      $\psi  \mapsto  \mathcal{L}_{\bign_{1}\cup  \bign_{2}}^0(\psi,  \lambda;
      \beta)$ over $\Psi$, using the Frank-Wolfe Algorithm~\ref{algo:FW}
      \While{$\sum_{i\in                        \bign_{2}}[\psi_{\bign_{1}\cup
            \bign_{2},\lambda,\beta}^{k}(X_{i})      -     \psi_{\bign_{1}\cup
            \bign_{2}, \lambda, \beta}^{k-1}(X_{i})]^2  > \gamma$ \textbf{ and
          } $k < K$} 
        \State \textbf{correction step:}
        \State compute  $\epsilon_{\mu,\bign_{1}\cup\bign_{2}}^{k+1} \in \argmin
        \ell_{\mu,\bign_{1}\cup    \bign_{2}}^{k+1}$     \eqref{eq:l:Y}    and
        $\epsilon_{\nu,\bign_{1}\cup\bign_{2}}^{k+1}        \in        \argmin
        \ell_{\nu,\bign_{1}\cup \bign_{2}}^{k+1}$ \eqref{eq:l:xi} 
        \State         define          the         targeted         estimators
        $\mu_{\bign_{1}}^{k+1}(\epsilon_{\mu,\bign_{1}\cup\bign_{2}}^{k+1})$ \eqref{eq:mu:update}
        and
        $\nu_{\bign_{1}}^{k+1}(\epsilon_{\nu,\bign_{1}\cup\bign_{2}}^{k+1})$
        \eqref{eq:nu:update}
        \State define  $P_{\bign_{1} \cup \bign_{2},X}^{k+1}  \in \mathcal{M}$
        such   that  $\mathcal{L}_{P_{\bign_{1}   \cup  \bign_{2}}^{k+1}}(\cdot,
        \lambda;\beta) = \mathcal{L}_{\bign_{1} \cup \bign_{2}}^{k+1}(\cdot,
        \lambda;\beta)$ simultaneously 

        \State \quad targets the $(k+1)$ landmark values
        $\mathcal{L}_{0}(\psi_{\bign_{1}\cup\bign_{2},\lambda,\beta}^{0},\lambda;\beta)$, 
        $\ldots$, $\mathcal{L}_{0}(\psi_{\bign_{1}\cup\bign_{2},\lambda,\beta}^{k},\lambda;\beta)$
        \State \textbf{minimization step:} 
        \State  compute $\psi_{\bign_{1}\cup  \bign_{2}, \lambda,\beta}^{k+1}$,
        an approximate minimizer of 
        $\psi   \mapsto   \mathcal{L}_{\bign_{1}\cup   \bign_{2}}^{k+1}(\psi,
        \lambda; \beta)$ over $\Psi$,
        \State \quad using the Frank-Wolfe Algorithm~\ref{algo:FW}
        \State $k\leftarrow k+1$
      \EndWhile
      \State  \Return  $\psi_{\bign_{1}\cup  \bign_{2},  \lambda,\beta}^{*}  =
      \psi_{\bign_{1}\cup \bign_{2}, \lambda,\beta}^{k}$ 
    \end{algorithmic}
    \label{algo:app:alternated:proced}
\end{algorithm}

\subsection{Main algorithm}

This section provides a unified  algorithm encompassing the ``Naive PLUC'' and
PLUC  procedures  presented  in Section~\ref{subsec:statistical}.   They  only
differ in step~2.

\begin{algorithm}[H]
  \caption{Main   algorithm:   ``Naive    PLUC''   \textbf{(a)}   and   PLUC
    \textbf{(b)}}
  \begin{scriptsize}
    \begin{algorithmic}
      \State Let $\{1, \ldots, n\} =\bign_{1} \cup \bign_{2}\cup \bign_{3}$ with
      mutually disjoint sets of the same cardinality up to 1
      \State \textbf{\underline{step 1:} estimation of nuisance parameters}
      \State estimate  $\mu_{0}$ with  $\mu_{\bign_{1}}^{0}$, using  $\{O_{i}: i
      \in  \bign_{1}\}$,  and  $\mu_{\bign_{3}}^{0}$,   using  $\{O_{i}:  i  \in
      \bign_{3}\}$;       let        $\Delta\mu_{\bign_{1}}^{0}(\cdot)       =
      \mu_{\bign_{1}}^{0}(1,\cdot) - \mu_{\bign_{1}}^{0}(0,\cdot)$ 
      \State estimate  $\nu_{0}$ with  $\nu_{\bign_{1}}^{0}$, using  $\{O_{i}: i
      \in  \bign_{1}\}$,  and  $\nu_{\bign_{3}}^{0}$,   using  $\{O_{i}:  i  \in
      \bign_{3}\}$;      let       $\Delta\nu_{\bign_{1}}^{0}(\cdot)      =
      \nu_{\bign_{1}}^{0}(1,\cdot)-\nu_{\bign_{1}}^{0}(0,\cdot)$ 
      \State estimate  $e_{0}$ with  $e_{\bign_{1}}^{0}$, using  $\{O_{i}: i
      \in \bign_{1}\}$, and $e_{\bign_{3}}^{0}$, using $\{O_{i}: i \in
      \bign_{3}\}$
      \State define $P_{\bign_{1}     \cup     \bign_{2}}^{0}    \in\mathcal{M}$     such     that
      $\Delta\mu_{P_{\bign_{1}  \cup  \bign_{2}}^{0}}  =  \Delta\mu_{\bign_{1}}^{0}$,
      $\Delta\nu_{P_{\bign_{1}          \cup          \bign_{2}}^{0}}          =
      \Delta\nu_{\bign_{1}}^{0}$, $e_{P_{\bign_{1}  \cup \bign_{2}}^{0}}  = e_{\bign_{1}}^{0}$,
      \State  \quad 
      and  $P_{\bign_{1} \cup  \bign_{2},X}^{0} =  (3/n) \sum_{i  \in \bign_{2}}
      \Dirac(X_{i})$ 
      \For{$\beta \in B$}
        \For{$\lambda \in \Lambda$}
          \State \textbf{\underline{step 2:} policy learning}
          \State \textbf{(a)}  compute $\psi_{\bign_{1}\cup  \bign_{2}, \lambda,
            \beta}$,   an   approximate   minimizer   of   $\psi   \mapsto
          \mathcal{L}_{P_{\bign_{1}   \cup    \bign_{2}}^{0}}(\psi,   \lambda;
          \beta)$ over     $\Psi$,    
          \State     \qquad    using     the    Frank-Wolfe
          Algorithm~\ref{algo:FW} 
          \State \textbf{(b)}  derive $\psi_{\bign_{1}\cup  \bign_{2}, \lambda,
            \beta}$     via     the     alternating     procedure     (see
          Algorithm~\ref{algo:app:alternated:proced})
          \State \qquad (the $*$-superscript is intentionally dropped to unify
          the rest of the algorithm) 
          \State  derive  $\pit_{\bign_{1}\cup  \bign_{2}, \lambda,  \beta}  =
          \sigma_{\beta} \circ \psi_{\bign_{1}\cup  \bign_{2}, \lambda,
            \beta}$
          \State \textbf{\underline{step 3:} policy evaluation}
          \State compute $\lNu_{\bign_{3}}^{*}(\psi_{\bign_{1}\cup
            \bign_{2},   \lambda,   \beta})$ and 
          $\overline{S}_{\bign_{3}}^{*}(\psi_{\bign_{1}\cup
            \bign_{2}, \lambda, \beta})$, 95\%-confidence targeted upper- and lower-bounds for
          \State \quad  $S_{0}(\pit_{\bign_{1}\cup 
            \bign_{2}, \lambda, \beta})$ and $\mathcal{V}_{P_{0}} (\pit_{\bign_{1}\cup
            \bign_{2},   \lambda,   \beta})$,   using   $\mu_{\bign_{3}}^{0}$,
          $\nu_{\bign_{3}}^{0}$, and $e_{\bign_{3}}^{0}$ 
          \If{$\overline{S}_{\bign_{3}}^{*}(\psi_{\bign_{1}\cup \bign_{2}, \lambda, \beta})\leq 0$}
            \State \textbf{break} from current $\lambda$ loop
          \EndIf
        \EndFor
      \EndFor
      \State \textbf{\underline{step 4:} best policy identification}
      \If{$\min\{\overline{S}_{\bign_{3}}^{*}(\psi_{\bign_{1}\cup \bign_{2}, \lambda, \beta}):(\lambda,\beta)\in\Lambda\times B \}>0$}
        \State \Return 0
      \Else
        \State  \Return $\pit_{\bign_{1}\cup  \bign_{2},\lambda^{*},\beta^{*}}$
        with     $(\lambda^{*},      \beta^{*})     \in      \argmax     \left\{
          \lNu_{\bign_{3}}^{*}(\pit_{\bign_{1}\cup  \bign_{2},\lambda,\beta})  :
          (\lambda,    \beta)\in    \Lambda    \times   B    \text{    s.t.    }
          \uS_{\bign_{3}}^{*}(\pit_{\bign_{1}\cup  \bign_{2},\lambda,\beta})\leq
          0 \right\}$
      \EndIf
    \end{algorithmic}
  \end{scriptsize}
  \label{algo:pluc}
\end{algorithm}


\newpage
\section{\centering \textsc{Proofs}}
\label{sec:proofs}
\subsection{Proofs for Sections~\ref{subsec:policy:learning},~\ref{subsec:challenge}, and beyond}
\label{subsec:more:arguments}

\paragraph{Lipschitz-continuity.} 

It is easily checked that, for any $\beta>0$, $\sigma_{\beta}$ is twice differentiable, with first and second derivatives characterized by
\begin{equation*}
    \sigma_\beta'(t) = c(\beta)^{-1} \frac{\beta e^{\beta t}}{(1+ e^{\beta t})} > 0 \quad\text{and}\quad \sigma_\beta''(t)= c(\beta)^{-1} \frac{\beta^2 e^{\beta t}}{(1+ e^{\beta t})^2} > 0.
\end{equation*}
Thus, the restriction of $\sigma_{\beta}$ to $[-1,1]$ is $\sigma_{\beta}'(1)$-Lipschitz and convex. Obviously, the linear function $\sigma_{0}$ is also $\sigma_{0}'(1)$-Lipschitz and convex. 

Fix arbitrarily  $\beta\geq 0$ and  $\psi_{1}, \psi_{2} \in \Psi$.   Using the
fact      that      $0      \leq     \Delta\nu_{0}      \leq      1$      (see
Assumption~\ref{assum:delet_effect}), we have
\begin{equation*}
    |S_{0}(\sigma_{\beta} \circ \psi_{1}) - S_{0}(\sigma_{\beta} \circ \psi_{2})| \leq \sigma_{\beta}'(1) \cdot P_{0}[|\psi_{1} - \psi_{2}| \cdot \Delta\nu_{0}]
    \leq \sigma_{\beta}'(1) \cdot \|\psi_{1} - \psi_{2}\|_{2,P_{0}},
\end{equation*}
revealing that $\psi \mapsto S_{0}(\sigma_{\beta} \circ \psi)$ is Lipschitz-continuous between $(\Psi, \|\cdot\|_{2,P_{0}})$ and $(\mathbb{R}, |\cdot|)$. Likewise,
\begin{equation*}
    |R_{0}(\psi_{1}) - R_{0}(\psi_{2})| \leq P_{0} [|\psi_{1} - \psi_{2}| \cdot |\psi_{1} + \psi_{2} - 2 \Delta\mu_{0}|] \leq 4 \|\psi_{1} - \psi_{2}\|_{2,P_{0}},
\end{equation*}
hence $R_{0}$ is also Lipschitz-continuous between $(\Psi, \|\cdot\|_{2,P_{0}})$ and $(\mathbb{R}, |\cdot|)$.

\paragraph{Convexity.} 

Set arbitrarily $\gamma_{1},\gamma_{2} \geq 0$ such that $\gamma_{1} + \gamma_{2}=1$, and define $\psi_{12} = \gamma_{1}\psi_{1} + \gamma_{2}\psi_{2}$. By 1-strong convexity of $u \mapsto u^{2} - 2mu$ from $\mathbb{R}$ to $\mathbb{R}$ for any $m\in\mathbb{R}$, it holds $P_{0}$-almost surely that
\begin{multline*}
    \psi_{12}(X)^{2} -2 \psi_{12}(X)\cdot \Delta\mu_{0}(X) - \tfrac{1}{2} \psi_{12}(X)^{2} 
    \\\leq \sum_{k=1}^{2}\gamma_{k} \left[\psi_{k}(X)^{2} -2 \psi_{k}(X)\cdot\Delta\mu_{0}(X) - \tfrac{1}{2} \psi_{k}(X)^{2}\right].
\end{multline*}
Integrating the above inequality with respect to $P_{0}$ leads to 
\begin{equation*}
    R_{0}(\psi_{12}) - \tfrac{1}{2} \|\psi_{12}\|_{2,P_{0}}^{2} \leq \gamma_{1}\left[R_{0}(\psi_{1}) - \tfrac{1}{2} \|\psi_{1}\|_{2,P_{0}}^{2}\right] + \gamma_{2} \left[R_{0}(\psi_{2}) - \tfrac{1}{2} \|\psi_{2}\|_{2,P_{0}}^{2}\right].
\end{equation*}
Since $\gamma_{1}, \gamma_{2}, \psi_{1}, \psi_{2}$ have been arbitrarily chosen, we conclude that $R_{0}$ is 1-strongly convex between $(\Psi, \|\cdot\|_{2,P_{0}})$ and $(\mathbb{R}, |\cdot|)$. Similarly, since $\sigma_{\beta}$ is a convex function and $\Delta\nu_{0}$ takes non-negative values, $\psi \mapsto S_{0}(\sigma_{\beta} \circ \psi)$ is  convex. 

\paragraph{Weak compactness.}

The set $\Psi$ enjoys the following properties. It is not empty. It is closed in $L^{2}(P_{0})$. It is  convex. It is a subset of $\bar{B}_{2,P_{0}}(0,1)$, the closed $\|\cdot\|_{2,P_{0}}$-ball centered at 0 and with radius 1, since $\|\psi\|_{\infty} \leq 1$ for all $\psi \in \Psi$.

Recall that $\Psi_{0,\beta} = \{\psi \in \Psi : S_{0}(\sigma_{\beta} \circ \psi) \leq 0\}$.  The set $\Psi_{0,\beta}$ enjoys the following properties. First, it is not empty since $S_{0}(\sigma_{\beta} \circ \psi) = -\alpha \leq 0$ when $\psi = -1$ (a choice for which $\sigma_{\beta} \circ \psi$ maps $\mathcal{X}$ to $\{0\}$). Second, it is closed in $(\Psi, \|\cdot\|_{2,P_{0}})$ since it is the preimage of $\mathbb{R}_{-}$ by the continuous mapping $\psi \mapsto S_{0}(\sigma_{\beta} \circ \psi)$. Third, it is convex since it is a sublevel set for the convex mapping $\psi \mapsto S_{0}(\sigma_{\beta} \circ \psi)$. Fourth, as a subset of $\Psi$, it is also a subset of $\bar{B}_{2,P_{0}}(0,1)$. 

The next arguments involve the weak topology associated with  $(L^{2}(P_{0}), \|\cdot\|_{2,P_{0}})$~\cite[][Sections~3.2 and 5.2]{Brezis2011}. In view of \cite[][Theorems~3.16 and 5.5]{Brezis2011}, $\bar{B}_{2,P_{0}}(0,1)$ is weakly compact. Moreover, by \cite[][Theorem~3.7]{Brezis2011}, $\Psi$ and $\Psi_{0,\beta}$ are closed in the weak topology since they are closed (in the strong topology) and convex. Therefore, as  closed subsets of a compact set, $\Psi$ and $\Psi_{0,\beta}$ are weakly compact. 

\paragraph{Existence and unicity of the minimizer.}

We now extend the restriction of $R_{0}$ to $\Psi_{0,\beta}$ into a mapping $\tilde{R}_{0}$ between $L^{2}(P_{0})$ and $\mathbb{R}\cup\{+\infty\}$ by setting $\tilde{R}_{0}(\psi) = R_{0}(\psi)$ if $\psi \in \Psi_{0,\beta}$ and $\tilde{R}_{0}(\psi) = +\infty$ otherwise. The mapping $\tilde{R}_{0}$ remains convex. It is moreover weakly-lower semicontinuous because, for any $t \in \mathbb{R}$, $\{\psi\in L^{2}(P_{0}) : \tilde{R}_{0}(\psi) \leq t\} = \{\psi\in \Psi_{0,\beta} : R_{0}(\psi) \leq t\}$ is both closed (in the strong topology, as the preimage of the closed set $(-\infty,t]$ by the continuous mapping $R_{0}$), and convex (since $R_{0}$ is convex on the convex set $\Psi_{0,\beta}$), hence closed in the weak topology by \cite[][Theorem~3.7]{Brezis2011}. In summary, $R_{0}$ is weakly-lower semicontinuous on the  weakly compact set $\Psi_{0,\beta}$; consequently, $R_{0}$ admits and achieves a minimum on $\Psi_{0,\beta}$~\cite[][Section~1.4]{Brezis2011}.  The strong-convexity of $R_{0}$ implies the unicity of the minimizer in \eqref{eq:actual:policy}.

\paragraph{Extensions for Section~\ref{subsec:oracular}.}

Fix $\lambda \geq 0$. The Lagrangian criterion $\mathcal{L}_{0}(\cdot, \lambda;\beta)$, defined in \eqref{eq:Lagrange}, is convex since it is the sum of two convex functions. By the  continuity proven above, it is also continuous. As above, we can extend  $\mathcal{L}_{0}(\cdot, \lambda;\beta)$ to $\tilde{\mathcal{L}}_{0}(\cdot, \lambda;\beta)$ defined between $L^{2}(P_{0})$ and $\mathbb{R}\cup \{+\infty\}$ which is weakly-lower semicontinuous. It then follows that $\mathcal{L}_{0}(\cdot, \lambda;\beta)$ admits a minimizer on $\Psi$. 

\paragraph{Extensions for Sections~\ref{subsec:statistical} and \ref{subsec:algo:fw}.}

We note  that the above convexity  argument readily extends to  the case where
$\Delta\mu_{0}$  and  $\Delta\nu_{0}$ are  replaced  by  estimators (with  the
latter restricted to  non-negative values), and the marginal  law of $P_{0,X}$
of    $X$    under   $P_{0}$    is    replaced    by   the    empirical    law
$P_{\bign_{1},X}   =  (3/n)   \sum_{i\in   \bign_{1}}  \Dirac(X_{i})$,   where
$\bign_{1}$      indexes     the      first      fold      of     data      in
Section~\ref{subsec:statistical}.   Consequently,  the   empirical  Lagrangian
criteria in Sections~\ref{subsec:statistical}  and \ref{subsec:algo:fw} remain
convex functions.

In addition, the  definition of $\Psi$ \eqref{eq:convex:hull}  can be slightly
altered  to  rigorously justify  the  validity  of the  empirical  Frank-Wolfe
algorithm   in  Sections~\ref{subsec:statistical}   and  \ref{subsec:algo:fw}.
Suppose  indeed  that  the  closure  appearing  in  \eqref{eq:convex:hull}  is
understood w.r.t.\  $L^{2}(Q)$ with $Q  = \tfrac{1}{2}[P_{0,X} +  P_{n,X}]$ as
opposed to  $L^{2}(P_{0})$. Then  the elements of  $\Psi$ are  defined without
ambiguity at  $X_{1}, \ldots, X_{n}$.  Moreover, the  above proof of  the weak
compactness  of  $\Psi$  is  still  valid  with  $Q$  substituted  for  $P_{0}$.
Therefore,  $\Psi$   is  compact   in  the   weak  topology   associated  with
$(L^{2}(Q), \|\cdot\|_{2,Q})$.

Furthermore,  for any  $(\lambda,\beta) \in  \Lambda \times  B$ and  iteration
$k            \geq           0$,            we           can            extend
$\mathcal{L}_{\bign_{1}\cup\bign_{2}}^{k}(\cdot,\lambda;\beta)$             to
$\tilde{\mathcal{L}}_{\bign_{1}\cup\bign_{2}}^{k}(\cdot,\lambda;\beta)$
defined  between  $L^{2}(Q)$  and   $\mathbb{R}  \cup  \{+\infty\}$  which  is
weakly-lower       semicontinuous.       It       then      follows       that
$\mathcal{L}_{\bign_{1}\cup\bign_{2}}^{k}(\cdot,\lambda;\beta)$    admits    a
minimizer on $\Psi$.

\subsection{Proofs for Section~\ref{subsec:algo:fw}} 
\label{app:proofs:fw}

\subsubsection{Differentiability of the Lagrangian criteria}
\label{app:proof:fw:diff}

Fix $P \in  \mathcal{M}$ and $\beta, \lambda\geq 0$.  For  instance, $P$ could
be equal to  $P_{0}$ or to one of the  laws $P_{\bign_{1} \cup \bign_{2}}^{k}$
from Section~\ref{subsec:statistical}.

Let $\phi: \mathbb{R}\times \mathcal{X} \rightarrow \mathbb{R}$ be given by
    $\phi(t,x) = t^2 -2t\cdot\Delta\mu_{P}(x) + \lambda[\sigma_{\beta}(t)\cdot \Delta\nu_{P}(x) -\alpha]$.
For any $\psi \in \Psi$, define 
    $\mathcal{L}_{P}(\psi, \lambda;\beta) = E_{P} \left[\phi(\psi(X), X)\right]$.
In particular,  $\mathcal{L}_{P_{0}}(\psi, \lambda;\beta) = \mathcal{L}_{0}(\psi, \lambda;\beta)$ \eqref{eq:oracular:Lagrangian}. 

For any $\psi \in \Psi$ and any perturbation $h \in L^{2}(P)$, we have
\begin{multline}
    \label{eq:expansion}
    \mathcal{L}_{P}(\psi + h, \lambda; \beta) - \mathcal{L}_{P}(\psi, \lambda; \beta) - 2 P (\psi - \Delta\mu_{P})\cdot h - \lambda P \sigma_{\beta}'\circ \psi \cdot \Delta\nu_{P}\cdot h\\
    =     P \{h^2 + \lambda [\sigma_{\beta}\circ (\psi+h) - \sigma_{\beta} \circ \psi - \sigma_{\beta}'\circ \psi] \cdot \Delta\nu_{P}\}.
\end{multline}
In      view     of      the     analysis      of     $\sigma_{\beta}$      in
Section~\ref{subsec:more:arguments}, a  second order Taylor  expansion ensures
that, pointwise,
\begin{equation}
\label{eq:second:order:sigma}
    |\sigma_{\beta}\circ (\psi+h) - \sigma_{\beta} \circ \psi - \sigma_{\beta}'\circ \psi| \leq \frac{1}{2}\sigma_{\beta}''(1) h^{2}. 
\end{equation}
Consequently, if
\begin{equation*}
  \nabla \mathcal{L}_{P}(\psi, \lambda;\beta) = 2 (\psi - \Delta\mu_{P}) + \lambda \sigma_{\beta}'\circ \psi \cdot \Delta\nu_{P},
\end{equation*}
then   \eqref{eq:expansion}   and  $0   \leq   \Delta\nu_{P}   \leq  1$   (see
Assumption~\ref{assum:delet_effect}) imply
\begin{equation}
    \label{eq:expansion:two}
    \mathcal{L}_{P}(\psi + h, \lambda; \beta) - \mathcal{L}_{P}(\psi, \lambda; \beta) = P \nabla \mathcal{L}_{P}(\psi, \lambda;\beta) \cdot h  + O(\|h\|_{2,P}^{2}).
\end{equation}
In                                                                 particular,
$\nabla      \mathcal{L}_{P_{0}}(\cdot,      \lambda;\beta)      =      \nabla
\mathcal{L}_{0}(\cdot, \lambda;\beta)$ defined in \eqref{eq:nabla:Lagrangian}.

\subsubsection{Convergence analysis of the Frank-Wolfe algorithm}
\label{app:proof:fw:jaggi}

Fix    arbitrarily    $\lambda,    \beta    \geq    0$.     The    proof    of
\eqref{eq:theoretical:bound}      is     a      straightforward     adaptation
of~\cite[][Theorem~1]{jaggi2013revisiting}.  It unfolds in three steps.

\paragraph*{Step~1: preliminary.}

Introduce $g_{0}:\Psi \to \mathbb{R}_{+}$ given by
\begin{equation*}
  g_{0}(\psi)    =    \max_{s\in     \Psi}    P_{0} \nabla
  \mathcal{L}_{0}(\psi,\lambda;\beta)\cdot   (\psi-s). 
\end{equation*}
By convexity of $\mathcal{L}_{0}(\cdot,\lambda;\beta)$ on $\Psi$,
\begin{equation*}
  \mathcal{L}_{0}(\psi',\lambda;\beta)                                    \geq
  \mathcal{L}_{0}(\psi,\lambda;\beta) + P_{0} \nabla
  \mathcal{L}_{0}(\psi,\lambda;\beta) \cdot (\psi'-\psi)
\end{equation*}
for all  $\psi,\psi'\in\Psi$. Consequently, it  holds for all $\psi  \in \Psi$
that
\begin{equation}
  \label{eq:certificate}
  g_{0}(\psi)        \geq         \mathcal{L}_{0}(\psi,\lambda;\beta)        -
  \mathcal{L}_{0}(\psi_{\lambda,\beta}^{\flat},\lambda;\beta). 
\end{equation}
In words,  $g_{0}$ can serve as  a ``certificate''. The key  inequality of the
proof              is              an              upper-bound              on
$\mathcal{L}_{0}(\psi^{j+1},\lambda;\beta)                                   -
\mathcal{L}_{0}(\psi^{j},\lambda;\beta)$ expressed in terms of $g_{0}$ and the
``curvature''    of    $\mathcal{L}_{0}(\cdot,\lambda;\beta)$:    for    every
$0 \leq j < J$, it holds that
\begin{equation}
  \label{eq:key:theoretical:bound}
  \mathcal{L}_{0}(\psi^{j+1},\lambda;\beta)                             -
  \mathcal{L}_{0}(\psi^{j},\lambda;\beta) \leq - \gamma_{j} g_{0}(\psi^{j}) +
  C(1+\tfrac{\delta}{2}) \gamma_{j}^{2},
\end{equation}
where $C = 4[1 + \tfrac{\lambda}{2} \sigma_{\beta}''(1)]$ is the same constant
as in Assumption~\ref{assum:FW}.

\paragraph*{Step~2: proof of \eqref{eq:key:theoretical:bound}.}

Fix  arbitrarily $0  \leq  j  < J$.   In  view  of \eqref{eq:expansion}  (with
$\psi                    =                    \psi^{j}$                    and
$h   =   \psi^{j+1}   -   \psi^{j}  =   \gamma_{j}   (s_{j}   -   \psi^{j})$),
\eqref{eq:second:order:sigma} and the fact that  $0 \leq \Delta\nu_{0} \leq 1$
by Assumption~\ref{assum:delet_effect}, we have
\begin{align}
  \notag
  & \mathcal{L}_{0}(\psi^{j+1},\lambda;\beta) -
    \mathcal{L}_{0}(\psi^{j},\lambda;\beta)\\
  \notag
  & \leq P_{0}
    \nabla\mathcal{L}_{0}(\psi^{j},  \lambda;\beta)\cdot (\psi^{j+1}-\psi^{j})
    +           [1          +           \tfrac{\lambda}{2}\sigma_{\beta}''(1)]
    P_{0}(\psi^{j+1}-\psi^{j})^{2}\\
  \label{eq:proof:key:theoretical:bound}
  & =  \gamma_{j} P_{0}  \nabla\mathcal{L}_{0}(\psi^{j}, \lambda;\beta)\cdot
    (s_{j}-\psi^{j}) 
    +         \gamma_{j}^{2}  [1          +           \tfrac{\lambda}{2}\sigma_{\beta}''(1)]
    P_{0}(s_{j}-\psi^{j})^{2}.
\end{align}
The      inequality~\eqref{eq:approx:SGD}      concerning      $s_{j}$      in
Assumption~\ref{assum:FW} is equivalent to
\begin{equation*}
  P_{0}  \nabla\mathcal{L}_{0}(\psi^{j}, \lambda;\beta)\cdot
  (s_{j}-\psi^{j}) \leq -g_{0}(\psi^{j}) +  C\tfrac{\delta}{2}\gamma_{j}. 
\end{equation*}
Moreover,   $P_{0}    (s_{j}   -   \psi^{j})^{2}   \leq    4$.    Therefore,
\eqref{eq:proof:key:theoretical:bound}       straightforwardly       implies
\eqref{eq:key:theoretical:bound}.

\paragraph*{Step~3: conclusion.}

We now  exploit $g_{0}$  as a  certificate.  For
notational                          simplicity,                          let
$\mathcal{H}_{0}(\psi,\lambda;\beta) = \mathcal{L}_{0}(\psi,\lambda;\beta) -
\mathcal{L}_{0}(\psi_{\lambda,\beta}^{\flat},\lambda;\beta)$    for    every
$\psi\in\Psi$.       In     view      of     \eqref{eq:certificate}      and
\eqref{eq:key:theoretical:bound}, it holds for all $0 \leq j < J$ that
\begin{equation}
  \label{eq:H0:ineq}
  \mathcal{H}_{0}(\psi^{j+1},\lambda;\beta)                             \leq
  (1     -     \gamma_{j})     \mathcal{H}_{0}(\psi^{j},\lambda;\beta)     +
  C(1+\tfrac{\delta}{2}) \gamma_{j}^{2}.
\end{equation}
The  conclusion follows  by recursion.   If $j=0$,  then $\gamma_{j}=1$  and
\eqref{eq:H0:ineq} implies
\begin{equation*}
  \mathcal{H}_{0}(\psi^{j+1},\lambda;\beta)  \leq C(1+\tfrac{\delta}{2})  \leq
  \frac{4C(1+\tfrac{\delta}{2})}{j+3},
\end{equation*}
thus proving the validity of \eqref{eq:theoretical:bound} for $j=0$. Suppose
now   that  \eqref{eq:theoretical:bound}   is  valid   $j=j^{*}$  for   some
$0  \leq j^{*}  <  (J-1)$  and let  us  prove that  it  then  also does  for
$j=(j^{*}+1)$.     By    \eqref{eq:H0:ineq}     and    the    validity    of
\eqref{eq:theoretical:bound} for $j=j^{*}$, we have
\begin{align*}
  \mathcal{H}_{0}(\psi^{j^{*}+2},\lambda;\beta)
  & \leq (1-\gamma_{j^{*}+1})  \mathcal{H}_{0}(\psi^{j^{*}+1},\lambda;\beta) + C(1 +
    \tfrac{\delta}{2}) \gamma_{j^{*}+1}^{2}\\
  & \leq (1-\gamma_{j^{*}+1}) \frac{4C(1+\tfrac{\delta}{2})}{j^{*}+3}  + C(1 +
    \tfrac{\delta}{2}) \gamma_{j^{*}+1}^{2}\\
  & = 4 C(1  + \tfrac{\delta}{2}) \frac{j^{*}+2}{(j^{*}+3)^{2}} \leq \frac{4
    C(1 +       \tfrac{\delta}{2})}{j^{*}+4}. 
\end{align*}
In  words, \eqref{eq:theoretical:bound}  is  valid  for $j=(j^{*}+1)$.  This
completes the proof.

\paragraph*{Extension      for      Sections~\ref{subsec:statistical}      and
  \ref{subsec:algo:fw}.}

In practice, we repeatedly use the Frank-Wolfe algorithm to minimize empirical
Lagrangian             criteria             of            the             form
$\mathcal{L}_{\bign_{1}\cup     \bign_{2}}^{k}(\cdot,\lambda;\beta)$,     with
$k\geq 0$ an integer,  rather than $\mathcal{L}_{0}(\cdot,\lambda;\beta)$ (see
Section~\ref{subsec:statistical}  and  the  \textit{Extensions}  paragraph  in
Section~\ref{subsec:more:arguments}).                Denoting               by
$\psi_{\bign_{1}  \cup  \bign_{2},\lambda,\beta}^{\flat,k}$   a  minimizer  of
$\mathcal{L}_{\bign_{1}\cup \bign_{2}}^{k}(\cdot,  \lambda;\beta)$ over $\Psi$
(where     the     closure     is     now     w.r.t.      $L^{2}(Q)$,     with
$Q =  \tfrac{1}{2}[P_{0,X} + P_{n,X}]$), analogs  of Assumption~\ref{assum:FW}
and    \eqref{eq:theoretical:bound}     are    obtained     by    substituting
$P_{\bign_{1}\cup            \bign_{2}}^{k}$           for            $P_{0}$,
$\mathcal{L}_{\bign_{1}    \cup    \bign_{2}}^{k}(\cdot,\lambda;\beta)$    for
$\mathcal{L}_{0}(\cdot,\lambda;\beta)$,
$\nabla\mathcal{L}_{\bign_{1}     \cup    \bign_{2}}^{k}(\cdot,\lambda;\beta)$
\eqref{eq:nabla:Lagrangian:emp}             in            place             of
$\nabla\mathcal{L}_{0}(\cdot,\lambda;\beta)$,                              and
$\psi_{\bign_{1}       \cup      \bign_{2},\lambda,\beta}^{\flat,k}$       for
$\psi_{\lambda,\beta}^{\flat}$.    With   these   substitutions,   the   above
three-step proof carries over without further modifications.

\subsubsection{Closed-form  expression  for  the outputs  of  the  Frank-Wolfe
  algorithm}
  \label{app:closed:form}

We  repeatedly  rely  on  the Frank-Wolfe  algorithm  to  compute  approximate
minimizers           of            \eqref{eq:oracular:Lagrangian}           or
\eqref{eq:empirical:Lagrangian},      see      Sections~\ref{subsec:oracular},
\ref{subsec:statistical} and Algorithm~\ref{algo:FW}.  It is easy to check, by
recursion,  that $\psi^{j}  =  2 \sum_{k=1}^{j}  ks_{k-1}/[j(j+1)]$ for  every
$1 \leq j \leq J$.

First, $\psi^{1} = (1-\gamma_{0}) \psi^{0} + \gamma_{0} s_{0} = s_{0}$, so the
expression is  valid when $j=1$. Suppose  now that it is  valid when $j=j^{*}$
for some $1 \leq j^{*} < J$. Then
\begin{align*}
  \psi^{j^{*}+1}
  & = (1 - \gamma_{j^{*}}) \psi^{j^{*}} + \gamma_{j^{*}} s_{j^{*}}\\
  &=  \frac{2j^{*}}{j^{*}+2}  \sum_{k=1}^{j^{*}} \frac{ks_{k-1}}{j^{*}(j^{*}+1)}  +
    \frac{2}{j^{*}+2} s_{j^{*}} \\
  &= 2 \sum_{k=1}^{j^{*}+1}  \frac{ks_{k-1}}{(j^{*}+1)(j^{*}+2)},
\end{align*}
so the expression is also valid when $j=(j^{*}+1)$. This completes the proof.


\section{Elements of targeted learning}
\label{app:TMLE}

Fix $(\lambda, \beta)\in \Lambda \times B$.

\subsection{Efficient influence curves}
\label{app:proof:EIC}

Fix $\psi \in \Psi$ and $\pit\in [0,1]^{\mathcal{X}}$. It is easy to check that the $\pit$-specific value function  $P \mapsto \mathcal{V}_{P}(\pit)$~\eqref{eq:value} and constraint function $P \mapsto S_{P}(\pit)$ \eqref{eq:S}, and the Lagrangian criterion $P \mapsto \mathcal{L}_{\psi, \lambda, \beta}(P)$~\eqref{eq:lagrangian} are pathwise differentiable at any $P \in \mathcal{M}$ with respect to $L_{0}^{2}(P)$.

Given $P \in \mathcal{M}$, the efficient efficient influence curve of the value at $P$, $\varphi_{\mathcal{V},\pit}(P)$, is characterized by
\begin{align*}
    \varphi_{\mathcal{V},\pit}(P)(O) =& \pit(X)\cdot \mu_{P}(1,X) + (1-\pit(X))\cdot \mu_{P}(0,X) - \mathcal{V}_{P}(\pit)\\
     &+ \frac{A\cdot\pit(X) + (1-A)\cdot(1-\pit(X))}{e_{P}(A,X)}\cdot (Y-\mu_{P}(A,X)).
\end{align*}
That of the constraint function at $P$, $\varphi_{S, \pit}(P)$, is characterized by
\begin{equation*}
    \varphi_{S,\pit}(P)(O) =  \pit(X) \cdot \Delta\nu_P(X) -\alpha - S_{P}(\pit) + \frac{2A-1}{e_P(A,X)}\cdot  \pit(X) \cdot \left(\xi - \nu_P(A,X) \right),
\end{equation*}
and that  of the Lagrangian criterion at $P$, $\varphi_{\mathcal{L}, \psi,\lambda,\beta}(P)$, is characterized by
\begin{align}
\notag 
\varphi_{\mathcal{L}, \psi,\lambda,\beta}(P)(O) =& \psi(X)^2 - 2\psi(X)\cdot\Delta\mu_P(X) + \lambda [\pit_{\beta,\psi}(X)\cdot \Delta\nu_P(X)-\alpha] - \mathcal{L}_{\psi,\lambda,\beta}(P) \\
\notag 
& + D_{\psi,\lambda,\beta}(P)(O), \quad \text{with}\\
\label{eq:EIC:L:bias}
D_{\psi,\lambda,\beta}(P)(O) =& \frac{2A-1}{e_P(A,X)}\left[-2\psi(X)\cdot (Y -\mu_P(A,X))+ \lambda \pit_{\beta,\psi}(X)\cdot(\xi - \nu_P(A,X))\right]. 
\end{align}

\subsection{Upper- and lower-bounds}
\label{subsec:TMLE:uS:lV}

This section details the construction of 95\% confidence upper- and lower-bounds for $S_{0}(\pit_{\lambda,\beta})$ and $\mathcal{V}_{P_{0}}(\pit_{\lambda,\beta})$ that we introduced in Sections~\ref{subsubsec:PLUC:0} and \ref{subsubsec:PLUC} for fixed $\pit_{\lambda,\beta}=\pit_{\bign_{1}\cup \bign_{2}, \lambda, \beta}\in [0,1]^{\mathcal{X}}$. Let $\mu_{\bign_{3}}^{0}, \nu_{\bign_{3}}^{0}, e_{\bign_{3}}^{0}$ denote initial estimators of $\mu_{0}$, $\nu_{0}, e_{0}$ built using data from $\{O_{i} : i \in \bign_{3}\}$. We update $\mu_{\bign_{3}}^{0}$ and $\nu_{\bign_{3}}^{0}$  using two parametric fluctuation models $\{\mu_{\bign_{3}, \lambda, \beta}^{*}(\epsilon) : \epsilon \in \mathbb{R}\}$ and $\{\nu_{\bign_{3}, \lambda, \beta}^{*}(\epsilon) : \epsilon \in \mathbb{R}\}$ characterized by
\begin{align*}
    \mu_{\bign_{3}, \lambda, \beta}^{*}(\epsilon)(A,X) &= \expit \left( \logit(\mu_{\bign_{3}}^{0}(A,X)) + \epsilon \cdot \frac{A \pit_{\lambda,\beta}(X) + (1-A) (1-\pit_{\lambda,\beta}(X))}{e_{\bign_{3}}^{0}(A,X)} \right),\\
    \nu_{\bign_{3}, \lambda, \beta}^{*}(\epsilon)(A,X) &= \expit \left( \logit(\nu_{\bign_{3}}^{0}(A,X)) + \epsilon\cdot \pit_{\lambda,\beta} \cdot \frac{2A-1}{e_{\bign_{3}}^{0}(A,X)} \right).
\end{align*} 

The     optimal     fluctuation      parameters     are     any     minimizers $\epsilon_{\mu,\bign_{3}}$ and $\epsilon_{\nu,\bign_{3}}$ of the convex criteria $\epsilon \mapsto \ell_{\mu, \lambda,\beta,\bign_{3}}^{*}$ and $\epsilon \mapsto \ell_{\nu, \lambda, \beta,\bign_{3}}^{*}$ characterized by
\begin{align*}
   \ell_{\mu, \lambda,\beta,\bign_{3}}^{*}(\epsilon) &= -\frac{3}{n}\sum_{i\in \bign_{3}}[Y_i\cdot \log(\mu_{\bign_{3}, \lambda, \beta}^{*}(\epsilon)(A_{i},X_{i})) + (1-Y_i)\cdot \log(1-\mu_{\bign_{3},\lambda, \beta}^{*}(\epsilon)(A_{i},X_{i}))]\\
   \ell_{\nu, \lambda, \beta,\bign_{3}}^{*}(\epsilon) &= -\frac{3}{n}\sum_{i\in \bign_{3}}[\xi_i\cdot \log(\nu_{\bign_{3},\lambda, \beta}^{*}(\epsilon)(A_{i},X_{i})) + (1-\xi_i)\cdot \log(1-\nu_{\bign_{3},\lambda, \beta}^{*}(\epsilon)(A_{i},X_{i}))].
\end{align*}
Given these minimizers, we then define
\begin{align*}
\Delta\mu_{\bign_{3}, \lambda, \beta}^{*}(\cdot) &=\mu_{\bign_{3},\lambda, \beta}^{*}(\epsilon_{\mu,\bign_{3}})(1,\cdot)-\mu_{\bign_{3},\lambda, \beta}^{*}(\epsilon_{\mu,\bign_{3}})(0,\cdot)\\
\Delta\nu_{\bign_{3},\lambda, \beta}^{*}(\cdot) &=\nu_{\bign_{3},\lambda, \beta}^{*}(\epsilon_{\nu,\bign_{3}})(1,\cdot)-\nu_{\bign_{3},\lambda, \beta}^{*}(\epsilon_{\nu,\bign_{3}})(0,\cdot),
\end{align*}
which we introduced in Section~\ref{subsubsec:PLUC:0}. Finally, we assume that the statistical model $\mathcal{M}$ contains a law $P_{\bign_{3}}^{*}$ such that $\Delta\mu_{P_{\bign_{3}}^{*}}=\Delta\mu_{\bign_{3}, \lambda, \beta}^{*}$, $\Delta\nu_{P_{\bign_{3}}^{*}}=\Delta\nu_{\bign_{3}, \lambda, \beta}^{*}$,  $e_{P_{\bign_{3}}^{*}}=e_{\bign_{3}, \lambda, \beta}^{0}$ and $P_{\bign_{3},X}^{*}=(3/n)\sum_{i \in \bign_{3}}\Dirac(X_{i})$.

\paragraph{Constraint upper-bound.}
\label{app:subsection:early-stop}

Defined in  \eqref{eq:updated:S},  $S_{\bign_{3}}^{*}(\pit_{\lambda,\beta})$ is a targeted estimator of $S_{0}(\pit_{\lambda,\beta})$. Under mild assumptions, it satisfies
\begin{equation*}
    \sqrt{n/3} \left(S_{\bign_{3}}^{*}(\pit_{\lambda,\beta}) - S_0(\pit_{\lambda,\beta}) \right) \overset{d}{\longrightarrow} \mathcal{N}(0, \sigma_{0,S}^2),
\end{equation*}
where $\sigma_{0,S}^{2} = \Var_{P_0}\left[ \varphi_{S,\pit_{\lambda,\beta}}(P_0)(O)\right]$, which we estimate with
\begin{align*}
   \sigma_{\bign_{3},S}^{2} = \frac{3}{n} \sum_{i\in \bign_{3}} \Bigg(&
    \frac{2A_{i} - 1}{e_{\bign_{3}}^{0}(A_{i}, X_{i})} \pit_{\lambda,\beta}(X_{i}) \cdot (\xi - \nu_{\bign_{3},\lambda,\beta}^{*}(A_{i}, X_{i}))
    \\
    &+ \pit_{\lambda,\beta}(X_{i}) \cdot \Delta \nu_{\bign_{3},\lambda,\beta}^{*}(X_{i}) 
    - S_{\bign_{3}}^{*}(\pit_{\lambda,\beta}) \Bigg)^{2}.
\end{align*}
Denoting by $q_{0.95}$ the 95\%-quantile of the standard normal distribution, the corresponding 95\% confidence upper-bound takes the form $(-\infty,\ \overline{S}_{\bign_{3}}^{*}(\pit_{\lambda,\beta})]$, with
\begin{equation*}
    \overline{S}_{\bign_{3}}^{*}(\pit_{\lambda,\beta})=S_{\bign_{3}}^{*}(\pit_{\lambda,\beta})+q_{0.95}\cdot\frac{\sqrt{\sigma_{n,S}^2}}{\sqrt{n/3}}.
\end{equation*}

\paragraph{Policy value lower-bound.}
\label{app:subsection:pv}

By substituting $\pit$ for $\pit_{\bign_{1}\cup\bign_{2}, \lambda,\beta}$ in \eqref{eq:updated:V}, we define $\mathcal{V}_{\bign_{3}}^{*}(\pit)$, a targeted estimator of $\mathcal{V}_{P_{0}}(\pit)$. Under mild assumptions, it satisfies
\begin{equation*}
    \sqrt{n/3}\left(\mathcal{V}_{\bign_{3}}^{*}(\pit) - \mathcal{V}_{P_{0}}(\pit)\right) \xrightarrow{d} \mathcal{N}(0, \sigma_{0,V}^2),
\end{equation*} 
where  $\sigma_{0,\mathcal{V}}^2=\Var_{P_{0}}[\varphi_{\mathcal{V}, \pit}(P_0)(O)]$, which we estimate with
\begin{align*}
\sigma_{\bign_{3},\mathcal{V}}^{2} &= \frac{3}{n}\sum_{i\in \bign_{3}} \Bigg( \frac{A_{i} \cdot \pit(X_{i}) + (1-A_{i}) \cdot (1-\pit_{\lambda, \beta}(X_{i}))}{e_{\bign_{3}}^{0}(A_{i},X_{i})} \cdot (Y_{i} - \mu_{\bign_{3},\lambda,\beta}^{*}(A_{i},X_{i})) \\ 
& \quad + \pit_{\lambda, \beta}(X_{i}) \cdot \mu_{\bign_{3},\lambda,\beta}^{*}(1,X_{i}) + (1-\pit_{\lambda, \beta}(X_{i})) \cdot \mu_{\bign_{3},\lambda,\beta}^{*}(0,X_{i}) - \mathcal{V}_{\bign_{3}}^{*}(\pit_{\lambda, \beta}) \Bigg)^2 .
\end{align*}
The        corresponding         95\%-confidence        lower-bound        for
$\mathcal{V}_{P_{0}}(\pit)$      takes     the      form
$[\lNu_{\bign_{3}}(\pit),+\infty)$,  with  
\begin{equation*}
    \lNu_{\bign_{3}}(\pit) = \mathcal{V}_{\bign_{3}}^{*}(\pit) - q_{0.95} \cdot \frac{\sqrt{\sigma_{\bign_{3},\mathcal{V}}^{2}}}{\sqrt{n/3}}. 
\end{equation*} 

\subsection{Proofs of Section \ref{subsubsec:PLUC}}
\label{app:local:oracle:efficiency}

As described in Sections~\ref{subsubsec:PLUC} and~\ref{app:algo:ap}, $P_{\bign_{1}\cup \bign_{2}}^{0}$ is iteratively updated into $P_{\bign_{1}\cup \bign_{2}}^{k}$ such that $\mathcal{L}_{\bign_{1}\cup \bign_{2}}^{k}(\cdot,\lambda,\beta)$ targets the $k$ landmark values $\mathcal{L}_{0}(\psi_{\bign_{1}\cup\bign_{2}}^{0},\lambda;\beta)$, $\ldots$, $\mathcal{L}_{0}(\psi_{\bign_{1}\cup\bign_{2}, \lambda,\beta}^{k-1},\lambda;\beta)$. Each update relies on the fluctuations models characterized in \eqref{eq:mu:update} and \eqref{eq:nu:update}.

\paragraph{Pointwise bias vanishing property. }

The optimal fluctuation parameters $\epsilon_{\mu,\bign_{1}\cup\bign_{2}}$ and $\epsilon_{\nu,\bign_{1}\cup\bign_{2}}$ are obtained by minimizing \eqref{eq:l:Y} and \eqref{eq:l:xi} over $\mathbb{R}^{k}$. Their existence follows from classical convexity and coercivity arguments. Choosing these particular values for the fluctuation parameters ensures that the corresponding $\mu_{\bign_{1}}^{k}(\epsilon_{\mu,\bign_{1}\cup\bign_{2}})$ and $\nu_{\bign_{1}}^{k}(\epsilon_{\nu,\bign_{1}\cup\bign_{2}})$ satisfy
\begin{equation*}
    \frac{3}{n} \sum_{i \in \bign_{2}} D_{\psi_{\bign_{1}\cup \bign_{2},\lambda,\beta}^{\ell}}(P_{\bign_{1}\cup \bign_{2}}^{k})(O_{i})=0,
 \end{equation*} 
simultaneously for all $\ell\in \{0,\ldots, k-1\}$. To establish this result, fix $\ell\in \{0,\ldots, k-1\}$, define $\pit_{\bign_{1} \cup\bign_{2},\lambda, \beta} = \sigma_{\beta} \circ \psi_{\bign_{1} \cup\bign_{2}, \lambda,\beta}$ and note that 
\begin{align*}
    \frac{\partial}{\partial \epsilon_{\ell}} \ell^{k}_{\mu,\bign_{1}\cup \bign_{2}}(\epsilon) &=  -\frac{3}{n} \sum_{i\in \bign_{2}} \frac{2A_i - 1}{e_{n}^{0}(A_i,X_i)} \psi_{\bign_{1}\cup \bign_{2}, \lambda, \beta}^{\ell}(X_{i}) \left( Y_i - \mu_{\bign_{1}}^{k}(\epsilon)(A_i,X_i) \right),\\
    \frac{\partial}{\partial \epsilon_{\ell}} \ell^{k}_{\nu,\bign_{1}\cup \bign_{2}}(\epsilon) &=  -\frac{3}{n} \sum_{i\in \bign_{2}} \frac{2A_i - 1}{e_{n}^{0}(A_i,X_i)}\pit_{\bign_{1}\cup \bign_{2}, \lambda, \beta}^{\ell}(X_{i})\left( \xi_i - \nu_{\bign_{1}}^{k}(\epsilon)(A_i,X_i) \right).
\end{align*}
Since $\epsilon_{\mu,\bign_{1}\cup\bign_{2}}$ and $\epsilon_{\nu,\bign_{1}\cup\bign_{2}}$ are roots of $\epsilon \mapsto \frac{\partial}{\partial \epsilon_{\ell}} \ell^{k}_{\mu,\bign_{1}\cup \bign_{2}}(\epsilon)$ and $\epsilon \mapsto \frac{\partial}{\partial \epsilon_{\ell}} \ell^{k}_{\nu,\bign_{1}\cup \bign_{2}}(\epsilon)$, the fluctuated nuisances $\mu_{\bign_{1}}^{k}(\epsilon_{\mu,\bign_{1}\cup\bign_{2}})$ and $\nu_{\bign_{1}}^{k}(\epsilon_{\nu,\bign_{1}\cup\bign_{2}})$ satisfy
\begin{align*}
    \frac{3}{n} 
    &\sum_{i \in \bign_{2}} D_{\psi_{\bign_{1}\cup \bign_{2}, \lambda, \beta}^{\ell}}(P_{\bign_{1}\cup\bign_{2}}^{k})(O_{i}) \\
    &= \frac{3}{n} \sum_{i \in I_{2}} \frac{2 A_{i} - 1}{e_{n}^{0}(A_{i}, X_{i})} [ -2 \, \psi_{\bign_{1}\cup \bign_{2}, \lambda, \beta}^{\ell}(X_{i}) \, (Y_{i} - \mu_{\bign_{1}}^{k}(\epsilon_{\mu,\bign_{1}\cup\bign_{2}})(A_{i}, X_{i})) \\
    & \quad + \lambda \, \pit_{\bign_{1}\cup \bign_{2}, \lambda, \beta}^{\ell}(X_{i}) 
    (\xi_{i} - \nu_{\bign_{1}}^{k}(\epsilon_{\nu,\bign_{1}\cup\bign_{2}})(A_{i}, X_{i}))]=0. 
\end{align*}

\section{\centering \textsc{About the numerical experiments}}
\label{app:about:num:exp}

The R code used to generate the data and implement the proposed algorithms is available through the dedicated R package \texttt{PLUCR}, which can be found on GitHub here: \texttt{\url{https://github.com/laufuentes/PLUCR.git}}

\subsection{Hyperparameters}
\label{app:exp:hyperaparam}
Throughout the simulation study presented in Section~\ref{sec:num:exp} several hyperparameters were to be fixed. 

For the Frank-Wolfe algorithm, we set the precision to $0.025$, corresponding to a total of $40$ iterations. The stochastic gradient descent (SGD) procedure was configured with a tolerance of
$10^{-3}$, a learning rate of $10^{-2}$, batch proportion of $0.2$ and maximum of $1.000$ iterations. 

For the PLUC method, an additional iterative bias-correction algorithm was employed. Within this algorithm, the tolerance on consecutive solutions was set to $0.025$, and the maximum number of iterations was set to $5$.

\subsection{Controlled setting}
\label{app:controlled:settings}
We generated $n \in \{3{,}000, 6{,}000, 18{,}000\}$ independent and identically distributed observations from the law $P_{0}$ on $\mathcal{O}=[0,1]^{10}\times \{0,1\} \times [0,1] \times \{0,1\}$. The covariate vector, $X=(X_1,\ldots, X_{10})$, is sampled from an uniform distribution on $[0,1]^{10}$. The binary treatment assignment $A\in \{0,1\}$ is generated conditionally on $X$, via a Bernoulli law, with the propensity score $e_{0}(X) = P_{0}(A=1|X)$ defined by two conditional mean functions: $e_{0}(X) = \expit(4(X_2-1/2))$ or $e_{0}(X) =\expit(4(X_5-1/2))$. The primary outcome $Y \in [0,1]$ is generated conditionally on $(A,X)$ under two distinct mechanisms for the conditional mean $\mu_{0}(A,X)=E_{P_{0}}[Y|A,X]$: one driven exclusively by the treatment effect, $\mu_0(A,X) = 0.95 \, \expit\bigl(f(A,X)\bigr) + 0.05 \, \expit(\epsilon)$, and a second that includes a covariate-dependent baseline component $\mu_0(A,X) = 0.55 \, \expit\bigl(f(A,X)\bigr) + 0.35 \expit(3X_3 - X_4) + 0.05 \, \expit(\epsilon)$, where $f:\{0,1\} \times X \rightarrow \mathbb{R}$ characterizes the treatment effect and $\epsilon$ is Gaussian noise. We examine two forms for $f$, which are detailed subsequently.
Finally, the adverse event indicator $\xi$ is generated from a Bernoulli law conditionally on $(A,X)$, using three distinct patterns. To enforce the monotonicity assumption $\Delta \nu_0(x) \geq 0$, we define the counterfactuals as follows. First, the untreated counterfactual $\xi(0)$ is sampled from a Bernoulli law with a conditional mean $\nu_{0}(0,x)=c>0$. The treated counterfactual $\xi(1)$ is constructed such that $\xi(1) = 1$ whenever $\xi(0) = 1$, otherwise, it is sampled from a Bernoulli law with conditional mean $p(x)$. This construction implies the treated conditional mean is $\nu_{0}(1,x) = \nu_{0}(0,x) + (1-\nu_{0}(0,x))p(x)$, thereby satisfying the monotonicity assumption by definition. We examine three distinct functional forms for $p(x)$, which will be detailed subsequently.
\subsubsection{Treatment effect patterns}
\begin{figure}[H]
    \centering
    \begin{minipage}{0.32\textwidth}
        \includegraphics[width=\textwidth]{images/Synthetic_data/Synthetic_data_plot_Linear-Linear.pdf}
        \subcaption{Linear}
         \label{fig:scenario:linear}
    \end{minipage}
    \hfill
    \begin{minipage}{0.32\textwidth}
        \includegraphics[width=\textwidth]{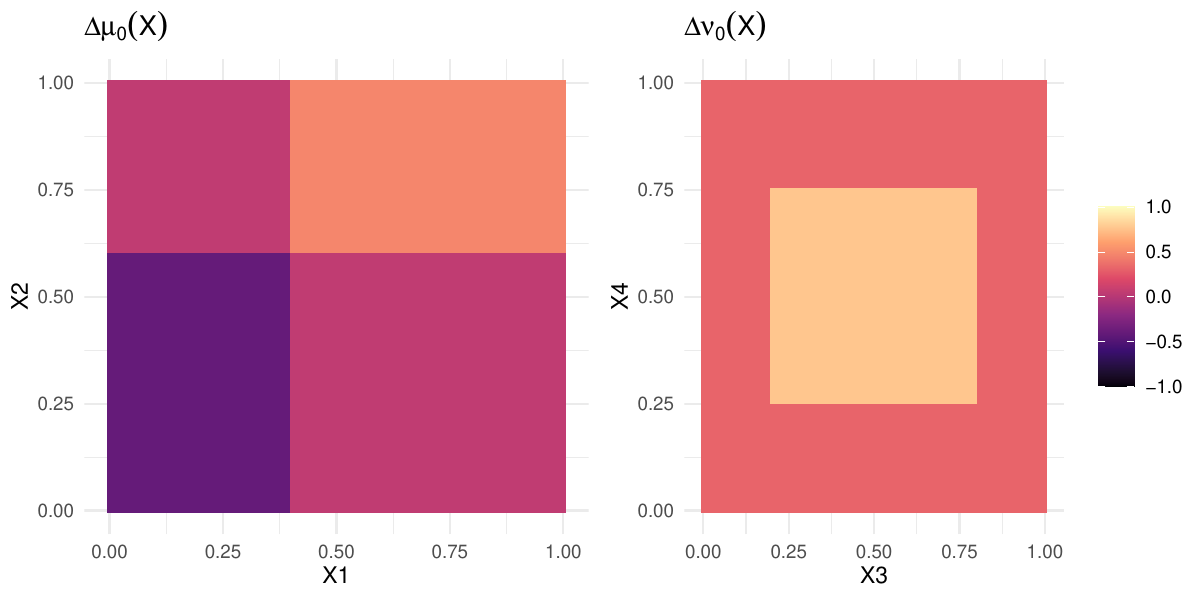}
        \subcaption{Threshold}
         \label{fig:scenario:threshold}
    \end{minipage}
    \hfill
    \begin{minipage}{0.32\textwidth}
    \centering
    \includegraphics[width=0.35\textwidth]{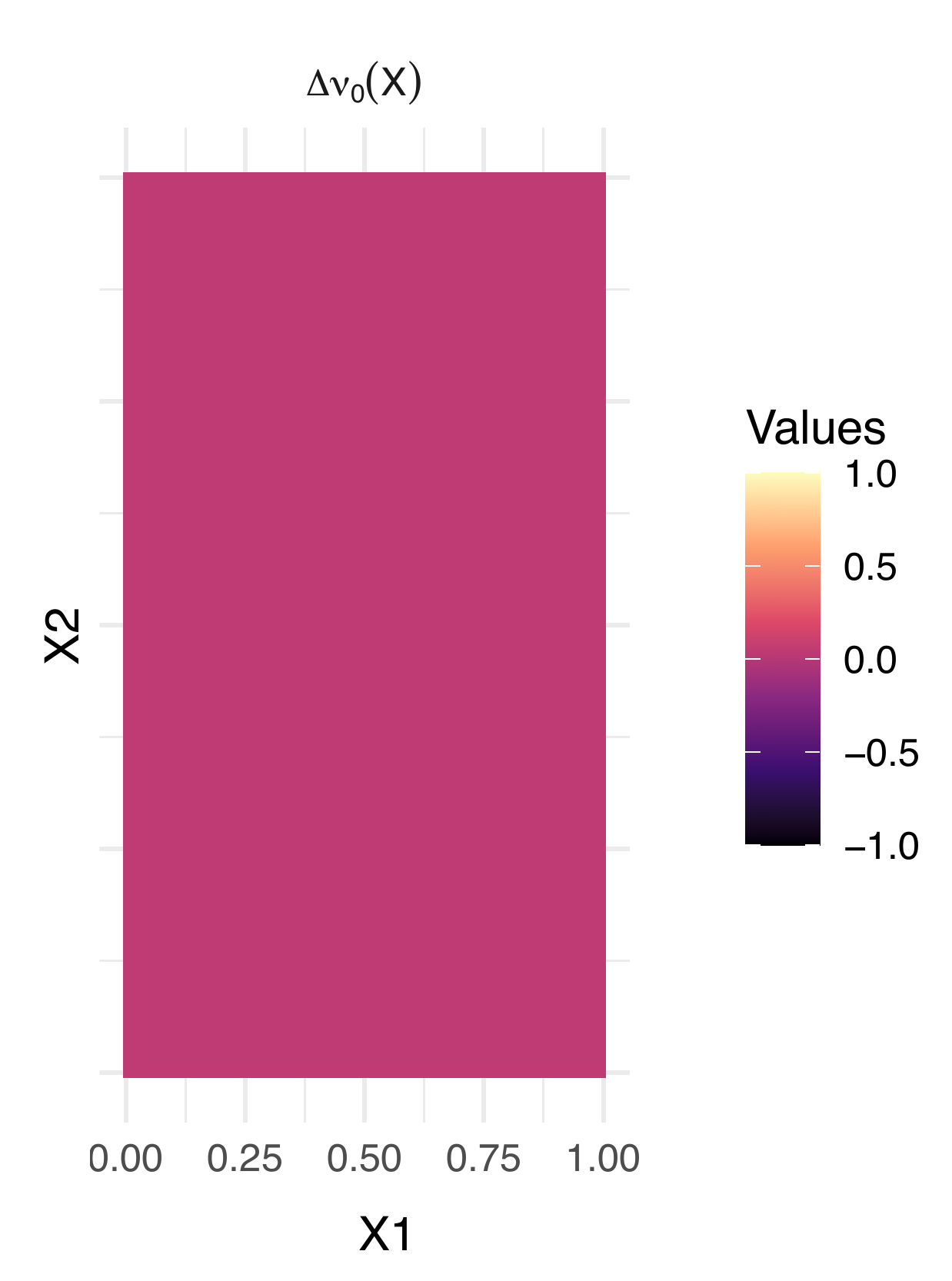}
    \subcaption{Small}
    \label{fig:scenario:small}
    \end{minipage}
    \caption{Treatment effect for primary outcome and adverse event for Linear scenario (a) and Threshold scenario (b) and small treatment effect on adverse event (c)
    }
    \label{fig:scenarios:l:t:s}
\end{figure}

\paragraph{Linear}\label{scenario:linear}
In the linear scenario represented in Figure~\ref{fig:scenario:linear} the treatment effect function $f_{\mathrm{linear}}(A,X)=2(2A-1)(1-X_{1}-X_{2})$, imposes a smooth linear gradient for the primary outcome $Y$. This function ensures that the treatment effect varies linearly with the first two covariates. For the adverse event indicator, a smooth linear gradient is imposed based on $X_{2}$: the untreated counterfactual conditional mean is set to a constant $\nu_{0}(0,X)=0.25$, and for the treated, the conditional probability is defined as $p_{\mathrm{linear}}(X)=\expit\left(4\left(X_2 - \frac{1}{2}\right)\right)$.

\paragraph{Threshold}\label{scenario:threshold}
In the tree-based scenario (Figure~\ref{fig:scenario:threshold}), the treatment effect on the primary outcome is governed by the piece-wise function
\begin{equation*}
    f_{\mathrm{threshold}}(A,X) = (2A-1)
    \begin{cases}
        1.1, & \text{if } X_1 > 0.4 \text{ and } X_2 > 0.6, \\
        -0.9, & \text{if } X_1 \leq 0.4 \text{ and } X_2 \leq 0.6, \\
        0.1, & \text{otherwise}.
    \end{cases}
\end{equation*}

This design introduces a region-wise treatment effect with abrupt changes, resulting in regions of strong positive, negative, and near-zero effects.

The adverse event probability exhibits as well a piece-wise structure. For the untreated group the conditional mean is set to $\nu_{0}(0,X)=0.1$ and for the treated the conditional probability is defined as $p_{\mathrm{threshold}}(X) = 0.1 + 0.9 \cdot \bone\{0.2<X_3<0.8,\ 0.25<X_4< 0.75\}$.

\paragraph{Small}
The small treatment effect pattern on the adverse event (Figure~\ref{fig:scenario:small}) is specifically designed to satisfy the constraint for any $\alpha>0.03$ effectively mimicking a scenario without significant adverse events. This pattern allows us to compare our algorithms directly against classic policy learning methods that do not incorporate constraints, thus facilitating an assessment of the complexity cost imposed by the constraint-handling procedure. For the untreated, we defined $\nu_{0}(0,X) = 0.01$, for the treated group, we fixed $p_{\mathrm{small}}(X) = 0.04$.

\subsection{Realistic example}
We generated $n\in\{3{,}000, 6{,}000, 18{,}000\}$ i.i.d. observations from the law $P_{0}$ on $\mathcal{O}=\mathbb{R}^{5}\times \{0,1\}\times \mathbb{R}\times \{0,1\}$. In this setting, the covariates $X$ and a primary outcome $Y$ were deliberately outside the $[0,1]$ range. This design allowed us to evaluate the influence of required pre-processing steps on the policy learning procedure's quality.

The covariate vector $X=(X_{1},\ldots, X_{5})$ is generated as follows: $X_1$ is sampled from a uniform distribution on the interval [16,65], $X_2$ is drawn from a Bernoulli law with constant conditional mean 0.5. The third covariate, $X_3$, is sampled via a Bernoulli law conditionally on $(X_{1},X_{2})$ with the conditional mean: $(X_{1},X_{2})\mapsto 0.3\cdot \bone\{18\le X_1 \le 45, X_2=1\}$. Finally, $X_4, X_5$ are independently sampled from a uniform law on $[0,10]$. The treatment is assigned conditionally on $X$ via a Bernoulli law with the propensity score $e_0(X)=\expit(-0.5X_2 + 0.2(X_5) + 0.6(X_4-5.5))$. 
\begin{figure}[H]
    \centering
    \includegraphics[width=0.45\textwidth]{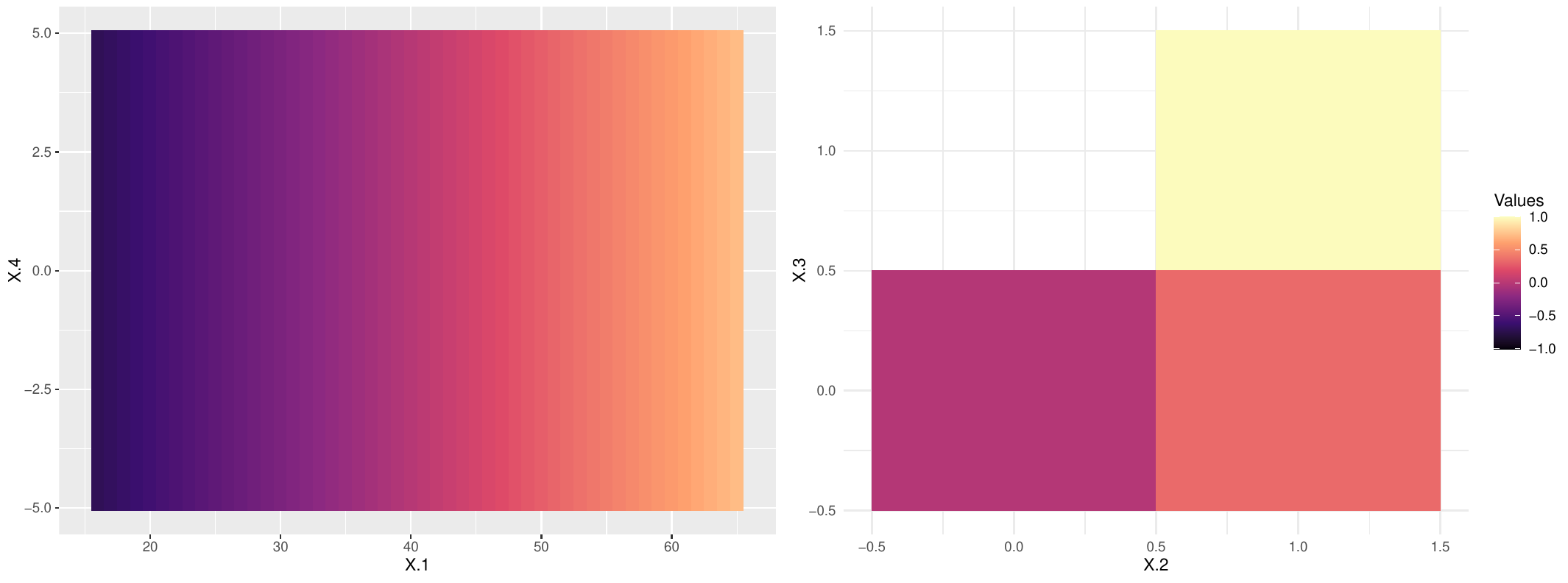}
    \caption{Treatment effect for primary outcome and adverse event for realistic sample}
    \label{fig:realistic:setting}
\end{figure}

The primary outcome $Y$ is defined conditionally on $(A,X)$ through $\mu_{0}(A,X)$ and includes a baseline effect, a treatment effect, and a Gaussian noise $\epsilon_Y$,
\begin{equation*}
\mu_0(A,X) = 0.4X_4 - 0.2X_5 + f_{\mathrm{realistic}}(A,X) + 0.5\epsilon_Y.
\end{equation*} The treatment effect function $f_{\mathrm{realistic}}(A,X) \rightarrow (2A-1)(-4 + 0.1X_{1})$ introduces an effect that increases linearly with $X_{1}$. For the adverse event $\xi$, the untreated  counterfactual $\xi(0)$ is sampled from a Bernoulli law with constant probability $\nu_{0}(0,X)=0.01$. The treated counterfactual $\xi(1)$ is $1$ if $\xi(0)=1$, and otherwise sampled from a Bernoulli distribution with conditional probability: $p_{\mathrm{realistic}}(X)= \bone\{X_3=1\} + 0.35\cdot \bone\{X_2=1, X_3=0\}$.

\subsection{Full simulation study results}
\label{app:full:simulation:results}
\begin{figure}[H]
    \centering
        \begin{minipage}[t]{0.48\textwidth}
            \centering
            \includegraphics[width=\textwidth]{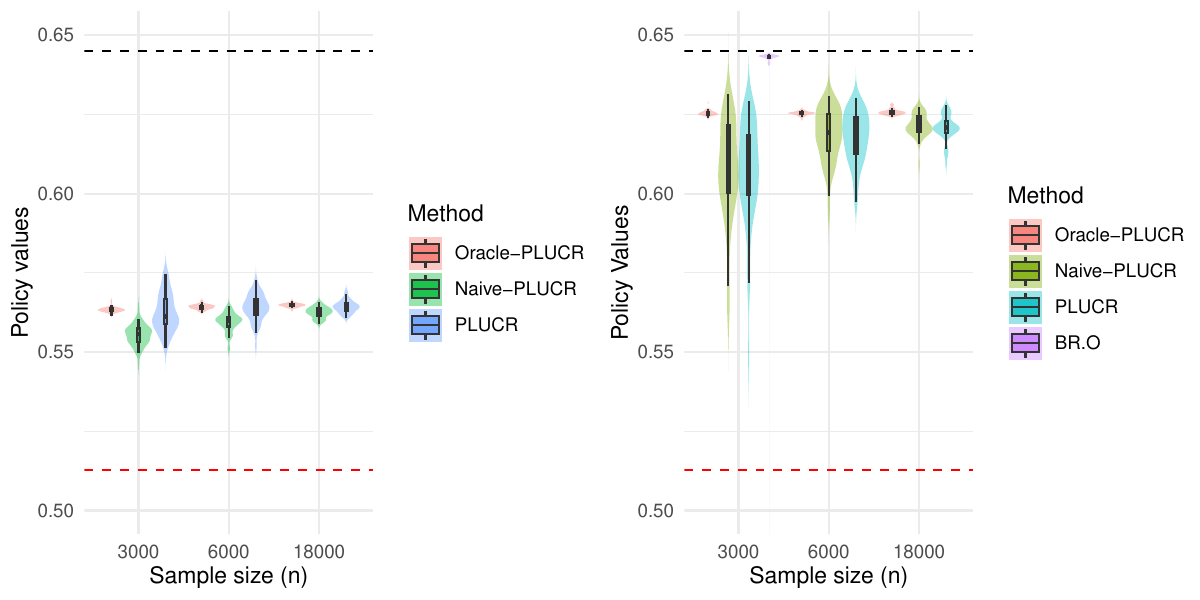}
            \subcaption{Policy values}
            \label{fig:boxplot:linear:pv}
        \end{minipage}
        \hfill
        \begin{minipage}[t]{0.48\textwidth}
            \centering
            \includegraphics[width=\textwidth]{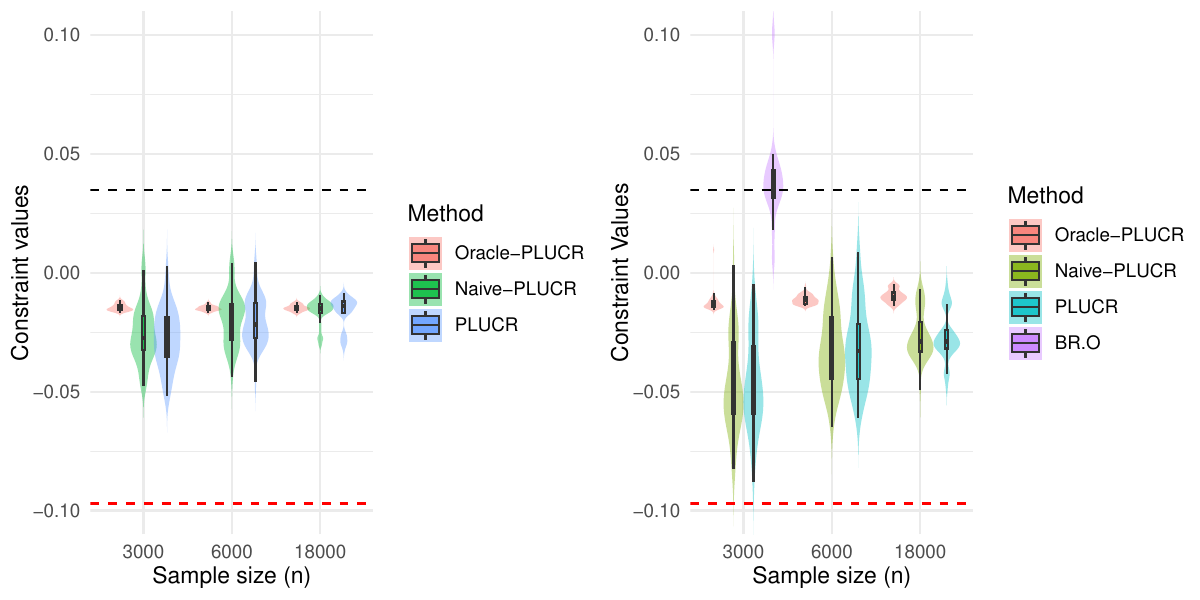}
            \subcaption{Constraint values}
            \label{fig:boxplot:linear:constraint}
         \end{minipage}
        \caption{Boxplots of (a) policy values and (b) constraint values for the linear scenario, shown separately for learned policies (left panels) and recommendations (right pannels). Dashed lines indicate the corresponding metrics of the surrogate functions $x\mapsto \bone\{\Delta\mu_{0}(x)>0\}$ (black) and respectively $x\mapsto \bone\{\Delta\mu_{0}(x)>0, \Delta\nu_{0}(x)\leq \alpha\}$ (red). BR-O did not converge 33 times.}
    \label{fig:linear:boxplot}
\end{figure}

A detailed exploration of the algorithms' performance across all simulation settings, discussed in Section~\ref{sec:num:exp}, is provided through three additional supplementary figures. These figures include:

The Policy-Constraint plots provided in the main text. Comparative boxplots (Figure~\ref{fig:linear:boxplot}), which illustrate the distribution, over the 50 simulations, of the oracular policy value (a) and constraint values (b) for each sample size $n$. These distributions are shown for both the policy form (left panels) and the decision counterpart (right panels) of the presented algorithms. 
The first surrogate function, $x\mapsto \bone\{\Delta\mu_{0}(x)>0\}$ corresponds to the unconstrained policy that assigns treatment whenever it improves the primary outcome. This rule is expected to yield high policy values, but since it disregards the constraint, it may fail to satisfy it. In contrast, the second surrogate, $x\mapsto \bone\{\Delta\mu_{0}(x)>0, \Delta\nu_{0}(x)\leq \alpha\}$ assigns treatment only when it both improves the primary outcome and keeps the effect on the adverse event below $\alpha$. This produces a more conservative strategy that achieves lower policy values but consistently satisfies the constraint. The expected optimal rule must therefore navigate between these extremes, maximizing policy value while ensuring that the population-level constraint is met on average.

\begin{figure}[H]
    \centering
    \includegraphics[width=\textwidth]{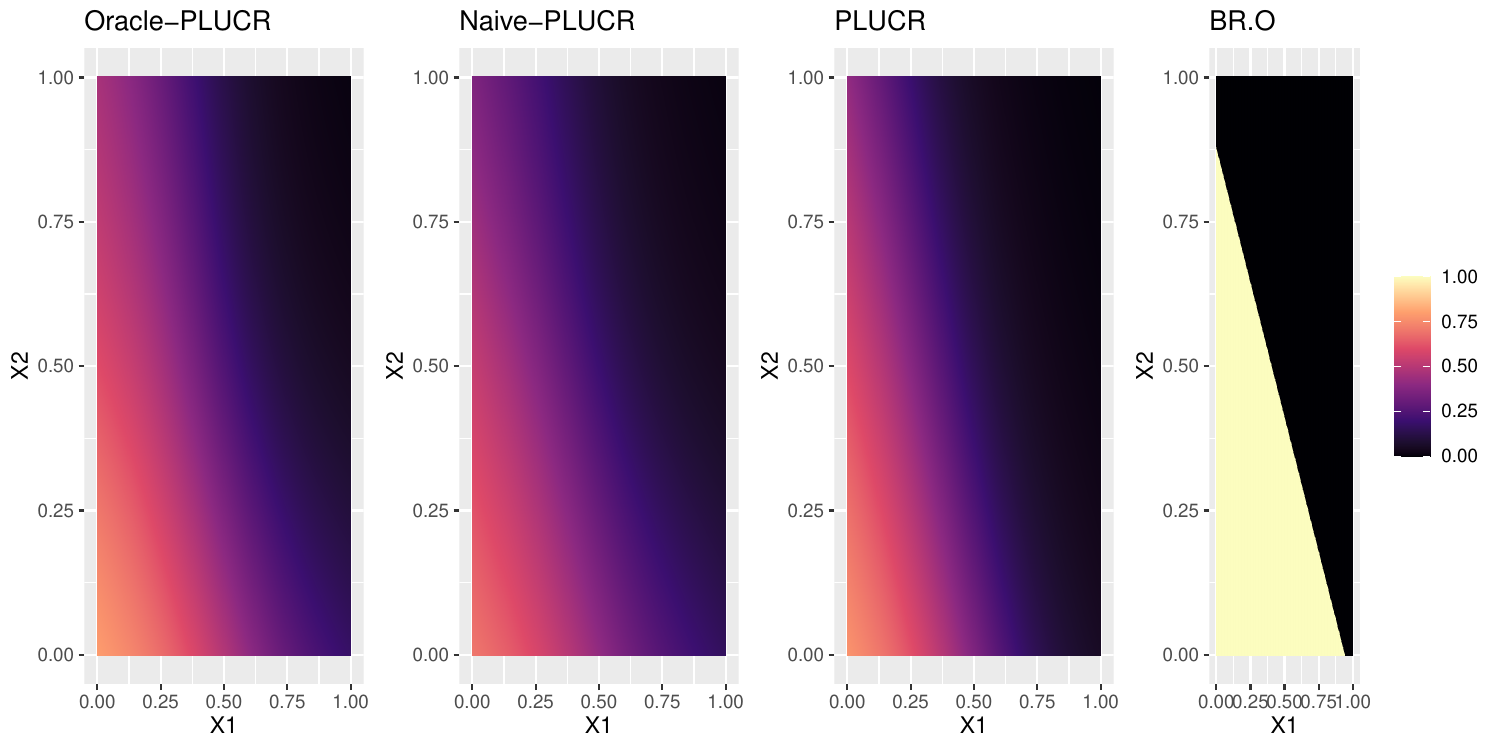}
\caption{Visual policies and recommendations for the linear scenario)}
\label{fig:linear:treatment}
\end{figure}

Figure \ref{fig:linear:treatment} provides a visual representation of the policies (Oracle PLUC, Naive PLUC and PLUC) and recommendations (Figure~\ref{fig:linear:treatment}). This comparison serves to highlight how the policies provide a crucial measure of confidence or uncertainty related to the recommended treatment assignment.

\begin{figure}[H]
\centering
\begin{minipage}[t]{0.45\textwidth}
  \includegraphics[width=\textwidth]{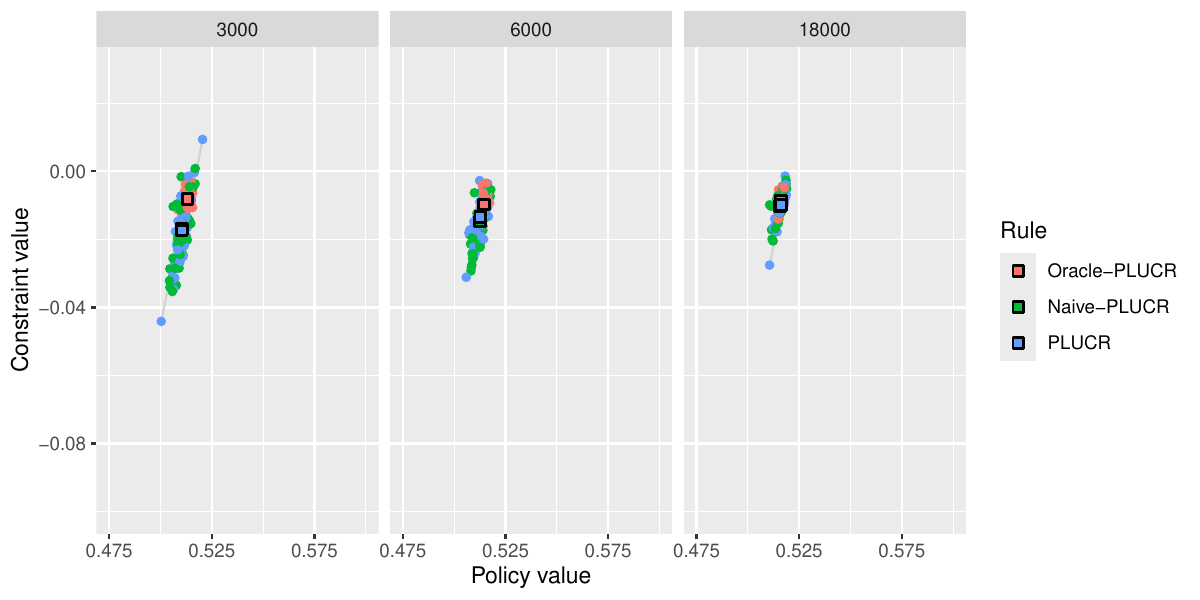}
    \subcaption{Policy value-Constraint plot}
    \label{fig:qq:threshold:proba}
\end{minipage}
\begin{minipage}[t]{0.45\textwidth}
    \includegraphics[width=\textwidth]{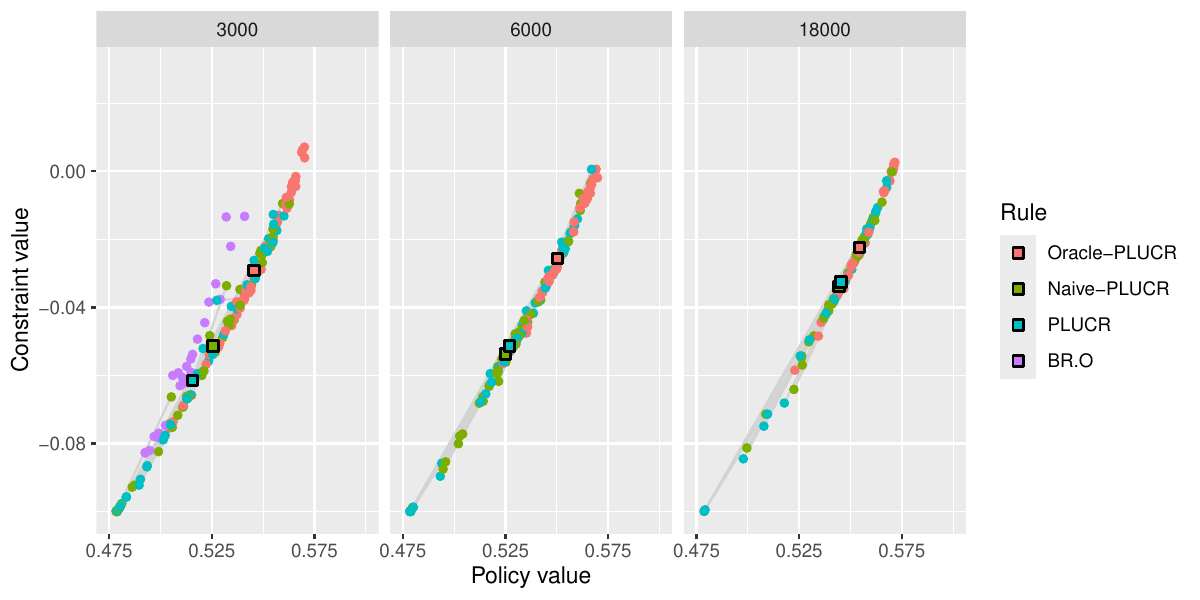}
    \subcaption{Policy value-Constraint plot}
     \label{fig:qq:threshold:decision}
\end{minipage}
\begin{minipage}[t]{0.45\textwidth} 
    \includegraphics[width=\textwidth]{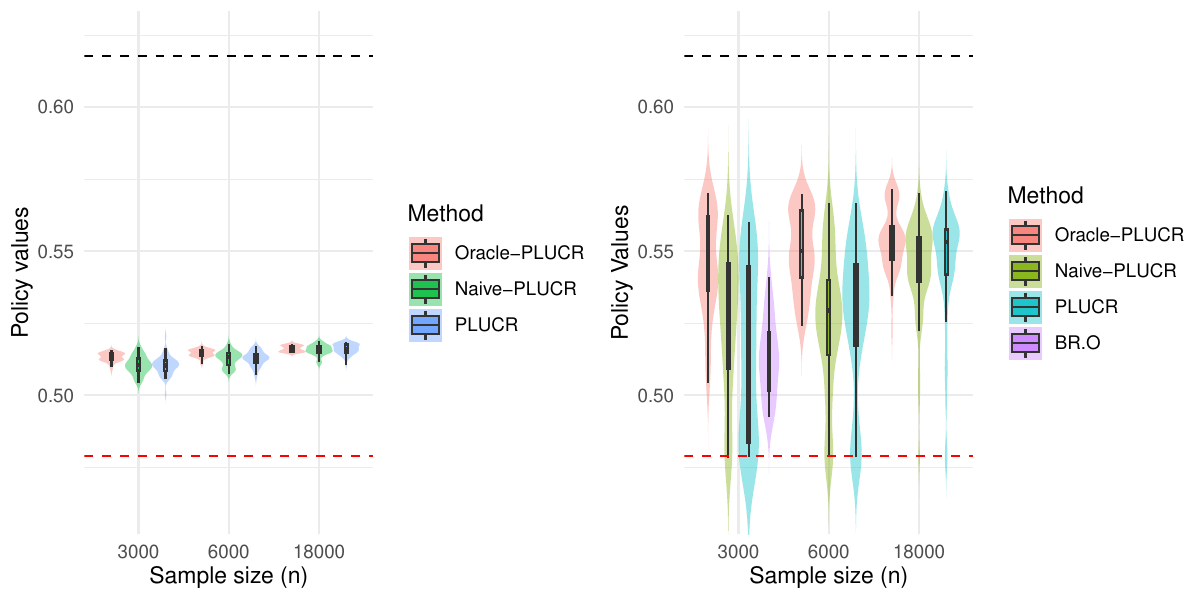}
    \subcaption{Policy values}
    \label{fig:boxplot:threshold:pv}
\end{minipage}
\begin{minipage}[t]{0.45\textwidth}
    \includegraphics[width=\textwidth]{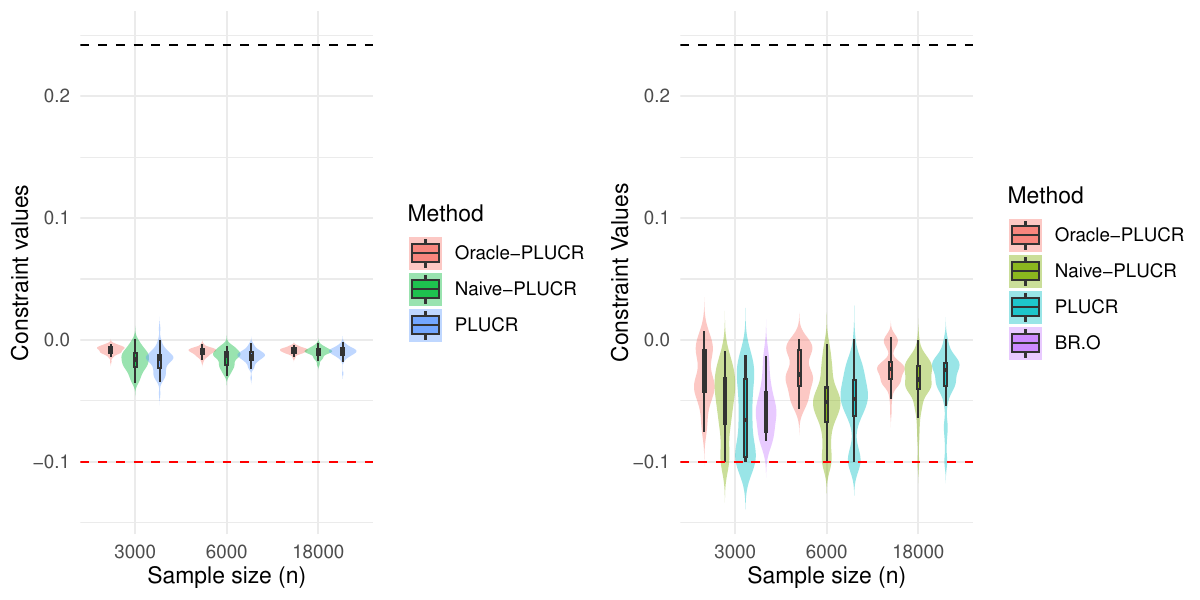}
    \subcaption{Constraint values}
    \label{fig:boxplot:threshold:constraint}
\end{minipage}
\begin{minipage}[t]{0.45\textwidth}
    \includegraphics[width=\textwidth]{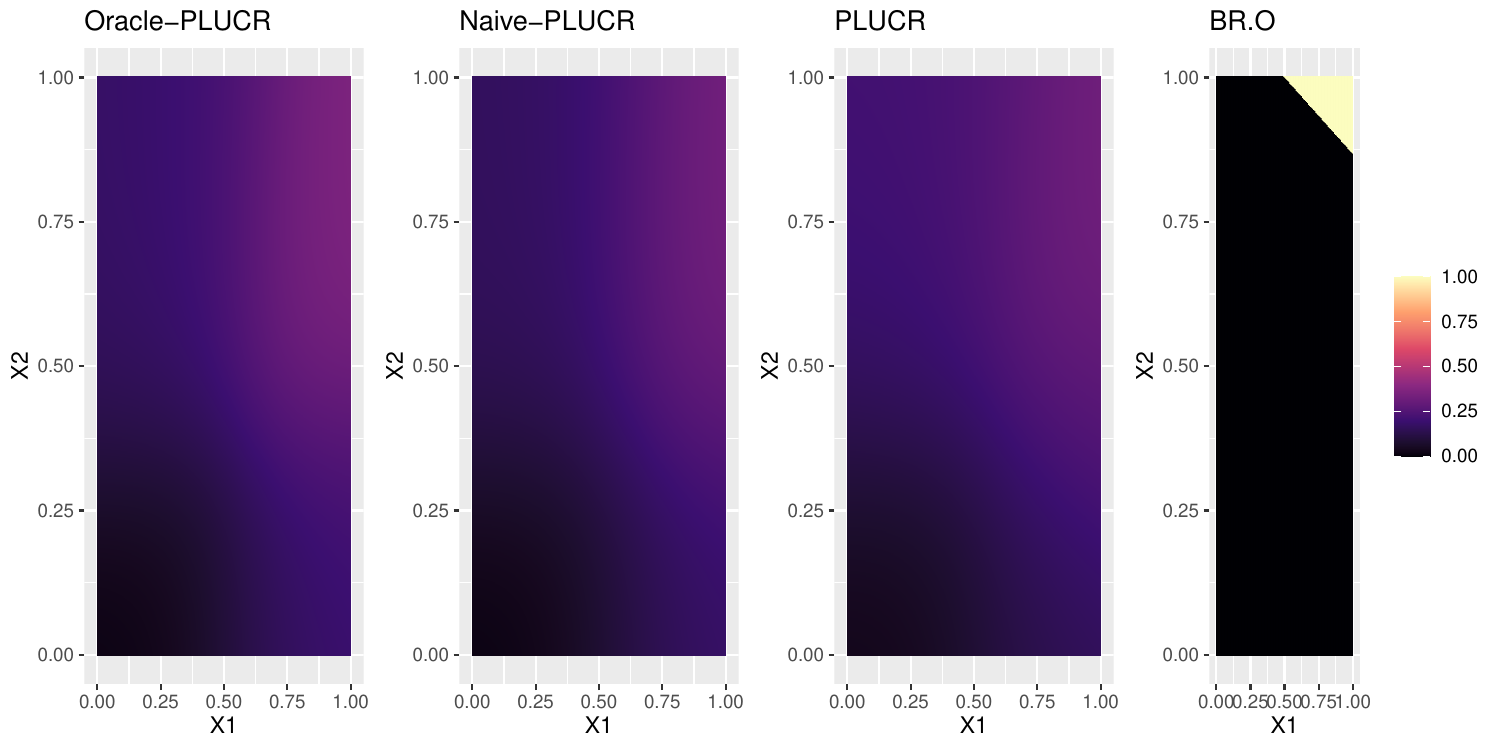}
    \subcaption{Treatment recommendations}
    \label{fig:threshold:treatment}
\end{minipage}
 \caption{Results for the threshold scenario (BR-O did not converge 26 times).}
\label{fig:threshold}
\end{figure}

\begin{figure}[H]
\centering
\begin{minipage}[t]{0.45\textwidth}
    \includegraphics[width=\textwidth]{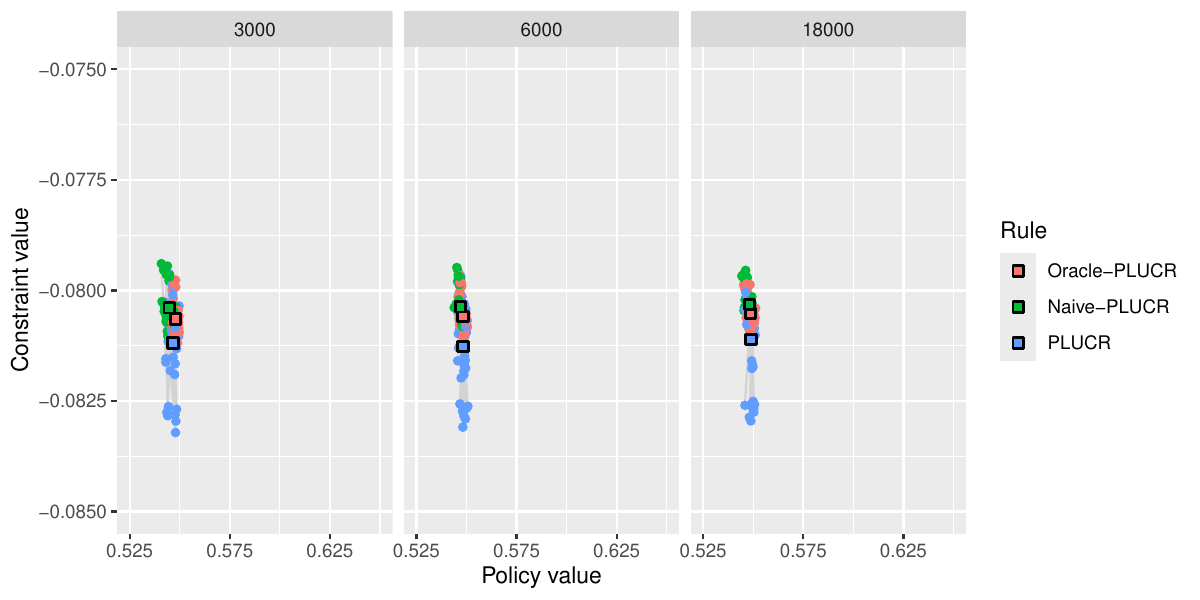}
    \subcaption{Policy value-Constraint plot}
     \label{fig:qq:satisfied:proba}
\end{minipage}
\begin{minipage}[t]{0.45\textwidth}
    \includegraphics[width=\textwidth]{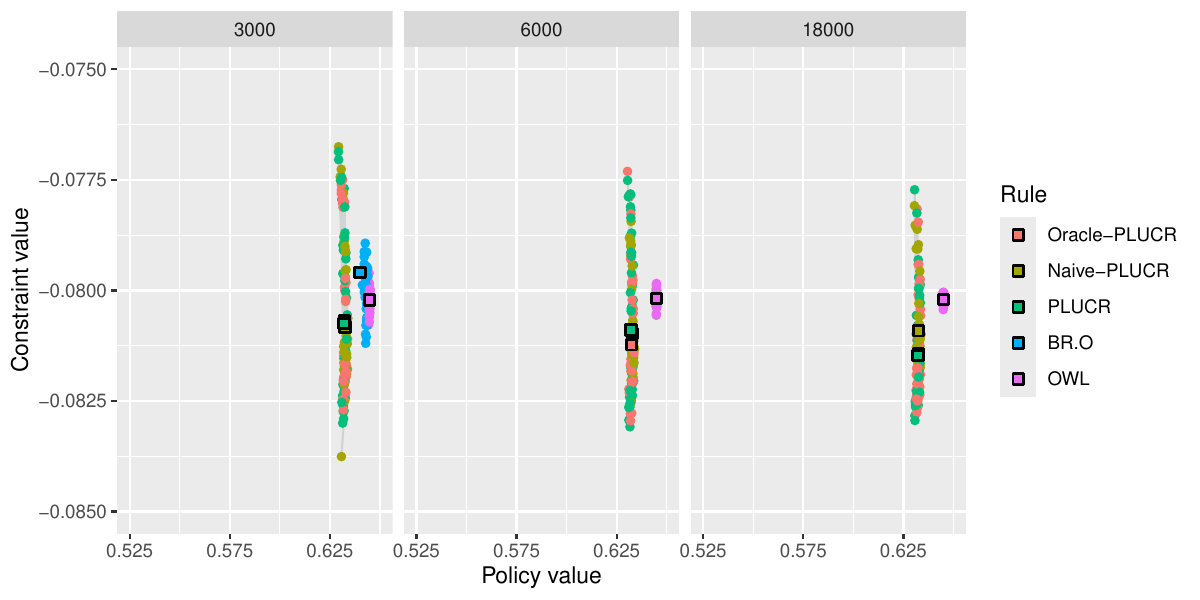}
    \subcaption{Policy value-Constraint plot}
   \label{fig:qq:satisfied:decision}
\end{minipage}
        \begin{minipage}[t]{0.45\textwidth} 
    \includegraphics[width=\textwidth]{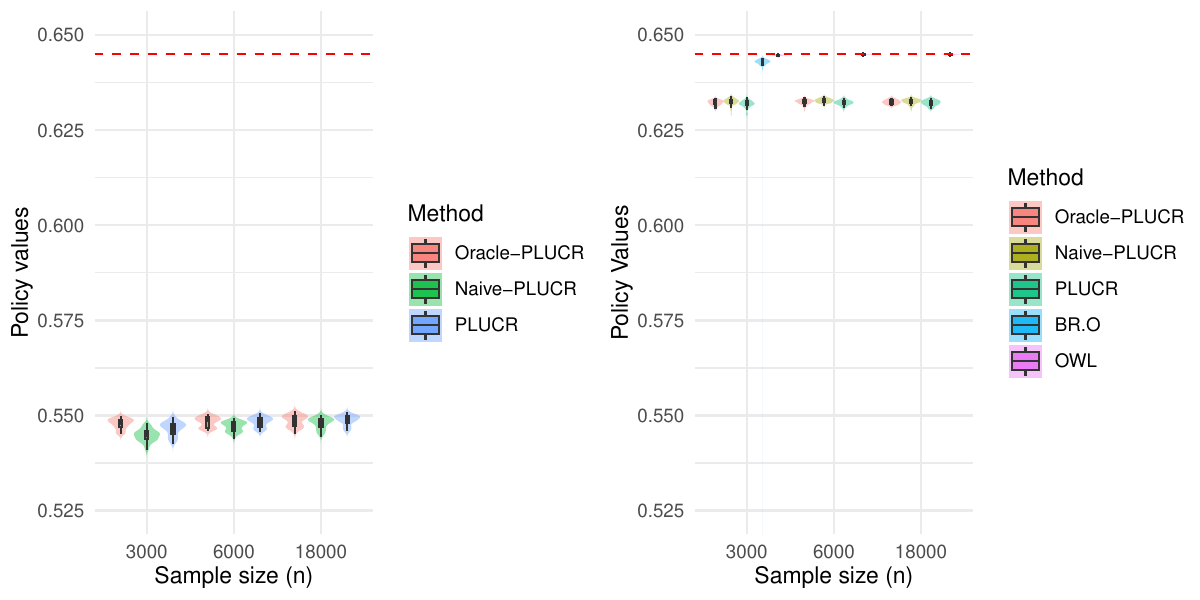}
    \subcaption{Policy values}
    \label{fig:boxplot:satisfied:pv}
\end{minipage}
\begin{minipage}[t]{0.45\textwidth}
    \includegraphics[width=\textwidth]{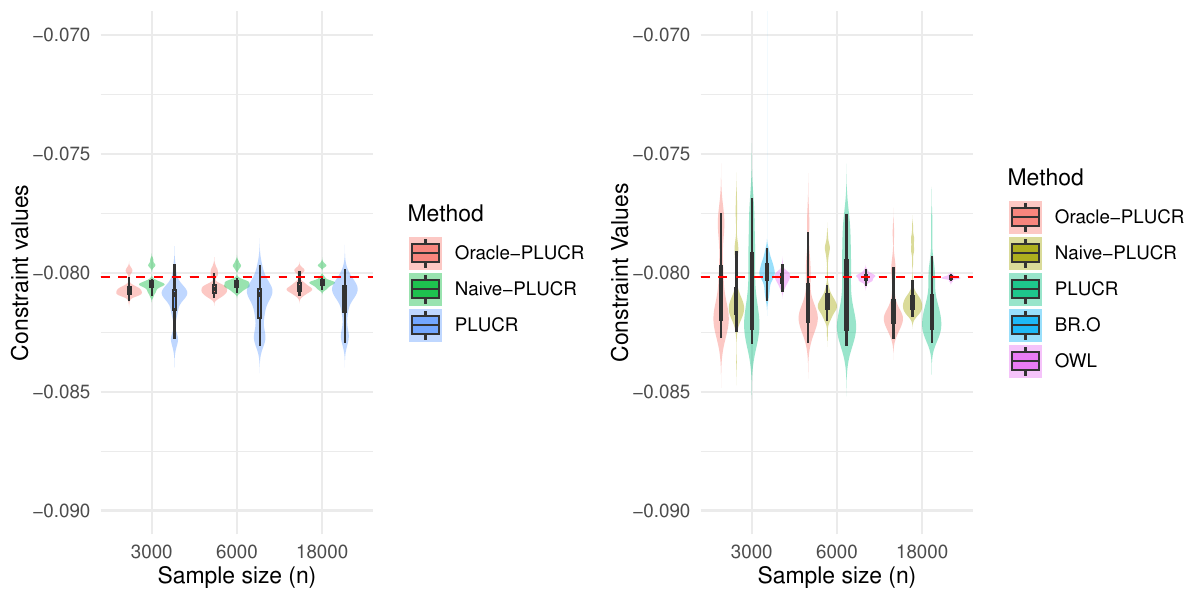}
    \subcaption{Constraint values}
     \label{fig:boxplot:satisfied:constraint}
\end{minipage}
\begin{minipage}[t]{0.45\textwidth}
    \includegraphics[width=\textwidth]{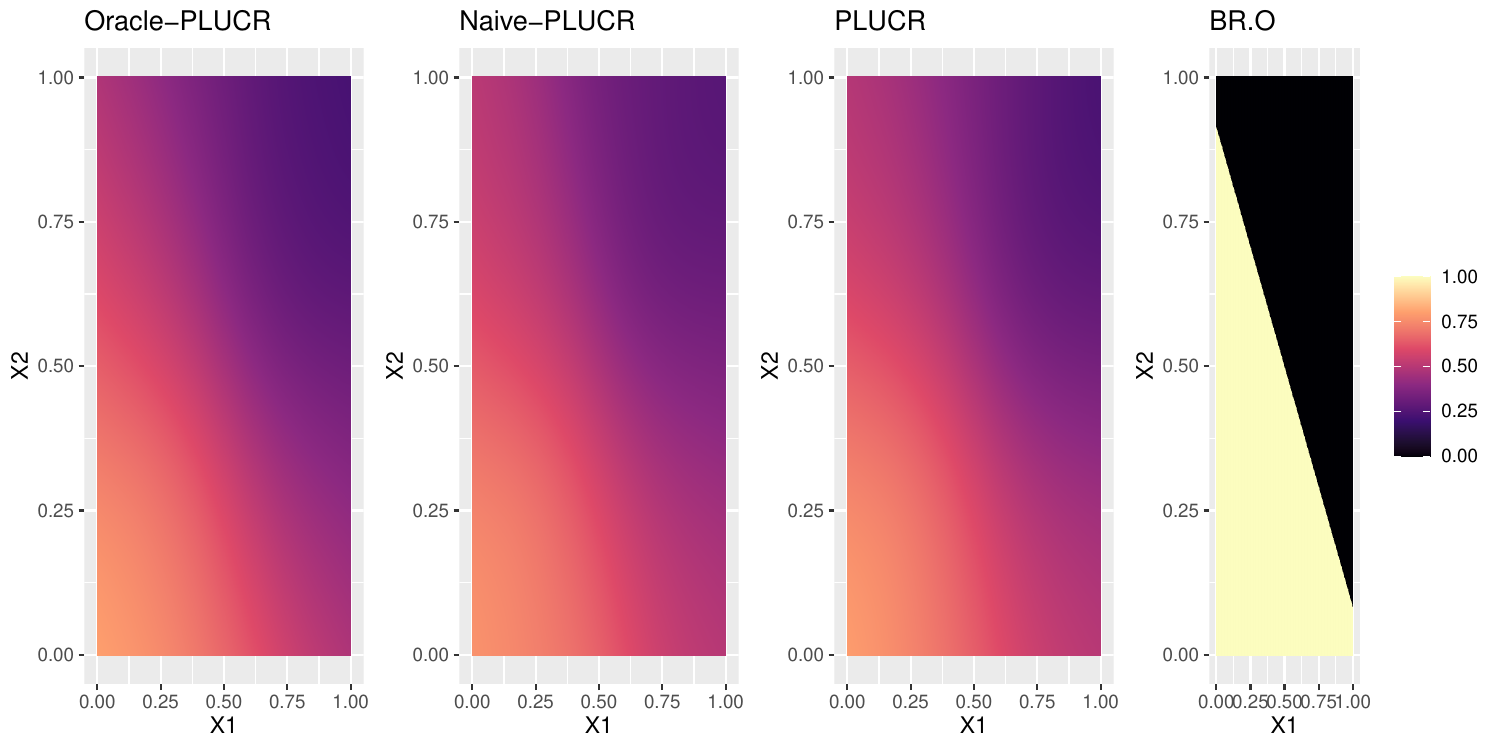}
    \subcaption{Treatment recommendations}
    \label{fig:satisfied:treatment}
\end{minipage}
 \caption{Results for the Linear ($Y$) and Small ($\xi$) scenario (BR-O converged everywhere).}
\label{fig:satisfied}
\end{figure}

\begin{figure}[H]
\centering
\begin{minipage}[t]{0.45\textwidth}
    \includegraphics[width=\textwidth]{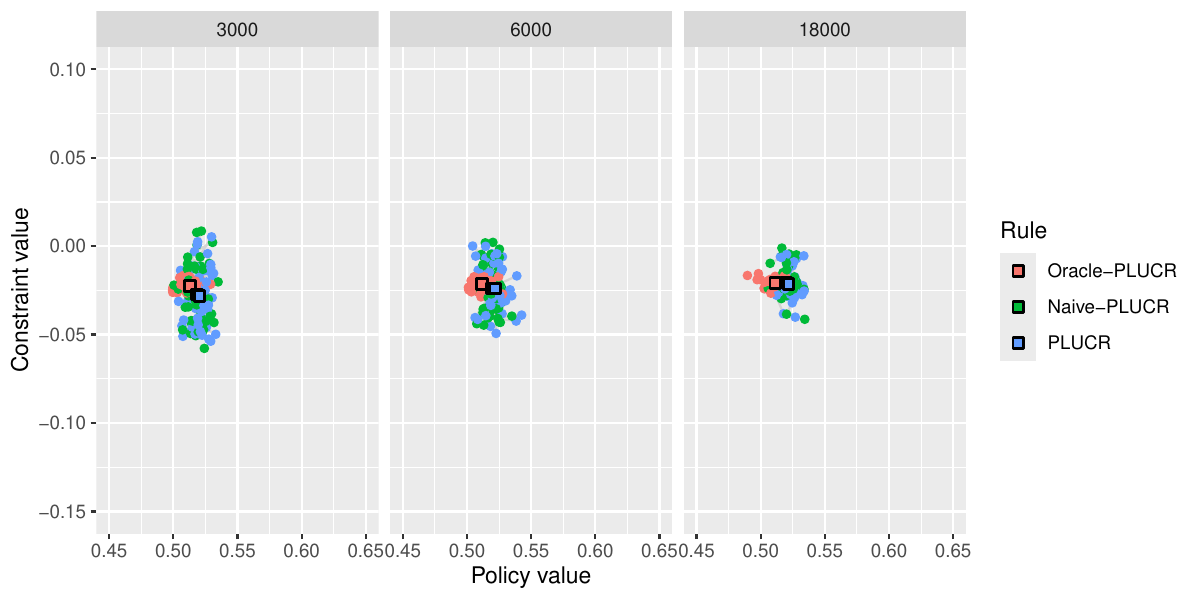}
    \subcaption{Policy value-Constraint plot}
    \label{fig:qq:realistic:proba}
\end{minipage}
\begin{minipage}[t]{0.45\textwidth}
    \includegraphics[width=\textwidth]{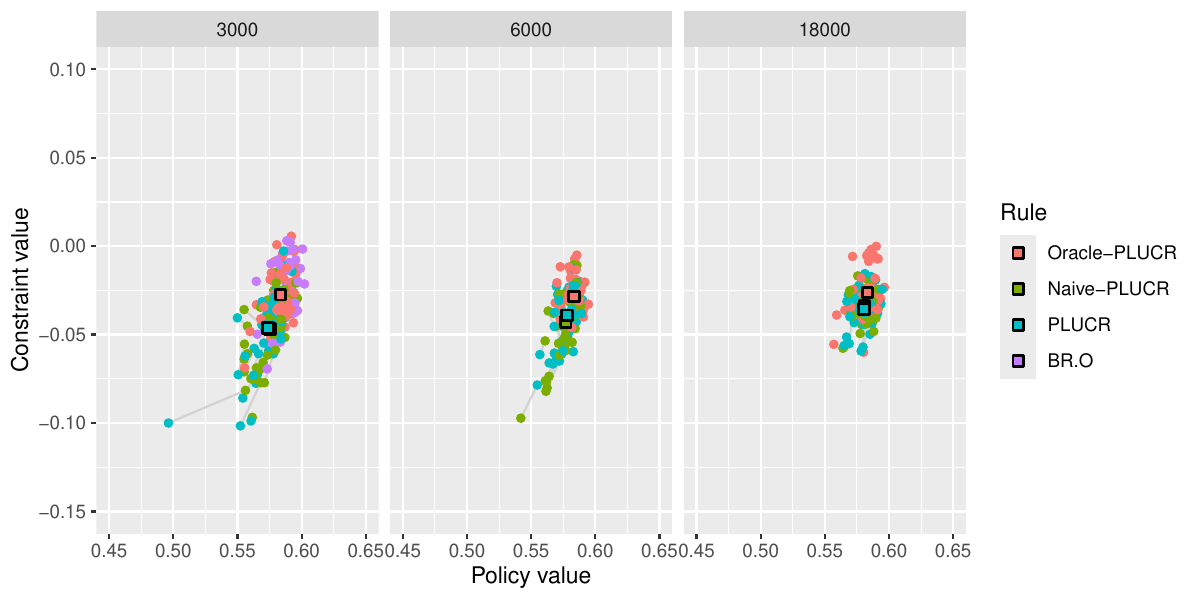}
    \subcaption{Policy value-Constraint plot}
    \label{fig:qq:realistic:decision}
\end{minipage}
\begin{minipage}[t]{0.45\textwidth} 
    \includegraphics[width=\textwidth]{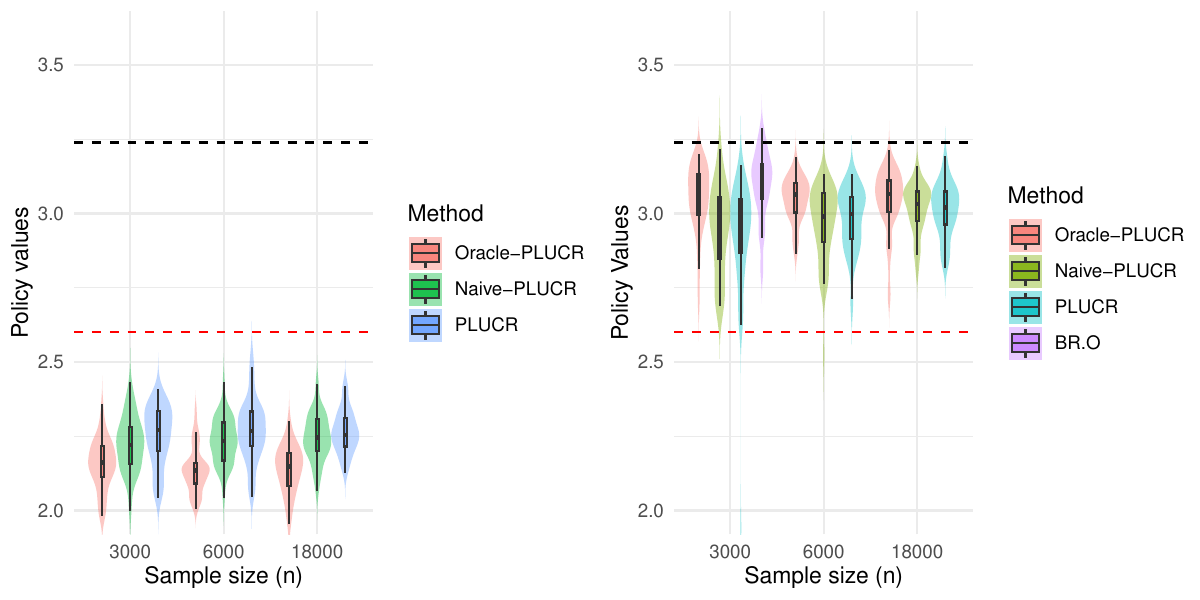}
    \subcaption{Policy values}
    \label{fig:boxplot:realistic:pv}
\end{minipage}
\begin{minipage}[t]{0.45\textwidth}
    \includegraphics[width=\textwidth]{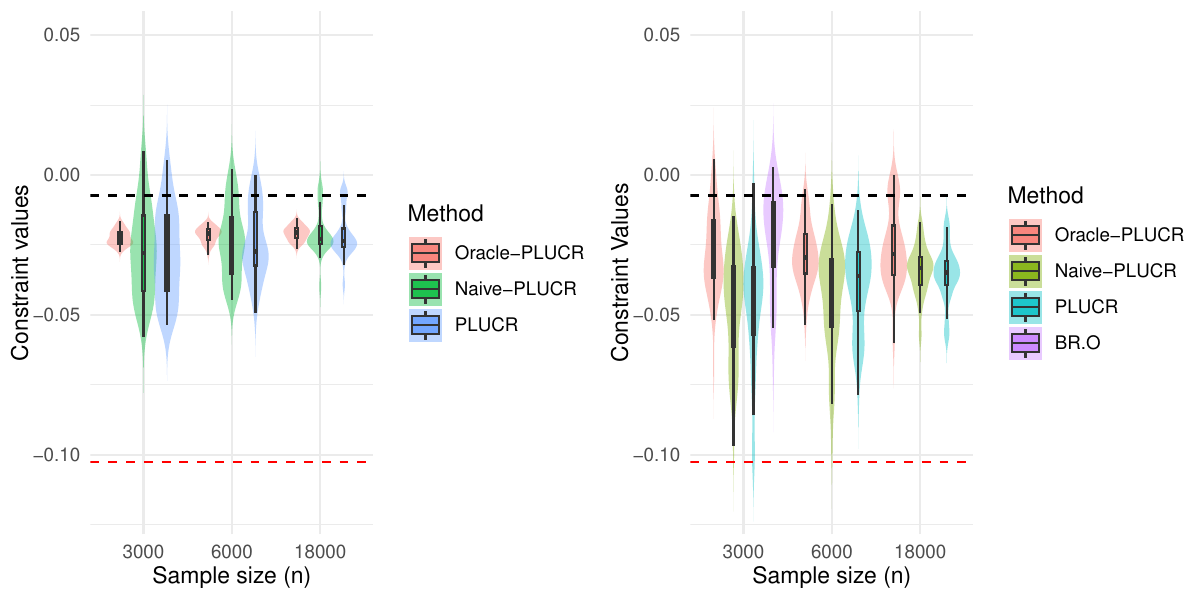}
    \subcaption{Constraint values}
    \label{fig:boxplot:realistic:constraint}
\end{minipage}
\begin{minipage}[t]{0.45\textwidth}
    \includegraphics[width=\textwidth]{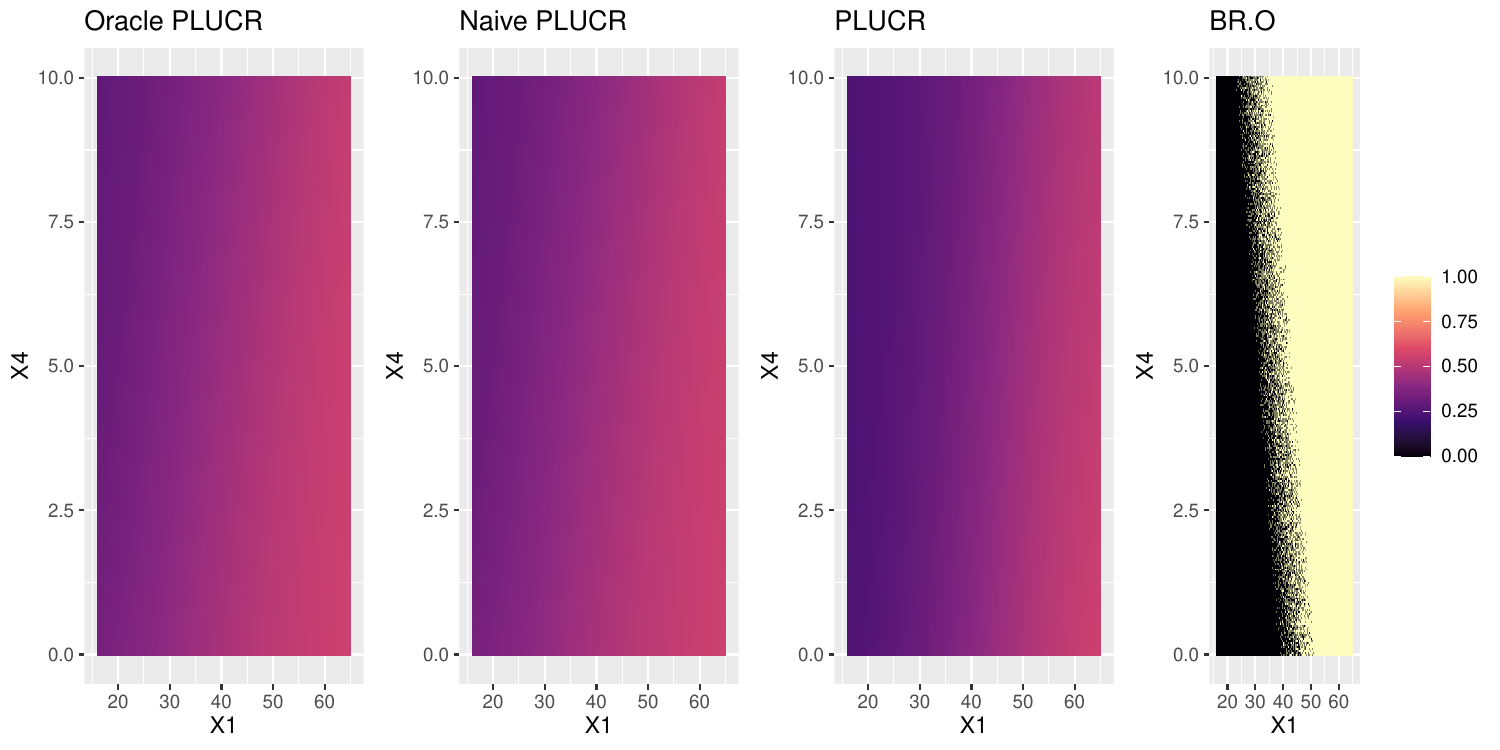}
    \subcaption{Treatment recommendations}
    \label{fig:realistic:treatment}
\end{minipage}
\caption{Results for the realistic scenario (BR-O did not converge 11 times)}
\label{fig:realistic}
\end{figure}

\paragraph{General remarks. } 
Throughout the simulation study, the variance of the policy values and constraint 
values decreased with sample size (Figures \ref{fig:linear:boxplot}, 
\ref{fig:boxplot:threshold:pv}, \ref{fig:boxplot:threshold:constraint}, 
\ref{fig:boxplot:satisfied:pv}, \ref{fig:boxplot:satisfied:constraint},
\ref{fig:boxplot:realistic:pv}, \ref{fig:boxplot:realistic:constraint}). 

Among the proposed algorithms, Oracle PLUC yields highest policy values 
while satisfying the constraint at lowest variance (Figures \ref{fig:linear:boxplot},
\ref{fig:threshold}, \ref{fig:satisfied}, \ref{fig:realistic}). While PLUC 
does not consistently improve upon naive PLUC's policy, it still shifts the barycenter to the right 
in the linear case. This shift indicates an improvement in policy value while still satisfying 
the constraint. In all other scenarios such shift is unnoticeable (Figures \ref{fig:qq:threshold:proba}, 
\ref{fig:qq:satisfied:proba}, \ref{fig:qq:realistic:proba}). This improvement is further subtle when 
comparing both recommendations (Figures \ref{fig:qq:threshold:decision}, \ref{fig:qq:satisfied:decision}, 
\ref{fig:qq:realistic:decision}).  In addition, such recommendations are consistently associated 
with higher policy values and constraint values close to the policy's counterpart with 
increased variances (Figures \ref{fig:linear:boxplot},\ref{fig:boxplot:threshold:pv}, 
\ref{fig:boxplot:threshold:constraint}, \ref{fig:boxplot:satisfied:pv},
\ref{fig:boxplot:satisfied:constraint},\ref{fig:boxplot:realistic:pv},
\ref{fig:boxplot:realistic:constraint} ). 

BR-O yielded unstable results, frequently failing to converge, likely, due to large sample size. 
Specifically, it failed in 33 simulations in linear example, 26 in the threshold simulation 
and 11 in the realistic simulations. Despite its instability, BR-O often achieves higher policy 
values (Figures \ref{fig:linear:boxplot}, \ref{fig:boxplot:realistic:pv}, \ref{fig:boxplot:realistic:constraint}) 
sometimes at the cost of violating the constraint. In fact, in Figure~\ref{fig:linear:boxplot}, the competitor is closest to 
the policy value maximizer ($x\mapsto \bone\{\Delta\mu_{0}(x)>0\}$), 
which confirms its disregard for the constraint. In the realistic example (Figure \ref{fig:realistic}), BR-O
attains comparatively high policy values, with performance approaching that of the first surrogate 
(which satisfies the constraint in this scenario). However, it violates the constraint in several isolated instances 
and fails to converge reliably (Figure \ref{fig:qq:realistic:decision}).

In the threshold scenario (Figure~\ref{fig:threshold}), the competitor
has a similar performance to Naive PLUC and PLUC. However, it outperforms 
them in the unconstrained setting (Figure~\ref{fig:satisfied}), 
achieving results closest to classical policy learning techniques, such 
as ``OWL" with double robust weights. 

The competitor's methodology was originally designed for RCTs, utilizing a DC algorithm suited for limited 
sample sizes and thus unadapted to our observational framework. Furthermore, fitting it to our study 
necessitated several critical adaptations, including providing the propensity score and estimating the 
baseline adverse event outcome. This extensive modification process is the likely reason the method 
occasionally fails to satisfy our constraint.

Although policies are associated with lower policy values with respect to recommendations, 
they enable an assessment of treatment assignment confidence. Across all scenarios 
(Figures \ref{fig:linear:treatment}, \ref{fig:threshold:treatment}, \ref{fig:satisfied:treatment},
\ref{fig:realistic:treatment}), we observe that the policies capture the complexity of 
the generated scenarios whereas the competitor draws a linear decision frontier. 
In further unconstrained settings with null treatment effects, our algorithms assign a 
0.5 probability to everyone, accurately capturing the inherent uncertainty. As a 
direct consequence, the recommendations assigned randomly the same treatment 
to everyone, whereas the competitor assigned treatment to half of the population. 
\end{document}